\documentclass[fleqn,usenatbib]{mnras}

\usepackage{txfonts}

\usepackage[T1]{fontenc}
\usepackage{ae,aecompl}

\newcommand{\iso}[2]{{}^{#2}\mbox{#1}}

\def\msun{$M_{\odot}$~}

\usepackage{graphicx}	

\title[MC uncertainties for $s$-process in low mass stars]{Uncertainties in
 $s-$process nucleosynthesis in low mass stars determined from Monte Carlo variations}

\author[G.Cescutti et al.]{G. Cescutti,$^{1}$\thanks{E-mail: cescutti@oats.inaf.it}\thanks{BRIDGCE UK Network; \url{www.bridgce.ac.uk}},
R. Hirschi,$^{2,3}$\footnotemark[2],
N. Nishimura, $^{4}$\footnotemark[2],
J. W. den Hartogh,$^{2,5}$\footnotemark[2],
\newauthor
T. Rauscher,$^{6,7}$\footnotemark[2],
A. St. J. Murphy$^{8}$\footnotemark[2]
and
S. Cristallo$^{9,10}$
\\
$^{1}$INAF, Osservatorio Astronomico di Trieste, I-34131 Trieste, Italy\\
$^{2}$Astrophysics group, Lennard-Jones Laboratories, Keele University, ST5 5BG, Staffordshire, UK\\
$^{3}$Kavli Institute for the Physics and Mathematics of the Universe (WPI),
University of Tokyo, 5-1-5 Kashiwanoha, Kashiwa, 277-8583, Japan\\
$^{4}$Yukawa Institute for Theoretical Physics, Kyoto University,
Kyoto 606-8502, Japan\\
$^{5}$Konkoly Observatory, Konkoly Thege Mikl\'{o}s \'{u}t 15-17, H-1121
Budapest, Hungary\\
$^{6}$Department of Physics, University of Basel, Klingelbergstr.\ 82, 4056 Basel, Switzerland\\
$^{7}$Centre for Astrophysics Research, University of Hertfordshire, College Lane, Hatfield AL10 9AB, UK\\
$^{8}$SUPA, School of Physics and Astronomy, University of Edinburgh, Edinburgh EH9 3FD, UK\\
$^{9}$INAF, Osservatorio Astronomico d'Abruzzo, I-64100 Teramo, Italy\\
$^{10}$INFN - Sezione di Perugia, I-06123, Perugia, Italy\\
}

\date{Accepted XXX. Received YYY; in original form ZZZ}

\pubyear{2018}

\begin{document}
\label{firstpage}
\pagerange{\pageref{firstpage}--\pageref{lastpage}}
\maketitle

\begin{abstract}
  The main s-process taking place in low mass stars produces about
  half of the elements heavier than iron. It is therefore very
  important to determine the importance and impact of nuclear physics
  uncertainties on this process. We have performed extensive nuclear
  reaction network calculations using individual and
  temperature-dependent uncertainties for reactions involving elements
  heavier than iron, within a Monte Carlo framework. 
  Using this
  technique, we determined the uncertainty in the main s-process
  abundance predictions due to nuclear uncertainties link to weak interactions and neutron captures on elements heavier than iron.
  We also identified the key nuclear reactions dominating these
  uncertainties.
We found that $\beta$-decay rate
uncertainties affect only a few nuclides near s-process branchings,
whereas most of the uncertainty in the final abundances is caused by
uncertainties in neutron capture rates, either directly producing or
destroying the nuclide of interest.  Combined total nuclear
uncertainties due to reactions on heavy elements are in general
small (less than 50\%). 
Three key
reactions, nevertheless, stand out because they significantly affect
the uncertainties of a large number of nuclides. These are
$^{56}$Fe(n,$\gamma$), $^{64}$Ni(n,$\gamma$), and
$^{138}$Ba(n,$\gamma$).  We discuss the prospect of reducing
uncertainties in the key reactions identified in this study with 
future experiments.

\end{abstract}




\section{Introduction}
\label{sec:intro}

Elements heavier than iron are mainly produced via neutron captures
because the significant Coulomb barrier of these elements
inhibits charged-particle captures.  It is well established that the
astrophysical origin of the majority of nuclides beyond Fe requires at
least two neutron-capture processes \citep{Cameron1957,burbidge1957},
the so-called slow process (s-process) and rapid
process (r-process): for the slow process the neutron-capture
timescale is generally longer than the $\beta$-decay time, whereas the
opposite is true for the rapid process.

\label{sec-evolution}
In this work, we focus on the main component of the s-process, which
takes place during the asymptotic giant branch (AGB) phase in low mass
stars, see, e.g., \citet{2001busso}, \citet{2002Abia},
\citet{2008sneden} and \citet{2009zamora}. The main neutron source for
the s-process is the reaction $^{13}$C($\alpha$,n)$^{16}$O \citep[for
a review of the main s-process, see][]{Kappeler2011}. This reaction is
activated during the thermally pulsing AGB phase, taking place after
central helium burning in low-mass stars. During this phase, energy
production is dominated by the burning hydrogen shell and the helium
shell flash events (thermal pulses, TPs), first described by
\cite{1965schw_tp}.  The thermal pulse starts when enough helium has
been deposited by the hydrogen burning shell on top of the degenerate
CO core and the helium shell becomes compressed and heated, see
\cite{falk_ARAA}. The helium shell ignites in an explosive way as the
layers are degenerate, leading to a large energy flux and the
extinction of the hydrogen burning shell. This large energy flux
creates the pulse driven convective zone (PDCZ) in the intershell, the
area in between the core and the helium shell, which is expanding as a
result of this energy flux. The expansion cools the region, allowing
the helium shell to cool. The helium shell is now burning helium in a
stable radiative manner until it runs out of fuel again. While the
intershell region expands and cools, the convective envelope
deepens. If the convective zone reaches sufficiently deep layers, it
dredges up material enriched by the last PDCZ, a process called third
dredge-up (TDU). Afterwards, the hydrogen shell re-ignites and the whole 
cycle repeats itself until the entire hydrogen-rich envelope has been
lost by stellar winds. At the deepest point of penetration of the
convective envelope, fresh protons are injected in the intershell,
which is rich in $\iso{C}{12}$. Incomplete CNO cycling leads to a
significant production of $\iso{C}{13}$ in a narrow region below the
convective envelope, which is often referred to as the 
$\iso{C}{13}$-pocket, see \cite{Gallino98}, \cite{falk_ARAA},
\cite{Straniero06} and the first description by \cite{1976Iben} for
more details. As this region later contracts as the thermal pulse (TP) cycle
proceeds, it heats up and a large number of neutrons are released by
the neutron source reaction $^{13}$C($\alpha$,n)$^{16}$O in a
radiative (non-convective) layer \citep{Straniero95}. A smaller contribution to the
neutron flux comes from the $^{22}$Ne($\alpha$,n)$^{25}$Mg neutron
source, which is activated in intermediate mass stars at the bottom of
the PDCZ \citep{Abia01} and, thus, releases neutrons in a convective environment. 
 We will refer to the PDCZ phase as ``TP'' phase in the rest of the paper.

Low mass AGB stars are the sites for the main component of the
s-process, i.e. elements between strontium and lead. The second
component of the s-process (called weak component) takes place at the
end of core helium burning and at the start of carbon (shell) burning
in massive stars. Typically, it produces elements up to the Sr peak
but depending on the metallicity and the mixing induced by rotation
can also produce heavier nuclides \citep[see][]{2016Frischknecht,Cescutti16,Prantzos18}.
The neutron source for the weak s-process is
$^{22}$Ne($\alpha$,n)$^{25}$Mg,

There are several well known uncertainties concerning the s-process
production in low-mass stars.  On the astrophysical side, the most
important one is the general properties of the $^{13}$C-pocket and in
particular its formation, see \cite{Cristallo15}, \cite{Battino16} and
\cite{Trippella16} for a discussion and references.  On the nuclear
reaction side, \citet{Koloczek16} (Ko16 hereinafter), recently
reviewed the impact of current nuclear uncertainties considering both
the $^{13}$C-pocket and TP conditions.  As expected, they identify the
neutron source reactions mentioned above as key reactions. They find
that their uncertainties strongly affect the s-process production as
do the competing reactions \citep[see, e.\,g., the discussion in][for
$^{22}$Ne$(\alpha,\gamma)^{26}$Mg]{Nishimura14}.  Neutron poison
reactions, such as $^{14}$N(n,p), $^{13}$C(n,$\gamma$),
$^{16}$O(n,$\gamma$), $^{22}$Ne(n,$\gamma$) for the $^{13}$C-pocket
conditions, and $^{22}$Ne(n,$\gamma$) and $^{25}$Mg(n,$\gamma$) for
the TP conditions, were also found to have a strong effect.  The Ko16
study also identified a wide range of neutron captures as well as a
few weak reactions on elements heavier than and including iron.  
  When varying charged-particle reactions on light nuclides (as done
  in the Ko16 study), it may be necessary to conduct these sensitivity
  studies using full stellar evolution models. For instance, the
  adoption of a lower rate for the $^{13}$C($\alpha$,n) $^{16}$O
  reaction could lead to the ingestion of some unburnt $^{13}$C in the
  PDCZ, with important consequences on the on-going s-process
  nucleosynthesis (Cristallo S. et al, ApJ submitted). In this study,
  we only explore uncertainties in neutron captures and beta decays on
  intermediate and heavy isotopes. We thus do not expect feedback
  effects from rate variations on the structure and the adopted
  post-processing approach is appropriate.  Our approach to vary
reaction rates is different from that of Ko16. We vary simultaneously
all reaction rates in a Monte Carlo (MC) framework rather than one
reaction at a time. Furthermore, we use temperature-dependent
uncertainties based both on experimental and theoretical studies as we
have already done for several other processes: the s-process in
massive star, $\gamma$-process in core collapse SNe and
$\gamma$-process in supernovae type Ia
\citep{Nishimura17a,Rauscher16,Nishimura18}. We will compare our
findings to those of Ko16 and comment further on similarities and
differences of methods and results in the discussion section.

The paper is organised in the following way. In Section~2, we describe
the astrophysical model used in this study as well as the MC framework
PizBuin. In Section~3, we present the results of our sensitivity study
and the list of key rates identified.  We also discuss these key rates
and the prospects to reduce their uncertainties with future
experiments. In Section~4, we give our conclusions.

\section{METHODS}

In this Section, we describe the main ingredients of our calculations:
the thermodynamic trajectories used for the $^{13}$C-pocket, the TP
phase and the Monte Carlo PizBuin framework.  The basic features of
s-process nucleosynthesis and the uncertainties of (n,$\gamma$) and
weak rates determination are also summarised.

\subsection{Astrophysical model}

The complete evolution of low-mass stars is complex, especially during
the TP-AGB phase. A full one-dimensional (1D) stellar model can
require more than 100,000 time steps and over one thousand spatial
zones to be simulated completely from start to finish. It is thus not
feasible to repeat such simulations 10000 times as required by the MC
procedure to complete a sensitivity study. We thus have to approximate
the thermodynamic conditions inside the star with a trajectory
following the key phase that we are studying. We start with the
$^{13}$C-pocket case. The fact that this phase occurs under radiative
conditions (rather than convective) makes it feasible to approximate
it with a carefully selected single trajectory. This
trajectory does not follow exactly what happens in real stars but, as
shown below, provides the conditions that lead to an s-process
production similar to that predicted using full stellar models.

\begin{figure}
\includegraphics[width=\columnwidth]{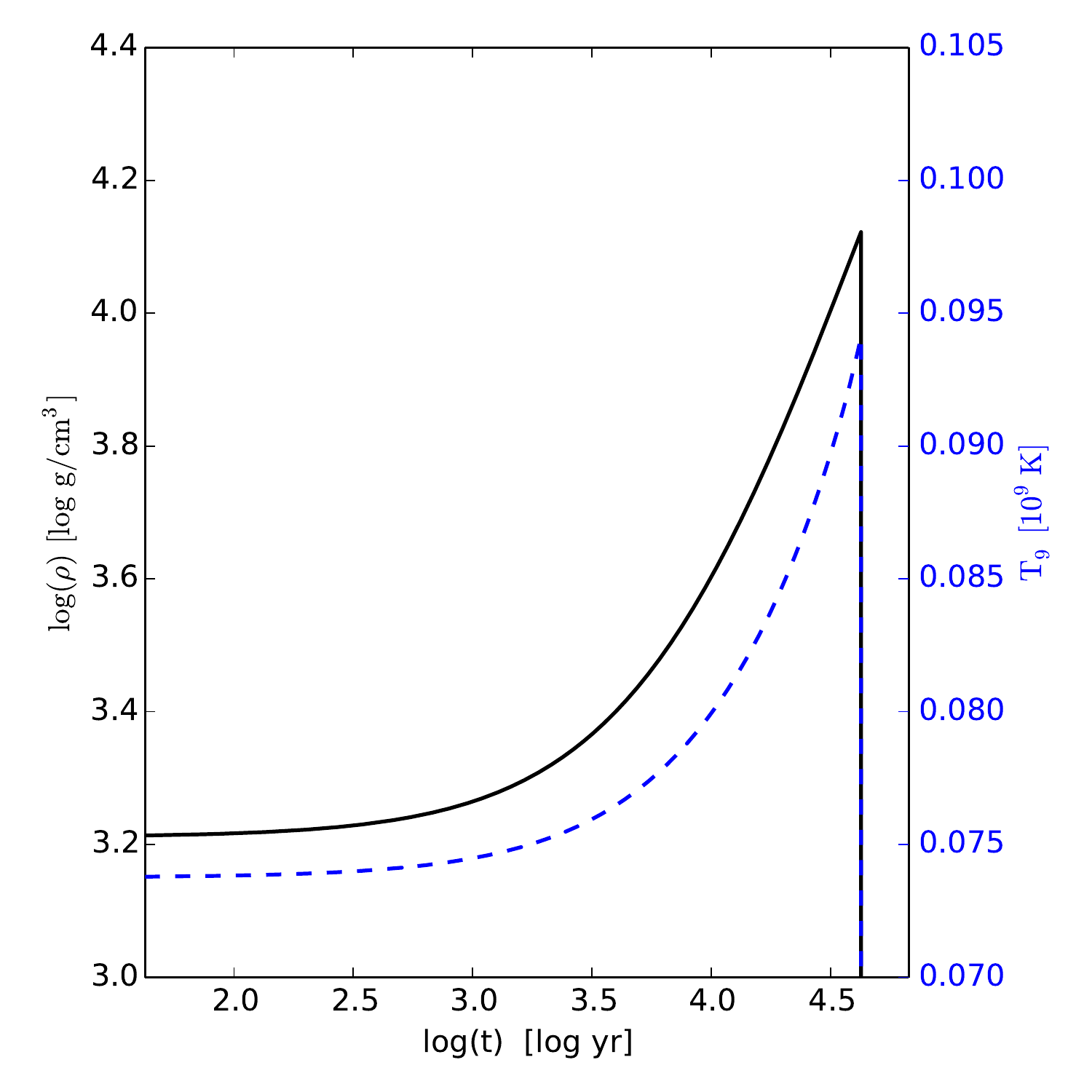}
\caption{Time evolution of the temperature [GK]  (blue dashed line)
 and density (solid black line)  of the trajectory used for
the $\iso{C}{13}$ pocket in this study.}
\label{traj}
\end{figure}

The trajectory used in this work was extracted from a 3 $M_{\odot}$,
$Z=0.014$ (solar metallicity) stellar evolution model, calculated with MESA, revision number 6208
\citep{2011_MESA_1}. The trajectory
was taken from the $^{13}$C-pocket following the 6$^\mathrm{th}$ TP.
The temperature and density profiles of the trajectory are shown in Fig. \ref{traj}.



\begin{figure}
\includegraphics[width=\columnwidth]{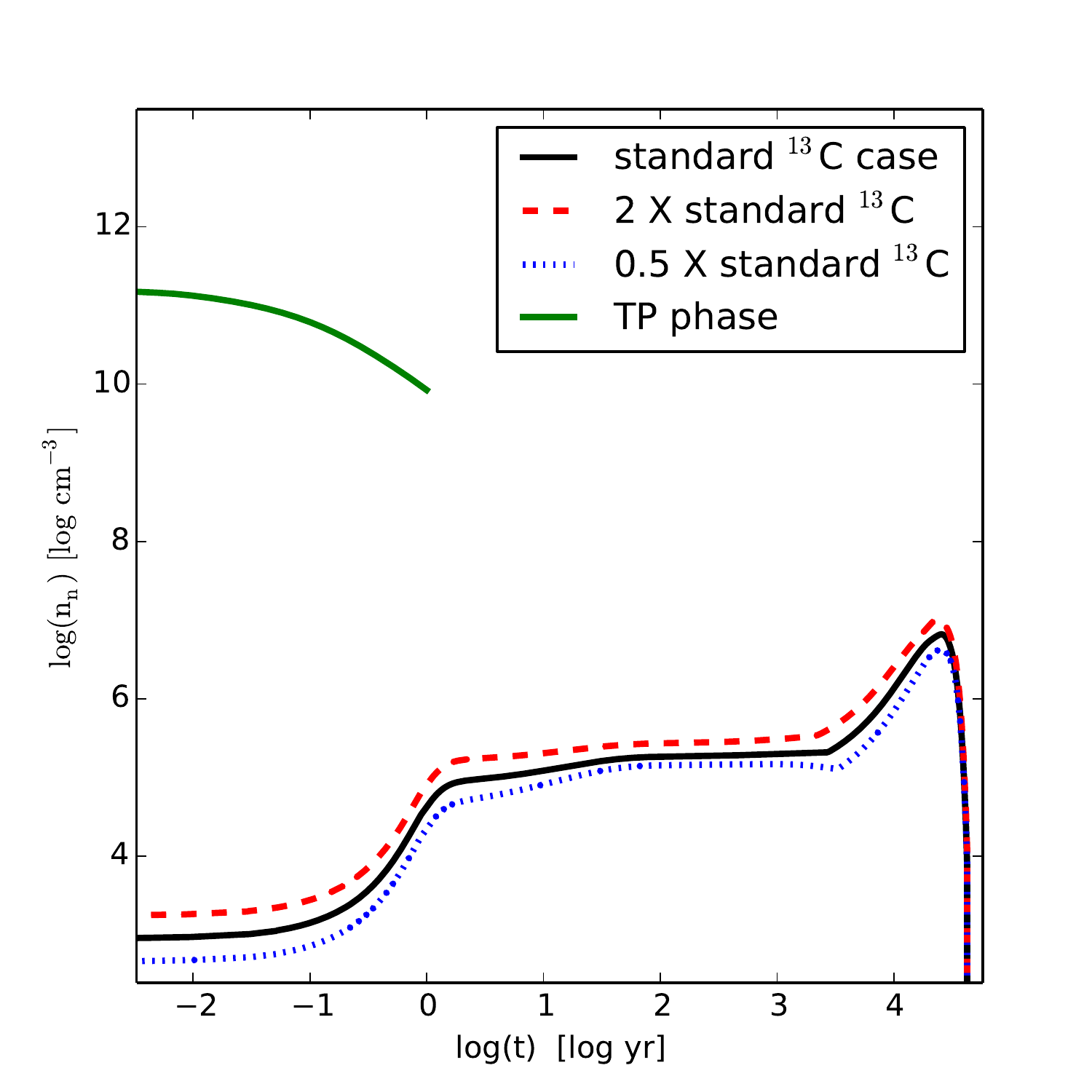}
\caption{Time evolution of the neutron density [cm$^{-3}$] for the
  three initial $\iso{C}{13}$ abundances considered in the
  $^{13}$C-pocket and the TP phase.}
\label{traj2}
\end{figure}

The trajectory starts after the $^{13}$C-pocket has formed. The
formation of the $^{13}$C-pocket is the main uncertainty on the
astrophysical side as mentioned above. The most advanced 3D models of
stellar evolution are starting to resolve this phase in detail and
\citet{Battino16} have shown that using prescriptions in 1D stellar
models guided by these 3D hydro simulations give promising results.
Most nucleosynthesis computations to-date, however, typically take
into account the $\iso{C}{13}$ pocket either by directly inserting a
specific proton abundance profile below the convective envelope
\citep[e.g.,][]{Karakas16} or by assuming the mixing process that
leads to it \citep{Cristallo11,Trippella16}. In this study, we
artificially increase the $\iso{C}{13}$ abundance, mimicking in this
way the enhancement of $\iso{C}{13}$ due to the injection of
protons. We explored variations in the initial content of
$\iso{C}{13}$ that lead to a s-process production similar to the one
predicted by full stellar models. Our tests revealed that an initial
mass fraction of $\iso{C}{13}$ of
$X_{^{13}\mathrm{C}} = 1.95\times 10^{-2}$ enables us to produce a
typical s-process pattern with the above trajectory. We call our
calculations using this value of $\iso{C}{13}$ our ``standard''
case. To fully explore the range of conditions found in low-mass
stars, we also used two additional initial $\iso{C}{13}$ abundances,
one in which the standard initial abundance of $\iso{C}{13}$ is halved
(``0.5 $\times$ $\iso{C}{13}$'' case), whereas in the other one it is doubled
(``2 $\times$ $\iso{C}{13}$'' case). The variations in the neutron densities for
the three cases are shown in Fig. \ref{traj2}.  The initial
composition for our calculations is given in Table 1 for nuclides for
which we do not use the standard solar composition. Besides the change
in $\iso{C}{13}$ explained above, the other initial abundances for our
calculations were extracted from the same stellar evolution model as
the trajectory.

Besides the $\iso{C}{13}$ pocket, neutron captures also take
  place at the bottom of the TP-driven convective zone (PDCZ) as
  explained in the introduction. In low mass AGB stars (M$<4$
  M$_{\odot}$, which dominate the overall s-process production given
  to the initial mass function), only a small production of
  neutron-rich isotopes is expected from the TP phase, such as
  $\iso{Zr}{96}$ (otherwise not produced during the
  radiative burning of the $\iso{C}{13}$ pocket).  Given that the TP
  only contributes a short neutron burst and has a very small
  contribution to the overall s-process prodcution, we approximated
  the TP conditions with a single-zone trajectory as in Ko16. The
  trajectory lasts for one year, with a constant temperature of 0.245
  GK and a constant density of 5$\times$10$^{3}$ [g/cm$^{3}$]. The
  initial abundances are summarised in Table \ref{Initial_comp_2} for
  light elements; for the other elements, the final abundances of the
  standard $\iso{C}{13}$ pocket were diluted by a factor of twenty to
  take into account the diluting effect of the PDCZ. The chosen
  trajectory, combined with the initial composition described above,
  is able to roughly reproduce typical isotopic compositions obtained
  during TP by more complex stellar evolution codes
  \citep{Cristallo15}.  The time evolution of the neutron density for
  the TP phase is also shown in Fig. \ref{traj2}.

\begin{table}
\caption{Initial composition for the nuclei that differ from the solar
  composition during the $\iso{C}{13}$ pocket.}\label{Initial_comp}
\begin{tabular}{l|r|l|r|}
\hline
nuclei & mass fraction & nuclei & mass fraction \\
\hline
$\iso{H}{1}$ & 1.08$\times 10^{-29}$        &    $\iso{O}{16}$& 4.32$\times 10^{-2     }$     \\       
$\iso{H}{2}$ & 1.43$\times 10^{-5}$         &    $\iso{O}{17}$& 2.80$\times 10^{-6     }$     \\       
$\iso{He}{3}$& 4.49$\times 10^{-5       }$  &    $\iso{O}{18}$& 4.83$\times 10^{-8     }$     \\       
$\iso{He}{4}$& 4.58$\times 10^{-1       }$  &    $\iso{F}{19}   $& 1.79$\times 10^{-9     }$  \\       
$\iso{Li}{6}$& 6.44$\times 10^{-10     }$   &    $\iso{Ne}{20}  $&  1.23$\times 10^{-3    }$  \\       
$\iso{Li}{7}$& 9.15$\times 10^{-9     }$    &    $\iso{Ne}{21}  $&  3.09$\times 10^{-6    }$  \\       
$\iso{Be}{9}$& 1.68$\times 10^{-10     }$   &    $\iso{Ne}{22}  $&  3.12$\times 10^{-2    }$  \\       
$\iso{B}{10}$& 7.75$\times 10^{-10     }$   &    $\iso{Na}{23}  $&  3.10$\times 10^{-5    }$  \\       
$\iso{B}{11}$& 3.43$\times 10^{-9     }$    &    $\iso{Mg}{24}  $&  5.86$\times 10^{-4    }$  \\       
$\iso{C}{12}$& 3.31$\times 10^{-1     }$    &    $\iso{Mg}{25}  $&  7.73$\times 10^{-5    }$  \\       
$\iso{C}{13}$& 1.95$\times 10^{-2     }$    &    $\iso{Mg}{26}  $&  8.86$\times 10^{-5    }$  \\       
$\iso{N}{14}$& 5.11$\times 10^{-3     }$    &    $\iso{Al}{27}  $&  5.91$\times 10^{-5    }$  \\       
$\iso{N}{15}$& 9.02$\times 10^{-8     }$    &    $\iso{Si}{28}  $&  6.49$\times 10^{-4    }$  \\       

\hline

\end{tabular}
\end{table}

\begin{table}
\caption{Initial composition for the light nuclei
during the TP phase.}\label{Initial_comp_2}
\begin{tabular}{l|r|l|r|}
\hline
nuclei & mass fraction & nuclei & mass fraction \\
\hline
$\iso{H}{1}$ & 5.35$\times 10^{-23}$        &    $\iso{O}{16}$& 6.00$\times 10^{-3     }$     \\       
$\iso{H}{2}$ & 1.37$\times 10^{-5}$         &    $\iso{O}{17}$& 1.00$\times 10^{-10     }$     \\       
$\iso{He}{3}$& 4.29$\times 10^{-5       }$  &    $\iso{O}{18}$& 1.00$\times 10^{-10    }$     \\       
$\iso{He}{4}$& 7.91$\times 10^{-1       }$  &    $\iso{F}{19}   $& 1.5$\times 10^{-5     }$  \\       
$\iso{Li}{6}$& 6.44$\times 10^{-10     }$   &    $\iso{Ne}{20}  $&  7.00$\times 10^{-4    }$  \\       
$\iso{Li}{7}$& 8.75$\times 10^{-9     }$    &    $\iso{Ne}{21}  $&  1.00$\times 10^{-5    }$  \\       
$\iso{Be}{9}$& 1.68$\times 10^{-10     }$   &    $\iso{Ne}{22}  $&  1.50$\times 10^{-2    }$  \\       
$\iso{B}{10}$& 7.41$\times 10^{-10     }$   &    $\iso{Na}{23}  $&  1.80$\times 10^{-4    }$  \\       
$\iso{B}{11}$& 3.28$\times 10^{-9     }$    &    $\iso{Mg}{24}  $&  7.00$\times 10^{-4    }$  \\       
$\iso{C}{12}$& 1.75$\times 10^{-1     }$    &    $\iso{Mg}{25}  $&  7.00$\times 10^{-5    }$  \\       
$\iso{C}{13}$& 1.50$\times 10^{-7     }$    &    $\iso{Mg}{26}  $&  1.00$\times 10^{-4    }$  \\       
$\iso{N}{14}$& 5.00$\times 10^{-3     }$    &    $\iso{Al}{27}  $&  7.00$\times 10^{-5    }$  \\       
$\iso{N}{15}$& 5.00$\times 10^{-6    }$    &    $\iso{Si}{28}  $&  5.00$\times 10^{-4    }$  \\       

\hline

\end{tabular}
\end{table}

\begin{figure*}
\includegraphics[width=\textwidth]{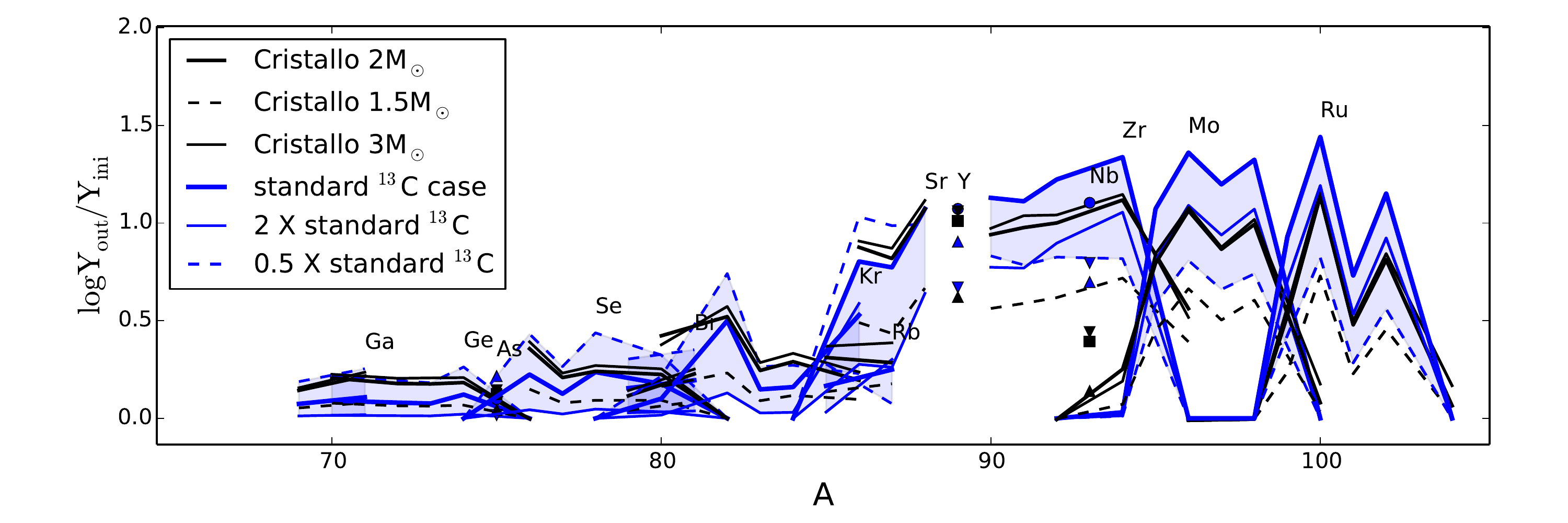}
\caption{Production factors as a function of the atomic mass for
  elements from Ga to Ru during the $\iso{C}{13}$ pocket.  Comparison
  between the results obtained for a 1.5 (dashed black line - black 
  triangle for monoisotopic elements), 2 (thick
  solid black line - black square) and 3\,\msun (thin solid black
  line - upside down black triangle) by C11 and
  the trajectory considered in this study with the 3 different initial
  abundance for $\iso{C}{13}$: standard case (thick solid blue line -
  blue dot),
  half (dashed blue line - blue triangle) and double (thin solid blue
  line - upside down blue triangle) the
  standard case. The blue area highlights the range of s-process
  production obtained using the three cases for the initial
  $\iso{C}{13}$ abundance.  Note that we applied a dilution factor,
  $f$, to our results to compare them to production factors of C11
  (see eq. \ref{dilution}).}
\label{fig:low_nc}
\end{figure*}

\begin{figure*}
\includegraphics[width=\textwidth]{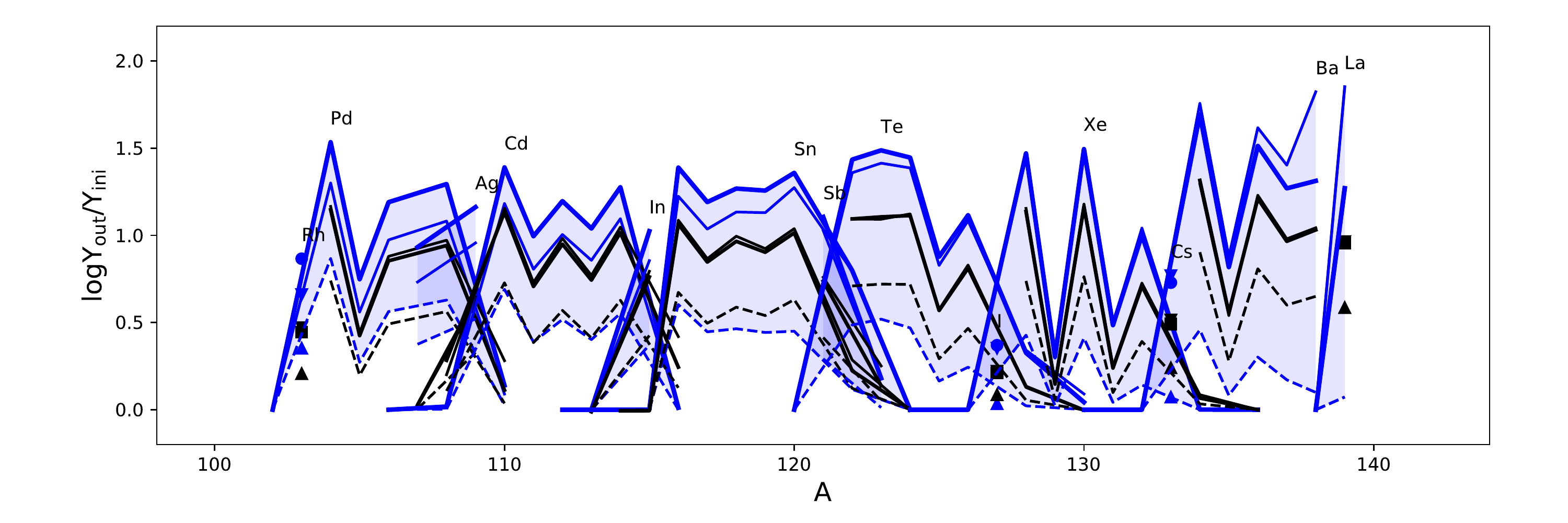}
\caption{Same as Fig.\,\ref{fig:low_nc} for elements from Rh to La.}
\label{fig:high_nc}
\end{figure*}

\begin{figure*}
\includegraphics[width=\textwidth]{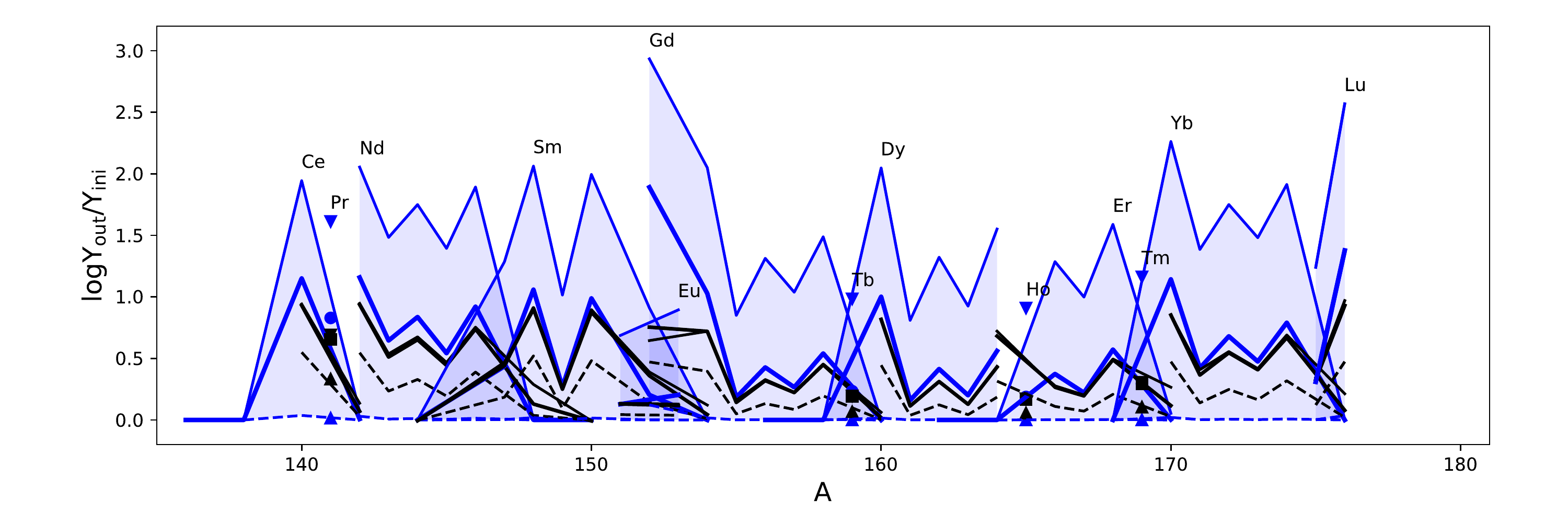}
\caption{Same as Fig.\,\ref{fig:low_nc} for elements from Ce to Lu.}
\label{fig:Eu_nc}
\end{figure*}
\newpage
\begin{figure*}
\includegraphics[width=\textwidth]{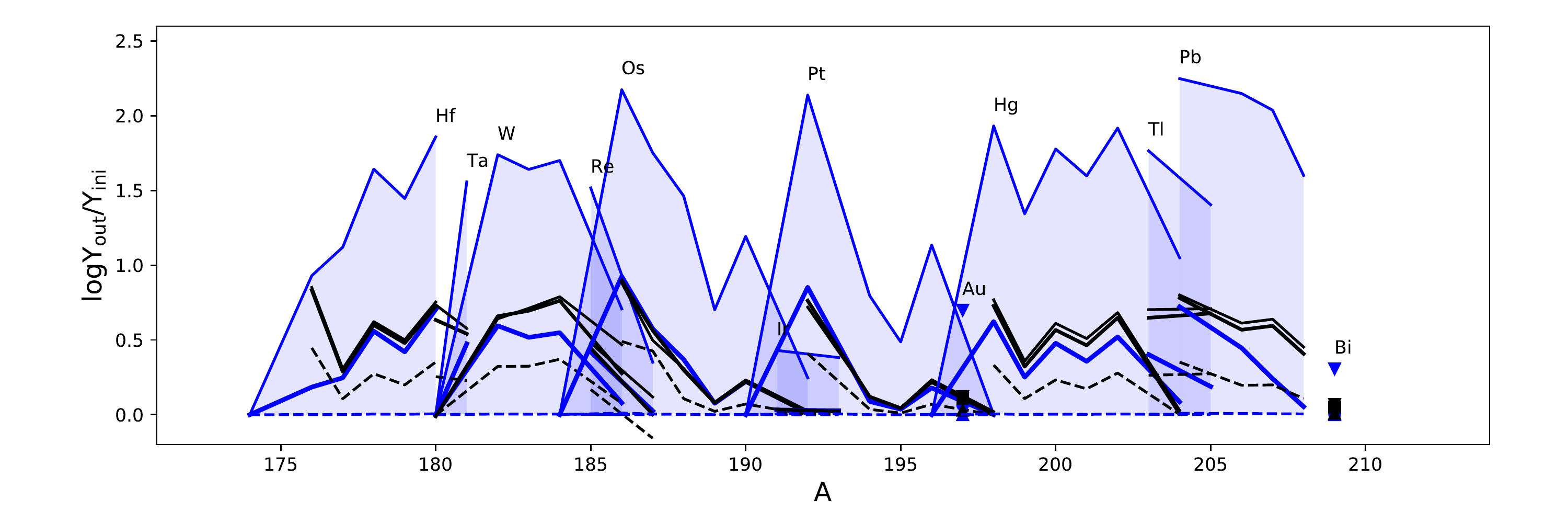}
\caption{Same as Fig.\,\ref{fig:low_nc} for elements from Hf to Bi.}
\label{fig:vhigh_nc}
\end{figure*}

\begin{figure*}
\includegraphics[width=\textwidth]{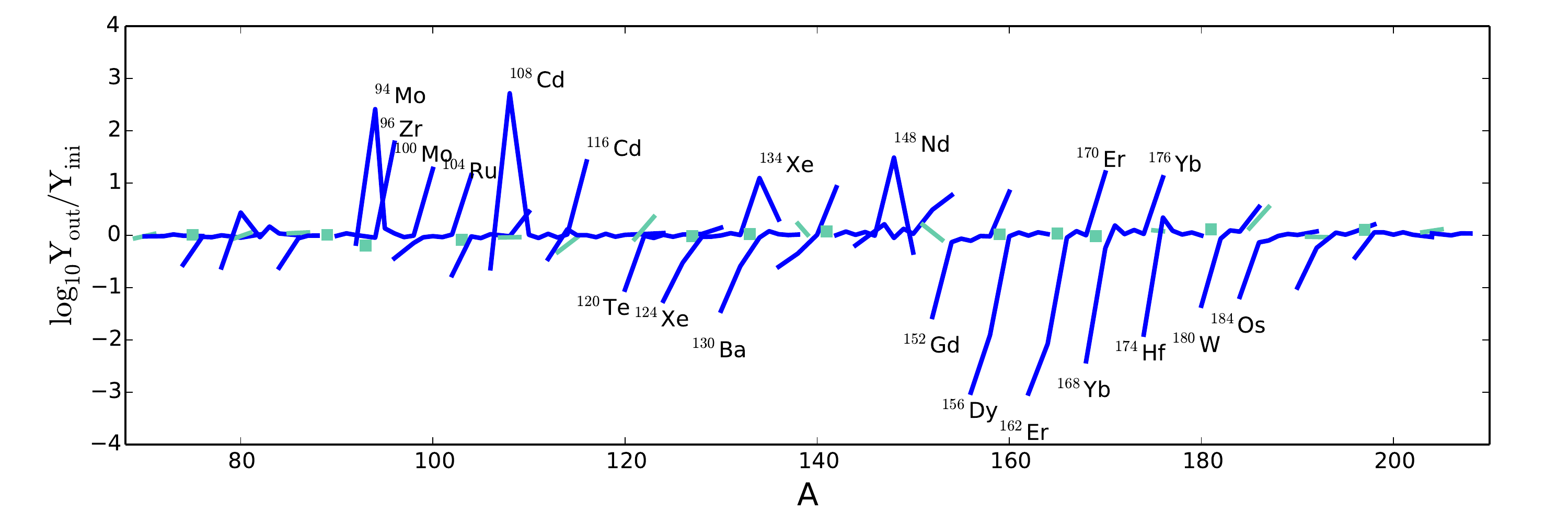}
\caption{Production factors (final abundances divided by the initial
  abundances used for the TP trajectory) as a function of the atomic
  mass for elements from Ga to Pb during the TP phase. The blue lines
  connect elements with even atomic number, other elements are
  presented in cyan.  Isotopes with a production factor greater than
  10 or less than 0.1 are indicated.}
\label{fig:TP}
\end{figure*}

\subsubsection{Comparison to the \citet{Cristallo11} yields}\label{sec:C11}

To validate our trajectory and initial composition combination, we
compared the final abundances of our calculations at solar metallicity
to the s-process pattern determined using full stellar models by
\citet{Cristallo11} \citep[C11 hereinafter, but see also for
details][]{Cristallo09,Cristallo07,Straniero06} for stars of 1.5, 2
and 3\msun at solar metallicity.  Our calculations used a single
trajectory covering a single 13C-pocket whereas the C11 computed full
stellar models. The s-process produced in the $^{13}$C-pocket in their
models is thus diluted into the convective envelope following the
TDU.  We thus used a dilution factor $f$ to compare our
final abundance to theirs.  We set the dilution factor $f$ to match
the production of $^{88}$Sr in our computations to the production in
the 2\,\msun model of C11:
\begin{equation}
\frac{f \, ^{\mathrm{^{88}Sr}} \rm Y^{\rm traj}_{\rm out}+(1-f) \,
  ^{\mathrm{^{88}Sr}} \rm Y^{traj}_{\rm ini}}{^{\mathrm{^{88}Sr}}\rm
  Y^{\rm traj}_{\rm ini}}=\frac {^{\mathrm{^{88}Sr}} \rm Y^{\rm
      C11}_{\rm out}} {^{\mathrm{^{88}Sr}} \rm Y^{\rm C11}_{\rm ini}}
\end{equation}
\label{dilution}
where $^{\mathrm{^{88}Sr}}Y^{\rm traj}_{\rm ini}$ is the initial abundance of Sr in our
trajectory, $^{\mathrm{^{88}Sr}}Y^{\rm traj}_{\rm out}$ is the final abundance (same for the C11 production factors).
Our final diluted abundances are compared to the C11 production factors in Figs.\ \ref{fig:low_nc},
\ref{fig:high_nc}, \ref{fig:Eu_nc}, and \ref{fig:vhigh_nc}.

The good overall agreement between our standard model and the C11
yields shows that our trajectory is adequate to determine the key
nuclear reactions that strongly affect s-process predictions. 
Nevertheless, it is also clear that a single trajectory -- as the one we adopt -- is
not able to reproduce the full range of conditions occurring in
low-mass stars. 
We thus added to our investigations two other initial
abundances for $\iso{C}{13}$, with the aim of  covering a wider
  range of conditions of the s-process. It also allows us to
determine the sensitivity of our results to the thermodynamic
conditions and neutron flux in particular.
Furthermore, since the main difference between models of main
s-process production is the ratio between seed (given by the
metallicity, mainly iron) and the neutrons (given by the $\iso{C}{13}$
present at the start of the calculation) we are also investigating in
some respect the metallicity dependence of this process. For the
purpose of determining key rates, it is not necessary to match exactly
the final results of \citet{Cristallo11}. More important is to
investigate the full range of neutron fluxes and the activated
branches.  Figs. \ref{fig:low_nc}, \ref{fig:high_nc}, \ref{fig:Eu_nc},
and \ref{fig:vhigh_nc} show that -- excluding rare cases -- the
results we obtain with the three initial $\iso{C}{13}$ contents cover
the full range of results obtained by C11 and thus prove that our
approach is suitable to determine the uncertain key rates for
thes-process in low-mass stars.   It is more difficult to apply a
  similar approach in the case of the TP phase. This is because the
  overall production during the TP is tiny compared to that occurring
  during the $\iso{C}{13}$ pocket. Therefore there is no way to
  directly compare the output from this phase and the final
  theoretical results. We show in Fig. \ref{fig:TP} the production
  factors of all the considered nuclei.  There is significant
  production of only a few isotopes, and of these, most are not
  produced by our $\iso{C}{13}$ trajectory (see Figs \ref{fig:low_nc} -
  \ref{fig:vhigh_nc}). These neutron-rich isotopes are in fact
  expected to mainly be produced during the TP phase in AGB stars
\citep{Gallino98}.

\subsection{Monte-Carlo procedure}\label{sec:MC}

The thermodynamic trajectory described above was post-processed using
the PizBuin code suite. This suite consists of a fast reaction network
and a parallelized Monte Carlo driver. We followed the same procedure
as presented in detail in \citet{Rauscher16}.  The nucleosynthesis
calculation was repeated 10,000 times, with different rate variation
factors each time, and the combined output was analysed
subsequently. The simultaneous variation of rates is superior to a
decoupled variation of individual rates as performed in the past and
in Ko16 because neglecting a combined change in rates may lead to an
overemphasis of certain reactions and an overestimation of their
impact on the total uncertainty \citep{Rauscher16,Rauscher17}.

In our method, we define key
rates to be those dominating the uncertainty of a given final
abundance. By this definition, reducing the uncertainty of a key rate
will also considerably decrease the uncertainty in the final
abundance of a given nuclide. The identification of key rates are obtained by examining
the correlation between a change in a reaction rate and the change of an
abundance. We used the
Pearson product moment correlation coefficient to
quantify these correlations. Positive values of the Pearson coefficients, $r$,
indicate a positive correlation between rate change and abundance
change, whereas negative values signify an inverse correlation, i.e.,
the abundance decreases when the rate is increased. The larger the
absolute value of the Pearson coefficient, the stronger the
correlation. As in \citet{Rauscher16}, \citet{Nishimura17a} and \citet{Nishimura18}, a level 1 key rate is identified by 
$r\geq 0.65$.  
Each astrophysical reaction rate involving elements from Fe to Bi was varied within its own uncertainty range. We used the same variation factor for forward and reverse rates as they are connected by detailed balance. The
uncertainty range used is temperature dependent and constructed
from a combination of the experimental uncertainty (if the rate has been measured) for
target nuclei in their ground states and a theoretical uncertainty for
predicted rates on nuclei in thermally excited states. Theory
uncertainties were different depending on the reaction type and can be
asymmetric. 
The reaction network consisted of 943 isotopes including all reactions
relevant to the s-process, i.e., fusion reactions of lighter
nuclei as well as (n,$\gamma$) reactions, electron captures, and $\beta$ decays for heavier nuclei.
The standard rate set and uncertainties used in this study are the
same as in \citet{Rauscher16} and \citet{Nishimura17a}. 
Rates for neutron-, proton-, and $\alpha$-induced reactions were a
combination of theoretical values by 
\citet{adndt00} supplemented by experimental rates taken from
\citet{kadonis} and \citet{cyburt}; decays and electron captures were taken from a
REACLIB file compiled by \citet{freiburghaus} and
supplemented by rates from \cite{taka} and \citet{gor99} 
as provided by \citet{NetGen05} and \citet{NetGen13}.

\subsubsection{Nuclide selection for the key rate determination}

Almost all the stable nuclides up to $\iso{Bi}{209}$ have an s-process
contribution. We might therefore present key rates for almost 250
isotopes. If an isotope, however, constitutes a negligible fraction of
the total elemental abundance, improving its key reaction rates would
not make a difference to the total production of an element. We thus
had to establish a selection procedure for nuclides to be presented in
our key rate determination.  One possible selection method is to
consider a threshold in the production factors. This method failed
because it was not possible to determine a suitable threshold for the
; either too few or too many nuclides were excluded and the resulting
exclusions were rather random.  Concerning the analysis of the
$\iso{C}{13}$ pocket phase, we therefore decided to analyse only
isotopes that contribute at least 10\% to the final total mass of the
element. This selection method yields a list of 109 nuclides, most of
which are listed in the Tables below (note that only nuclides with a
key rate are listed). In addition, we considered seven more nuclides
$\iso{Sr}{86}$, $\iso{Sr}{87}$, $\iso{Cd}{110}$, $\iso{Te}{123}$,
$\iso{Ba}{134}$, $\iso{Sm}{148}$, and $\iso{Hf}{176}$. Although their
total production factor is below the 10\% threshold explained above,
they are s-only nuclides and are thus worth investigating.  
  Regarding the analysis of the TP phase, a similar procedure fails to
  select all the isotopes that characterised this production; we
  therefore select all the isotopes not destroyed whose production is
  above 1\% of their production during the $\iso{C}{13}$ pocket
  phase. In this way, we exclude isotopes that have negligible
  production during the TP phase. To the isotopes selected in this
  way, we have added the s-only nuclides and the final list contains 
  58
  nuclei (see Table \ref{tabTPunc}).

\section{RESULTS AND DISCUSSION}
As explained above, we used the PizBuin code suite to determine the
uncertainty in the final s-process abundances due to uncertainties of
reactions involving heavy elements as well as the key reactions
dominating these uncertainties. The total uncertainty of the final
abundances are given in Table\,\ref{tab-C13-unc} for the $\iso{C}{13}$
pocket and in Table\,\ref{tabTPunc} for the TP phase and shown in
Figs. \,\ref{L1} and \ref{MC_TP}, respectively.

\subsection{Total uncertainties}
 
\begin{table}
\centering
\caption{ Uncertainties in the final abundance of
  {\it s}-process nuclides from the MC calculation for the standard 
$\iso{C}{13}$ pocket phase.
  The column labeled ``Level'' indicates the level of the first key
  reaction found. The remaining columns show 
uncertainty factors for variations Up and Down, the values of which
are  $Y(95\%)/Y_{\rm peak}$ and $Y(5\%)/Y_{\rm peak}$, respectively. 
They enclose a 90\% probability interval, as shown in Fig.~\ref{L1}. } 
\label{tab-C13-unc}
\begin{tabular}{@{}lcrrlcrr}
  \hline                                                           
Nuclide & Level & Up & Down             & Nuclide & Level & Up & Down \\                
\hline                                       
$\iso{Ga}{69}$ &  1 &  1.13 & 0.896    &            $\iso{Ba}{138}$ &  2 &  1.08 & 0.941 \\
$\iso{Ga}{71}$ &  1 &  1.24 & 0.918    &              $\iso{La}{139}$ &  1 &  1.34 & 0.922 \\
$\iso{Ge}{70}$ &  1 &  1.18 & 0.888    &              $\iso{Ce}{140}$ &  2 &  1.12 & 0.877 \\
$\iso{Ge}{72}$ &  1 &  3.23 & 0.944    &              $\iso{Pr}{141}$ &  2 &  1.09 & 0.854 \\
$\iso{Ge}{74}$ &  1 &  1.51 & 0.966    &              $\iso{Nd}{142}$ &  3 &  1.17 & 0.886 \\
$\iso{As}{75}$ &  1 &  1.14 & 0.936    &              $\iso{Nd}{144}$ & -- &  1.14 & 0.860 \\
$\iso{Se}{76}$ &  1 &  1.17 & 0.939    &              $\iso{Nd}{146}$ &  3 &  1.17 & 0.880 \\
$\iso{Se}{78}$ &  1 &  1.98 & 0.971    &              $\iso{Sm}{147}$ &  3 &  1.14 & 0.858 \\
$\iso{Se}{80}$ &  1 &  1.29 & 0.939    &              $\iso{Sm}{148}$ &  3 &  1.19 & 0.889 \\
$\iso{Br}{79}$ &  1 &  2.79 & 0.962    &              $\iso{Sm}{150}$ &  3 &  1.17 & 0.878 \\
$\iso{Br}{81}$ &  1 &  1.08 & 0.942    &              $\iso{Eu}{151}$ &  1 &  1.23 & 0.810 \\
$\iso{Kr}{80}$ &  1 &  2.57 & 0.782    &              $\iso{Eu}{153}$ & -- &  1.14 & 0.842 \\
$\iso{Kr}{82}$ &  1 &  1.25 & 0.940    &              $\iso{Gd}{152}$ &  3 &  1.18 & 0.768 \\
$\iso{Kr}{84}$ &  1 &  1.52 & 0.970    &              $\iso{Gd}{154}$ &  3 &  1.15 & 0.854 \\
$\iso{Kr}{86}$ &  1 &  1.78 & 0.472    &              $\iso{Gd}{156}$ &  3 &  1.15 & 0.852 \\
$\iso{Rb}{85}$ &  1 &  1.07 & 0.943    &              $\iso{Gd}{158}$ & -- &  1.15 & 0.848 \\
$\iso{Rb}{87}$ &  1 &  1.94 & 0.514    &              $\iso{Tb}{159}$ &  1 &  1.37 & 0.833 \\
$\iso{Sr}{86}$ &  1 &  1.17 & 0.945    &              $\iso{Dy}{160}$ & -- &  1.20 & 0.878 \\
$\iso{Sr}{87}$ &  1 &  1.15 & 0.957    &              $\iso{Dy}{162}$ & -- &  1.17 & 0.855 \\
$\iso{Sr}{88}$ &  1 &  1.06 & 0.950    &              $\iso{Dy}{164}$ &  3 &  1.20 & 0.861 \\
$\iso{Y}{89}$  &  1 &  1.10 & 0.926    &              $\iso{Ho}{165}$ &  1 &  1.29 & 0.844 \\
$\iso{Zr}{90}$ &  1 &  1.12 & 0.907    &              $\iso{Er}{166}$ &  1 &  1.40 & 0.818 \\
$\iso{Zr}{92}$ &  1 &  1.21 & 0.932    &              $\iso{Er}{167}$ &  1 &  1.39 & 0.846 \\
$\iso{Zr}{94}$ &  1 &  1.13 & 0.923    &              $\iso{Er}{168}$ &  1 &  1.57 & 0.826 \\
$\iso{Nb}{93}$ &  1 &  1.46 & 0.945    &              $\iso{Tm}{169}$ &  1 &  1.76 & 0.806 \\
$\iso{Mo}{95}$ &  1 &  1.13 & 0.927    &              $\iso{Yb}{170}$ & -- &  1.21 & 0.873 \\
$\iso{Mo}{96}$ &  1 &  1.32 & 0.967    &              $\iso{Yb}{172}$ & -- &  1.17 & 0.836 \\
$\iso{Mo}{97}$ &  1 &  1.12 & 0.910    &              $\iso{Yb}{174}$ & -- &  1.19 & 0.847 \\
$\iso{Mo}{98}$ &  1 &  1.26 & 0.927    &              $\iso{Lu}{175}$ &  3 &  1.21 & 0.871 \\
$\iso{Ru}{99}$ &  1 &  1.20 & 0.943    &              $\iso{Lu}{176}$ &  3 &  1.19 & 0.848 \\
$\iso{Ru}{100}$ &  1 &  1.19 & 0.908   &               $\iso{Hf}{176}$ &  3 &  1.27 & 0.833 \\
$\iso{Ru}{102}$ &  1 &  1.13 & 0.926   &              $\iso{Hf}{178}$ & -- &  1.22 & 0.866 \\
$\iso{Rh}{103}$ &  1 &  1.28 & 0.939   &              $\iso{Hf}{180}$ & -- &  1.19 & 0.841 \\
$\iso{Pd}{104}$ &  1 &  1.46 & 0.968   &              $\iso{Ta}{181}$ &  1 &  1.52 & 0.788 \\
$\iso{Pd}{106}$ &  1 &  1.42 & 0.943   &              $\iso{W}{182}$  & -- &  1.20 & 0.837 \\
$\iso{Pd}{108}$ &  1 &  1.37 & 0.918   &              $\iso{W}{183}$  &  3 &  1.20 & 0.800 \\
$\iso{Ag}{107}$ &  1 &  1.11 & 0.936   &              $\iso{W}{184}$  & -- &  1.23 & 0.859 \\
$\iso{Ag}{109}$ &  1 &  1.08 & 0.914   &              $\iso{Re}{185}$ & -- &  1.19 & 0.820 \\
$\iso{Cd}{110}$ &  2 &  1.05 & 0.939   &              $\iso{Os}{186}$ & -- &  1.25 & 0.852 \\
$\iso{Cd}{112}$ &  2 &  1.06 & 0.952   &              $\iso{Os}{187}$ &  1 &  1.72 & 0.820 \\
$\iso{Cd}{114}$ &  2 &  1.06 & 0.953   &              $\iso{Os}{188}$ &  3 &  1.22 & 0.825 \\
$\iso{In}{115}$ &  1 &  1.39 & 0.912   &              $\iso{Os}{190}$ & -- &  1.22 & 0.827 \\
$\iso{Sn}{116}$ &  1 &  1.05 & 0.938   &              $\iso{Ir}{191}$ & -- &  1.20 & 0.820 \\
$\iso{Sn}{118}$ &  2 &  1.07 & 0.948   &              $\iso{Ir}{193}$ &  3 &  1.31 & 0.815 \\
$\iso{Sn}{120}$ &  2 &  1.06 & 0.953   &              $\iso{Pt}{192}$ &  1 &  2.31 & 0.871 \\
$\iso{Sb}{121}$ &  1 &  1.19 & 0.954   &              $\iso{Pt}{194}$ &  1 &  2.91 & 0.850 \\
$\iso{Te}{122}$ & -- &  1.06 & 0.957   &              $\iso{Pt}{196}$ &  3 &  1.32 & 0.795 \\
$\iso{Te}{123}$ & -- &  1.04 & 0.945   &              $\iso{Au}{197}$ & -- &  1.24 & 0.838 \\
$\iso{Te}{124}$ &  3 &  1.06 & 0.955   &              $\iso{Hg}{198}$ &  2 &  1.31 & 0.782 \\
$\iso{Te}{126}$ &  1 &  1.07 & 0.950   &              $\iso{Hg}{200}$ &  1 &  1.36 & 0.774 \\
$\iso{I}{127}$  &  1 &  1.16 & 0.945   &              $\iso{Hg}{202}$ & -- &  1.34 & 0.858 \\
$\iso{Xe}{128}$ &  1 &  1.04 & 0.908   &              $\iso{Tl}{203}$ &  3 &  1.30 & 0.779 \\
$\iso{Xe}{130}$ &  2 &  1.06 & 0.958   &              $\iso{Tl}{205}$ &  1 &  2.40 & 0.772 \\
$\iso{Xe}{132}$ &  1 &  1.33 & 0.957   &              $\iso{Pb}{204}$ & -- &  1.27 & 0.797 \\
$\iso{Cs}{133}$ &  1 &  1.13 & 0.949   &              $\iso{Pb}{206}$ & -- &  1.30 & 0.763 \\
$\iso{Ba}{134}$ &  1 &  1.08 & 0.935   &              $\iso{Pb}{207}$ & -- &  1.43 & 0.792 \\
$\iso{Ba}{136}$ &  1 &  1.12 & 0.954   &              $\iso{Pb}{208}$ & -- &  1.39 & 0.784 \\
$\iso{Ba}{137}$ &  1 &  1.09 & 0.950   &              $\iso{Bi}{209}$ &  3 &  1.38 & 0.746 \\
\hline
\end{tabular}
\medskip
\end{table}

\begin{table}

\centering

\caption{ Uncertainties in the final abundance of
  {\it s}-process nuclides from the MC calculation for the  
TP phase.
  The column labeled ``Level'' indicates the level of the first key
  reaction found. The remaining columns show 
uncertainty factors for variations Up and Down, the values of which
are  $Y(95\%)/Y_{\rm peak}$ and $Y(5\%)/Y_{\rm peak}$, respectively. 
They enclose a 90\% probability interval, as shown in Fig.~\ref{MC_TP}. } 
\label{tabTPunc}

\begin{tabular}{@{}lcrrlcrr}
  \hline                                                           
Nuclide & Level & Up & Down             & Nuclide & Level & Up & Down \\                
\hline    

$\iso{Ge}{70}$ &  1 &  1.04 & 0.946 &      $\iso{Ba}{134}$ &  2 &  1.08 & 0.923 \\
$\iso{Se}{76}$ &  1 &  1.06 & 0.901 &      $\iso{Ba}{136}$ &  1 &  1.03 & 0.988 \\
$\iso{Se}{82}$ &  1 &  1.14 & 0.941 &      $\iso{La}{138}$ &  1 &  2.56 & 0.919 \\
$\iso{Kr}{80}$ &  1 &  1.11 & 0.789 &      $\iso{Ce}{142}$ &  1 &  3.42 & 0.619 \\
$\iso{Sr}{86}$ &  1 &  1.04 & 0.977 &      $\iso{Nd}{142}$ & -- &  1.01 & 0.974 \\
$\iso{Sr}{87}$ &  1 &  1.05 & 0.960 &      $\iso{Nd}{148}$ &  1 &  2.09 & 0.442 \\
$\iso{Zr}{96}$ &  1 &  3.73 & 0.469 &      $\iso{Sm}{148}$ &  2 &  1.10 & 0.862 \\
$\iso{Mo}{94}$ &  1 &  1.21 & 0.879 &      $\iso{Sm}{150}$ &  1 &  1.11 & 0.956 \\
$\iso{Mo}{96}$ &  1 &  1.08 & 0.900 &      $\iso{Sm}{152}$ &  1 &  1.10 & 0.918 \\
$\iso{Mo}{100}$ &  1 &  1.73 & 0.466 &     $\iso{Sm}{154}$ & -- &  3.40 & 0.602 \\
$\iso{Ru}{100}$ &  1 &  1.11 & 0.912 &     $\iso{Gd}{152}$ &  1 &  1.41 & 0.263 \\
$\iso{Ru}{104}$ &  1 &  1.81 & 0.456 &     $\iso{Gd}{154}$ &  1 &  1.16 & 0.886 \\
$\iso{Pd}{104}$ &  1 &  1.31 & 0.956 &     $\iso{Gd}{160}$ &  1 &  2.14 & 0.500 \\
$\iso{Pd}{110}$ &  1 &  1.39 & 0.388 &     $\iso{Dy}{160}$ &  1 &  1.29 & 0.884 \\
$\iso{Cd}{108}$ & -- &  1.11 & 0.926 &     $\iso{Er}{170}$ &  1 &  3.52 & 0.627 \\
$\iso{Cd}{110}$ &  1 &  1.06 & 0.927 &     $\iso{Yb}{170}$ &  1 &  1.94 & 0.727 \\
$\iso{Cd}{116}$ &  1 &  1.39 & 0.256 &     $\iso{Yb}{176}$ &  1 &  1.34 & 0.414 \\
$\iso{Sn}{114}$ &  1 &  1.04 & 0.928 &     $\iso{Lu}{176}$ &  1 &  1.12 & 0.867 \\
$\iso{Sn}{115}$ &  1 &  1.05 & 0.923 &     $\iso{W}{186}$ &  1 &  2.14 & 0.878 \\ 
$\iso{Sn}{116}$ &  1 &  1.02 & 0.970 &     $\iso{Re}{187}$ &  3 &  1.85 & 0.843 \\
$\iso{Sn}{124}$ &  1 &  1.07 & 0.783 &     $\iso{Os}{186}$ &  1 &  1.20 & 0.674 \\
$\iso{Te}{122}$ &  1 &  1.04 & 0.944 &     $\iso{Os}{187}$ &  1 &  2.13 & 0.762 \\
$\iso{Te}{123}$ &  2 &  1.06 & 0.932 &     $\iso{Os}{192}$ &  1 &  1.28 & 0.680 \\
$\iso{Te}{124}$ &  1 &  1.01 & 0.973 &     $\iso{Pt}{192}$ &  1 &  1.92 & 0.686 \\
$\iso{Te}{130}$ &  1 &  1.40 & 0.927 &     $\iso{Pt}{195}$ &  1 &  2.96 & 0.841 \\
$\iso{Xe}{128}$ &  1 &  1.04 & 0.883 &     $\iso{Pt}{198}$ &  1 &  1.39 & 0.448 \\
$\iso{Xe}{130}$ &  1 &  1.02 & 0.961 &     $\iso{Hg}{198}$ &  1 &  1.13 & 0.890 \\
$\iso{Xe}{134}$ &  1 &  2.38 & 0.639 &     $\iso{Pb}{204}$ &  1 &  1.04 & 0.951 \\
$\iso{Xe}{136}$ &  1 &  2.06 & 0.835 &     $\iso{Bi}{209}$ &  1 &  1.02 & 0.996 \\
\hline
\end{tabular}
\medskip
\end{table}


\begin{figure*}
\includegraphics[width=\textwidth]{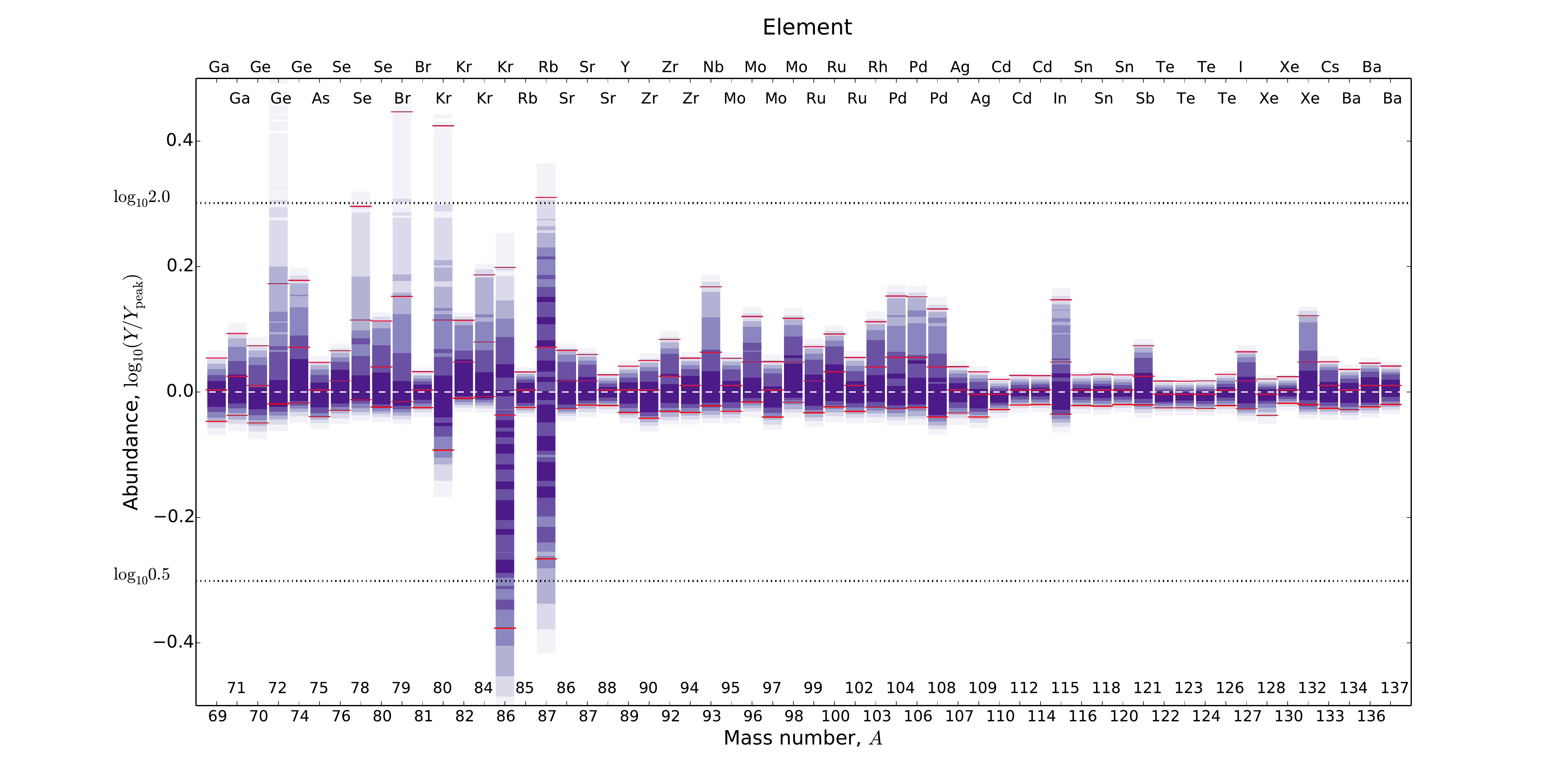}
\includegraphics[width=\textwidth]{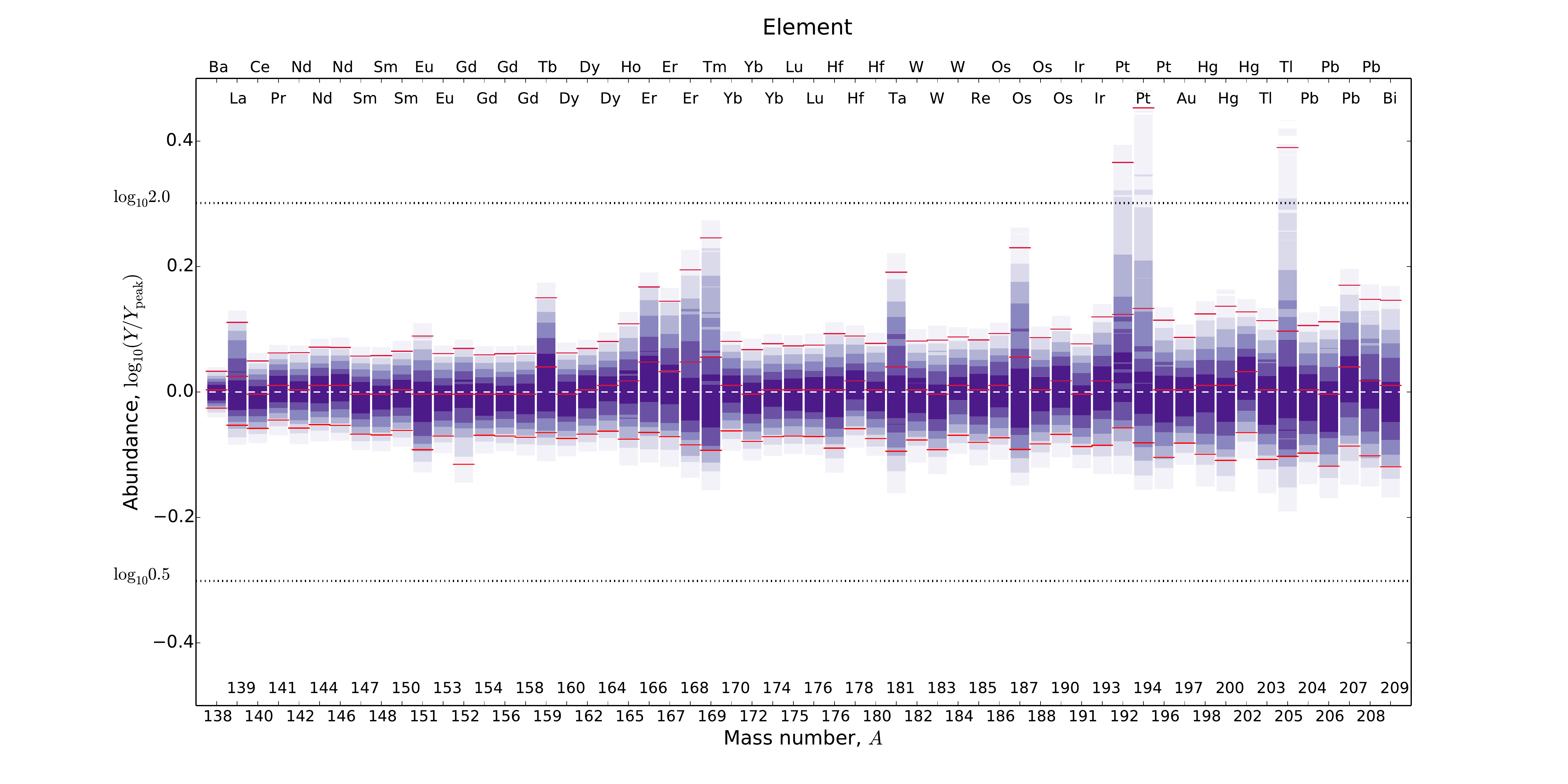}
\includegraphics[width=5cm]{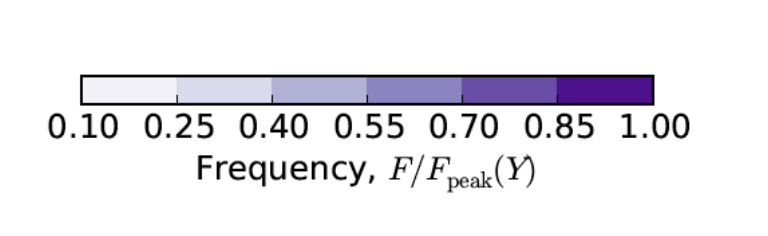}

\caption{Total production uncertainties in the final s-process
  abundances obtained with the trajectory described in the previous 
section for the $\iso{C}{13}$ pocket with the standard initial abundance. The color shading denotes the probabilistic frequency and the
 90\% probability intervals up and down are marked for each nuclide with the red lines. The final abundances are normalised by the final abundance at the peak of the distribution.
 Horizontal dotted lines indicate a factor of two uncertainties.}
\label{L1}

\end{figure*}

\begin{figure*}
\includegraphics[width=\textwidth]{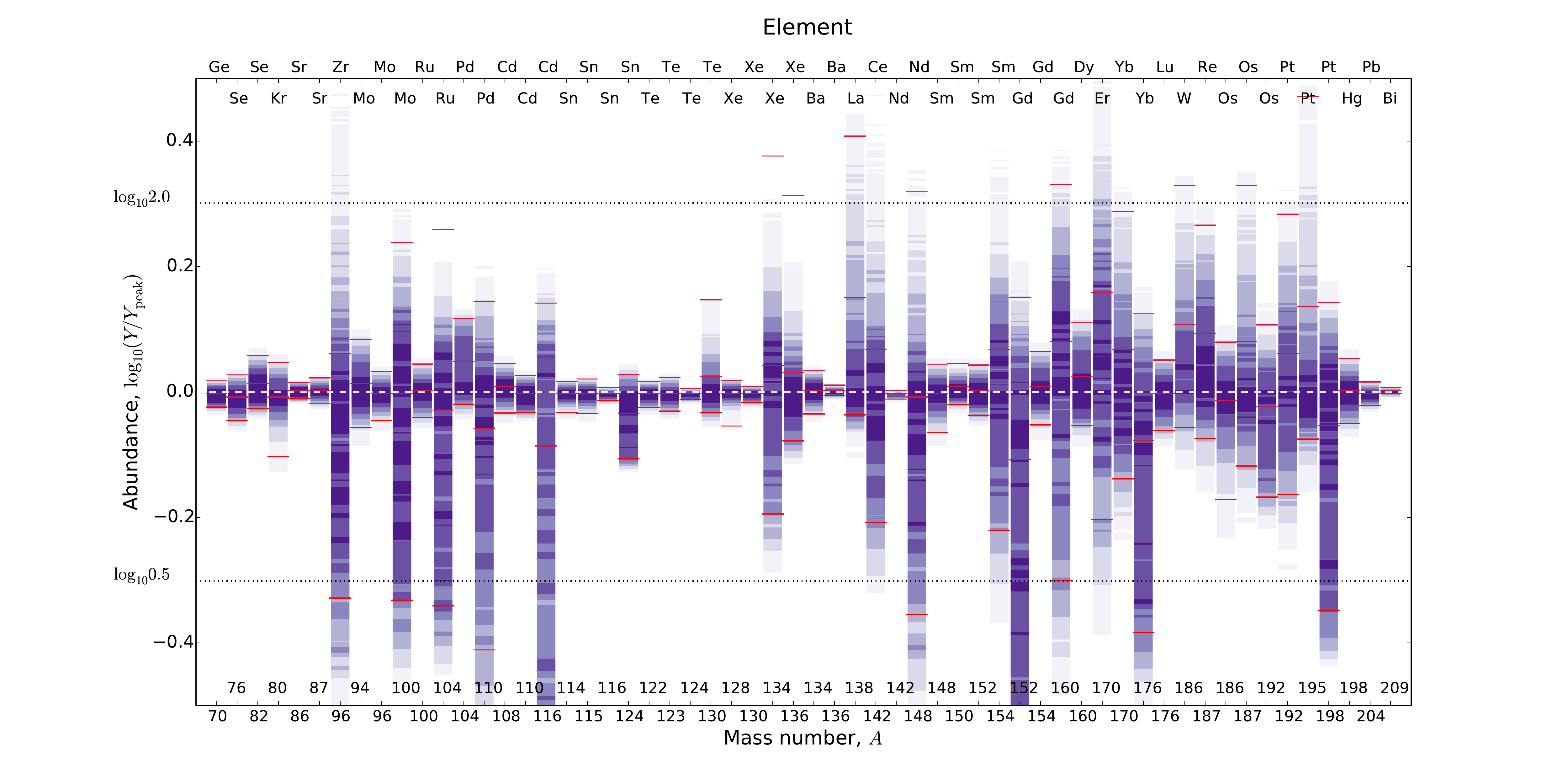}

\caption{Total production uncertainties in the final s-process
  abundances obtained with the trajectory described in the previous 
section for the TP phase. The color shading denotes the probabilistic frequency and the
 90\% probability intervals up and down are marked for each nuclide
 with the red lines. 
The final abundances are normalised by the final abundance at the peak of the distribution.
 Horizontal dotted lines indicate a factor of two uncertainties.}
\label{MC_TP}

\end{figure*}

As can be seen in Table\,\ref{tab-C13-unc} and Fig.\,\ref{L1}, the
overall uncertainties  during the $\iso{C}{13}$ pocket are
generally small. Indeed, most of them are smaller than 50\%. This is
not too surprising since the relevant temperature range ($\sim$8\,keV)
is accessible to experimental measurements so many of the relevant
rates, which are along the valley of stability, have already been
measured experimentally. Furthermore, excited states generally have a
weak contribution in this temperature range so the nuclear
uncertainties are generally small to start with. There are
nevertheless several nuclides, for which uncertainties are larger than
a factor of two. These are generally nuclides around branching points
such as $\iso{Kr}{86}$. We also notice a propagation effect for
nuclides more massive than $\iso{Ba}{138}$. This is due to the
combined effect of uncertainties in neutron capture rates above
$\iso{Ba}{138}$.   For the TP case, we find somewhat larger
  uncertainties (see Table\,\ref{tabTPunc} and Fig.\,\ref{MC_TP}), in
  several cases greater than a factor of 2, and in four cases reaching
  a factor of 3.  This is due both to the higher temperatures
  encountered, and the effect that at branching points there is a
  stonger sensitivity to the
  ratio between beta decay rates and capture rates.


\subsection{Key rates}
\label{sec:key}
As explained in Sect.\,\ref{sec:MC}, key rates are obtained by
examining the correlation between a change in a reaction rate and the
change of an abundance. The key reaction rates are listed in Table
\ref{tab:key} for levels 1, 2, and 3 (for an explanation of key rate
levels, see Sect.\ \ref{sec:uncertlevels}).  Most of them are neutron
capture reactions either directly producing or destroying the nuclide
in question. This is not surprising because steady-flow
  equilibrium applies to most of the s-process path between the
  peaks. We nevertheless list all of them in the Appendix for
  completeness.  Moreover, not all of the selected isotopes and key
  reactions appear at same level or with the same correlation, thus
  indicating the impact of specific reactions, or the impact of
  different degrees of constraint in experimental uncertainty on final
  abundances.  Notable exceptions  for the $\iso{C}{13}$ pocket 
are neutron captures on $\iso{Fe}{56}$, $\iso{Ni}{64}$, and
$\iso{Ba}{138}$, which are level 2 key rates for many nuclides. We
will come back to these three reactions in
Sect.\,\ref{sec:trio}. There are also a few key weak reactions at
branching points, $\iso{Se}{79}$, $\iso{Kr}{85}$, and $\iso{I}{128}$.
We will discuss the possibility of reducing the uncertainties of the
key reactions linked to the most uncertain final abundances in
Sect.\,\ref{opp-exp}.  For the TP phase, there are two
  exceptions, the neutron capture reactions
  $^{56,57}$Fe(n,\,$\gamma$).  While this may be surprising at first,
  $^{56,57}$Fe act as ``poisons'' in the TP phase. Indeed, they
  compete for neutrons with heavier nuclides and the neutron burst in
  the TP phase is too short for iron to act as a seed for the heavy
  elements produced during the TP.

%

\subsubsection{Uncertainties for the different key reaction levels}
\label{sec:uncertlevels}

As in our previous studies, we determined level 2 key reactions by using the standard rates for all previously identified (level 
1) key reaction rates and performing another MC variation without varying those rates. This shows the effect when the level 
1 key rates would have been determined. Level 2 key rates are then key to the remaining uncertainties. Similarly, level 3 
key rates were determined by exempting level 1 \textit{and} level 2 key rates from the MC variation. It has to be emphasised 
that level 2 and level 3 key reactions are only important \textit{provided that level 1 and level 2 rates, respectively, 
have been constrained}.

\begin{figure*}
\includegraphics[width=\textwidth]{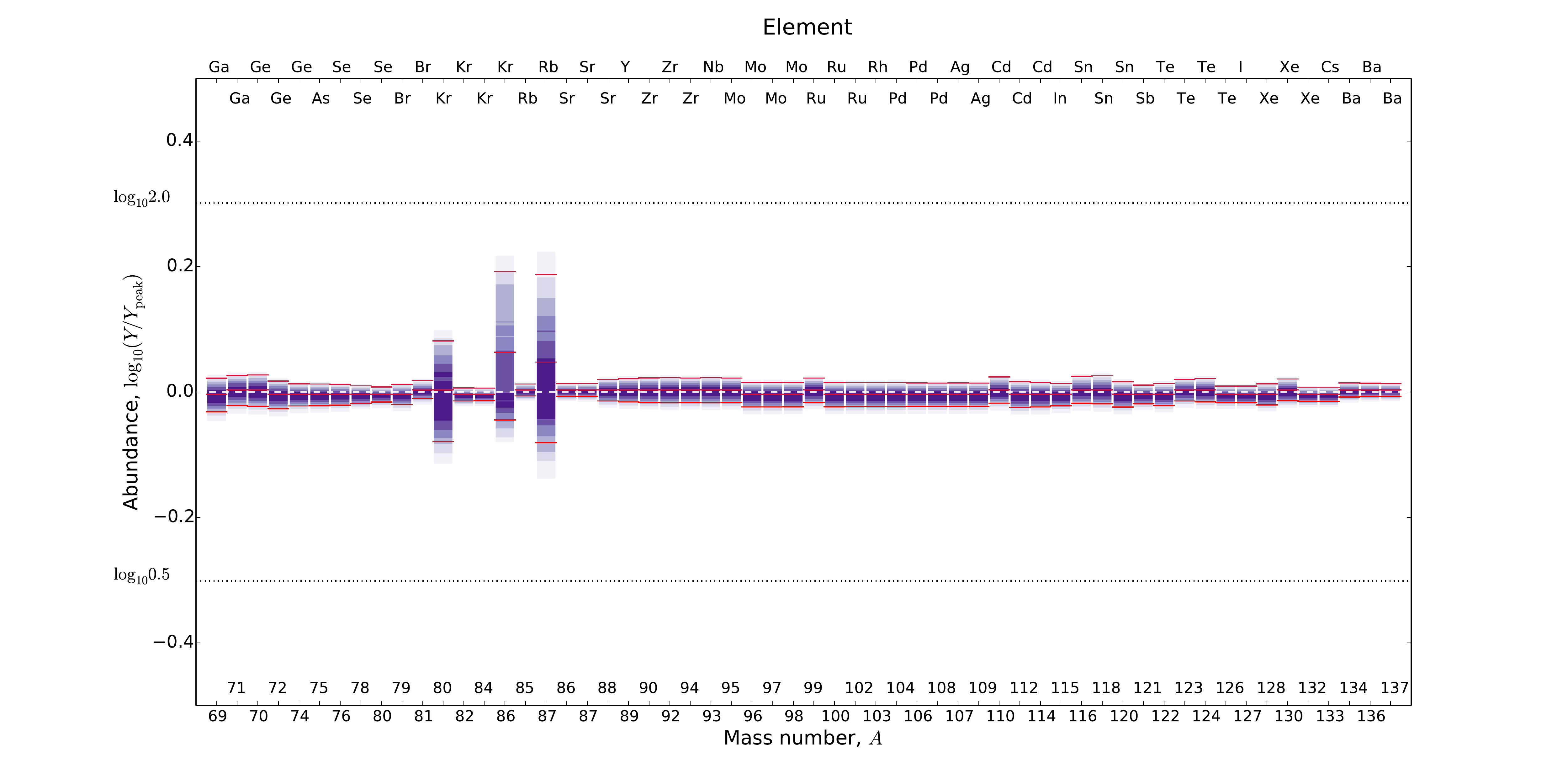}
\includegraphics[width=\textwidth]{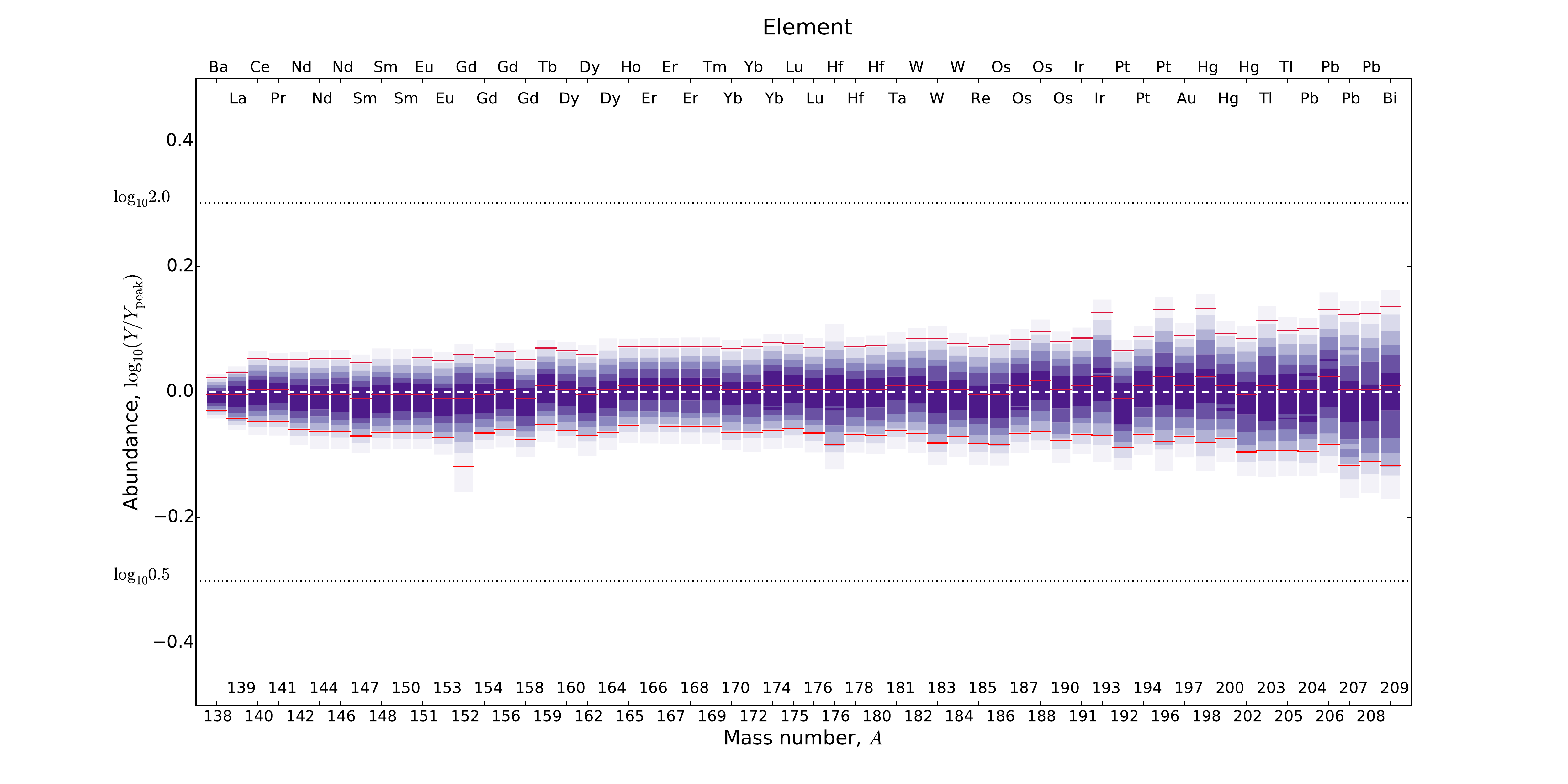}
\includegraphics[width=5cm]{colorbar7.pdf}
\caption{Same as Fig.\ref{L1} except that all the level 1 key reactions are now fixed to show the improvements that determining all level 1 rates would make.}

\label{L2}
\end{figure*}

\begin{figure*}
\includegraphics[width=\textwidth]{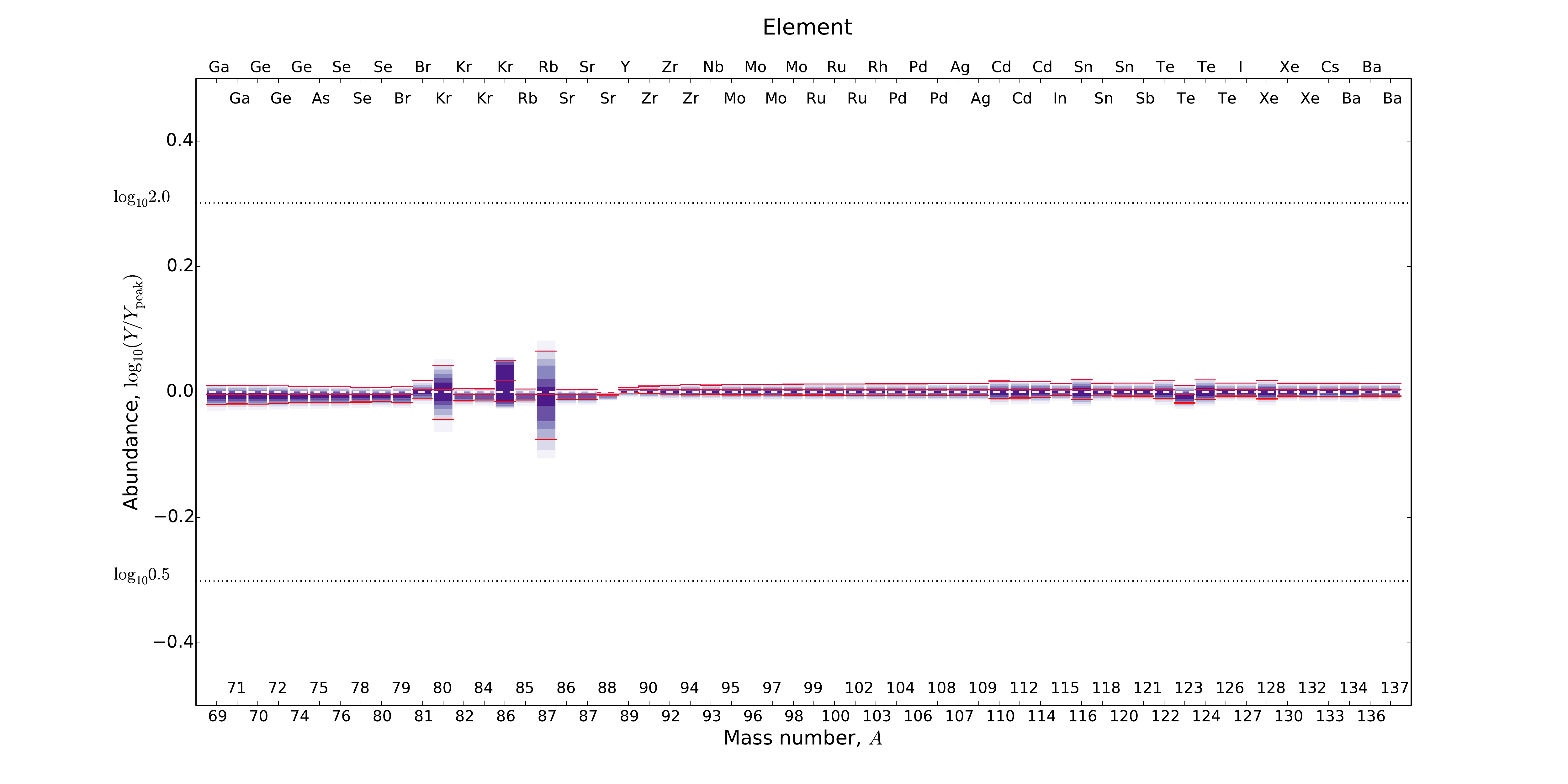}
\includegraphics[width=\textwidth]{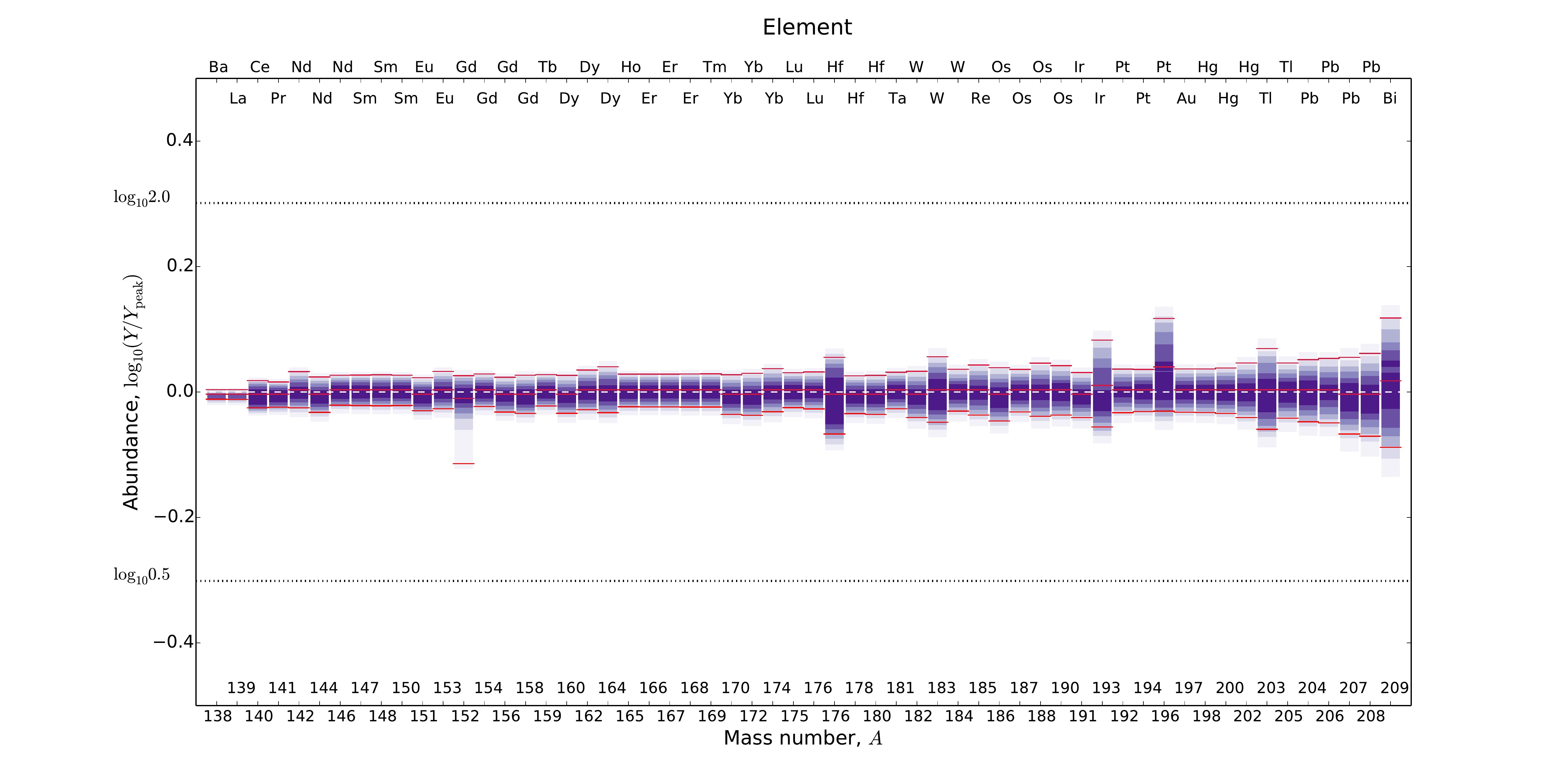}
\includegraphics[width=5cm]{colorbar7.pdf}
\caption{Same as Fig.\ref{L1} except that all the level 1 and 2 key reactions are now fixed.}

\label{L3}
\end{figure*}

Figures \ref{L2} and \ref{L3} show the total uncertainties obtained for levels 2 and 3, respectively. We see that 
already at level 2, uncertainties are tiny for nuclides lighter than $\iso{Ba}{138}$. Exceptions are a few isotopes at branching 
points ($\iso{Kr}{80}$, $\iso{Kr}{86}$, and $\iso{Rb}{87}$). The propagation effect for nuclides more massive than $\iso{Ba}{138}$ 
remains. The total uncertainty has already significantly decreased compared to level 1 so limited improvements can be made by 
future measurements of the level 2 rates. The level 3 uncertainties shown in Fig.\,\ref{L3} show that all key rates were identified 
at level 1 or 2 and that the uncertainties are negligible once these
have been determined. 
 The same is true for the TP phase.

\begin{figure*}

\includegraphics[width=\textwidth]{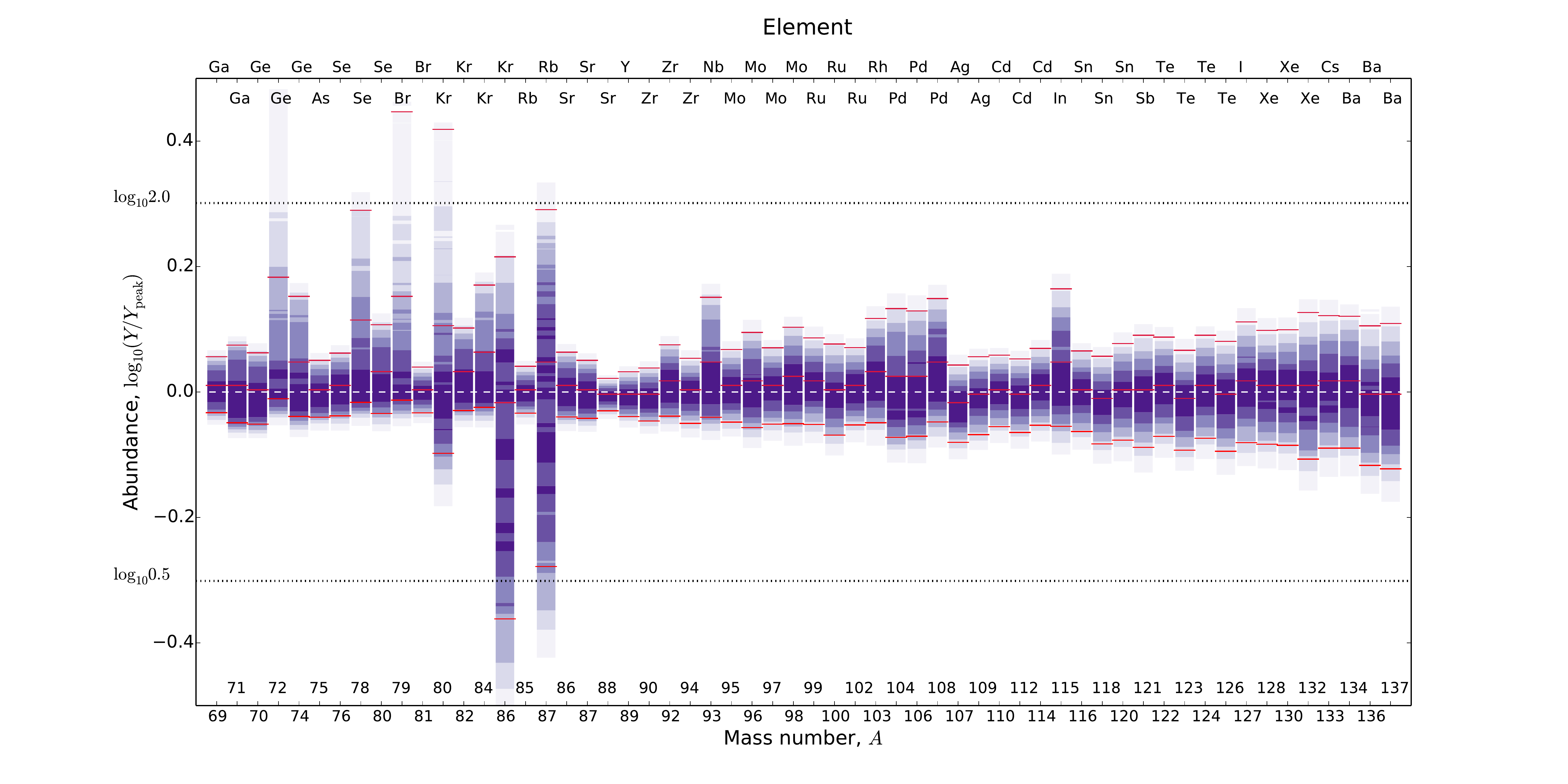}
\includegraphics[width=\textwidth]{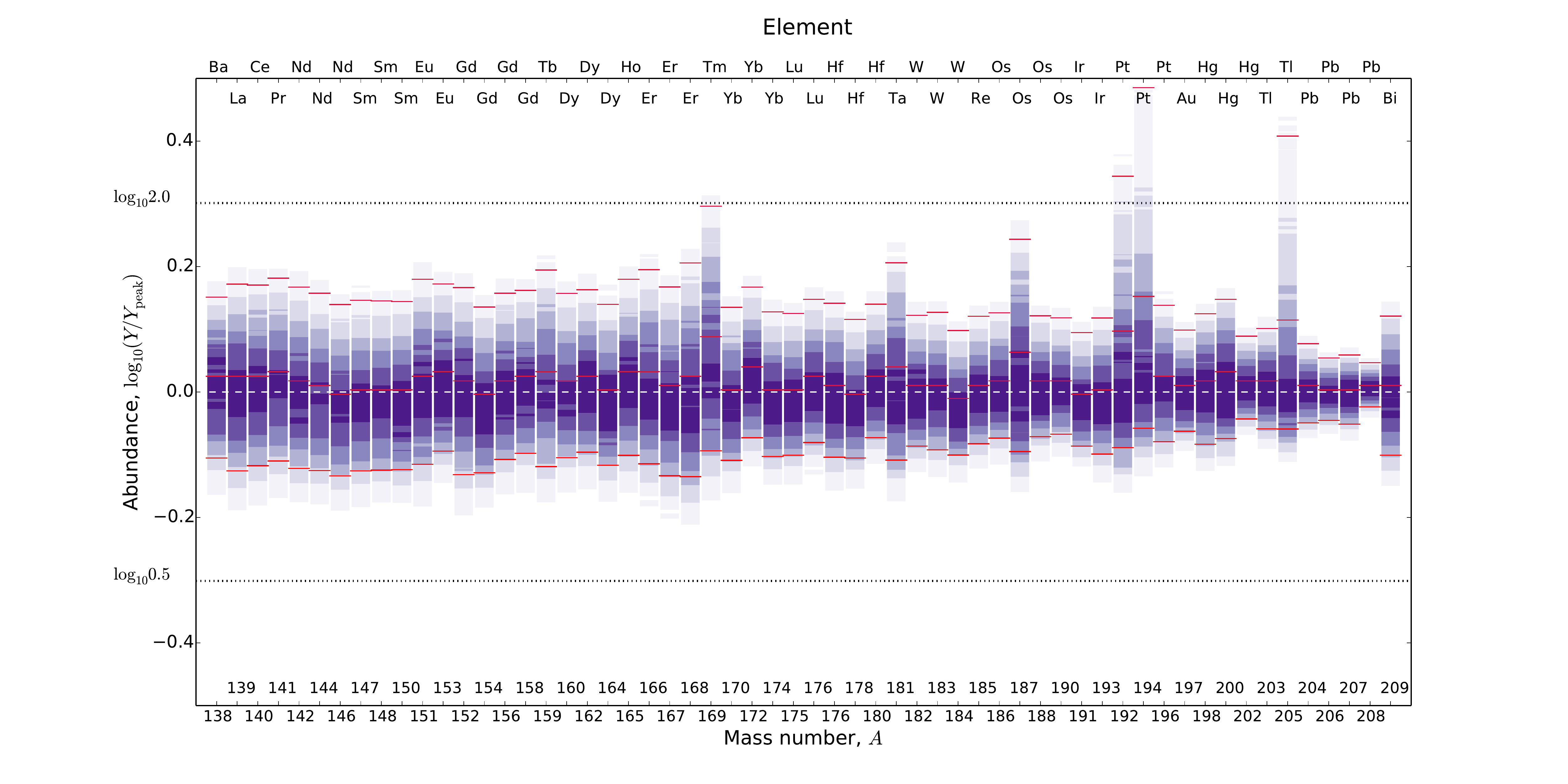}
\includegraphics[width=5cm]{colorbar7.pdf}

\caption{Total production uncertainties (same as Fig.\,\ref{L1}) for the case with half of the standard initial $\iso{C}{13}$ abundance (``0.5 $\times$ $\iso{C}{13}$ '' case).}
\label{HC}
\end{figure*}

\begin{figure*}
\includegraphics[width=\textwidth]{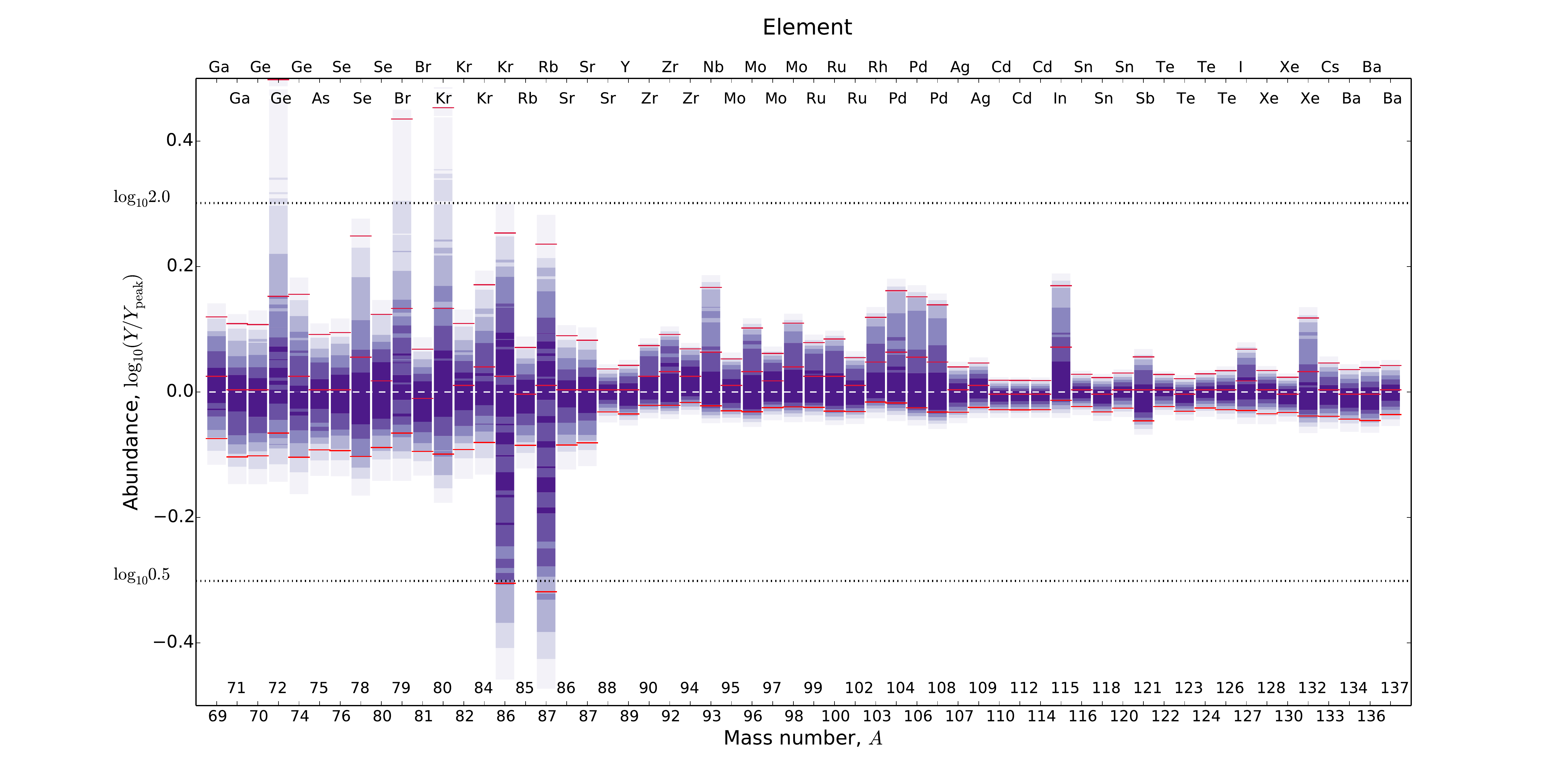}
\includegraphics[width=\textwidth]{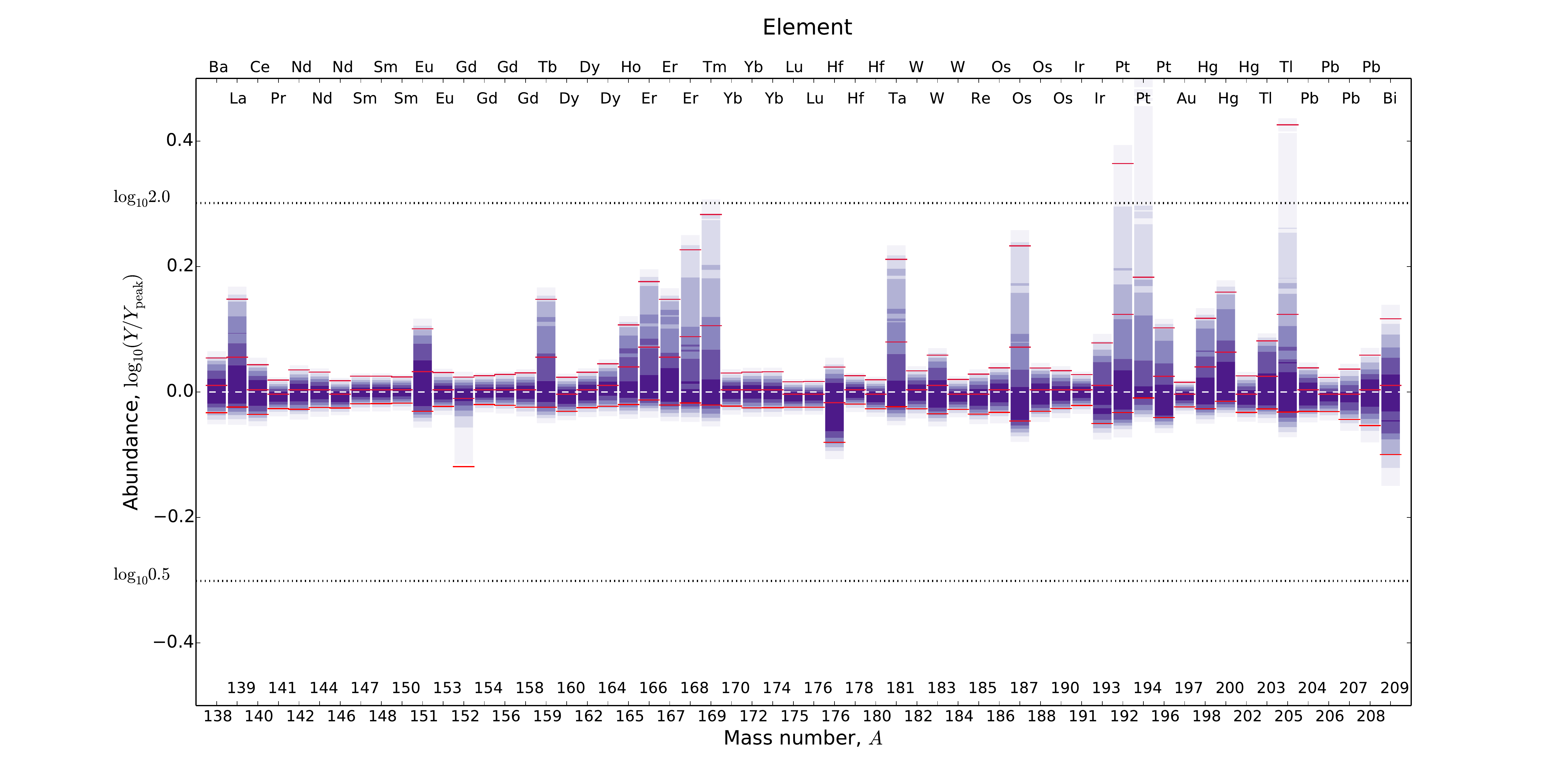}
\includegraphics[width=5cm]{colorbar7.pdf}
\caption{Total production uncertainties (same as Fig.\,\ref{L1}) for the case with double the standard initial $\iso{C}{13}$ abundance (``2 $\times$ $\iso{C}{13}$'' case).}
\label{DC}
\end{figure*}

\subsection{Dependence of uncertainties and key rates on astrophysical conditions}
\label{sec:sens}

We used a single-zone trajectory to mimic the astrophysical conditions
taking place in the TP-AGB phase of low-mass stars in our MC
calculations. A single-zone trajectory cannot capture the full
conditions found in stars. As explained in Sect.\,\ref{sec-evolution},
conditions vary in stars and there are still major uncertainties in
the modelling of the TP-AGB phase and in particular concerning the
formation of the $^{13}$C-pocket. Nevertheless, as our comparison to
the yields of C11 in Sect.\,\ref{sec:C11} shows, using an initial
$^{13}$C abundance divided (``0.5 $\times$ $\iso{C}{13}$'' case) and multiplied
(``2 $\times$ $\iso{C}{13}$'' case) by a factor of two compared to the standard
case samples the variations in the s-process production in stars of
different masses. It also samples different neutron to seed ratios and
thus to some extent the metallicity dependence of our results. More
generally, it allows us to determine the sensitivity of our results to
the astrophysical conditions found in the $^{13}$C-pocket.  Figures
\ref{fig:low_nc}, \ref{fig:high_nc}, \ref{fig:Eu_nc}, and
\ref{fig:vhigh_nc} show that in the ``0.5 $\times$ $\iso{C}{13}$'' case, the
production stops around $\iso{Ba}{138}$ and therefore that this case
underestimates the neutron flux needed to produce the main
s-process. This leads to a stronger production for elements between
iron and strontium. In the ``2 $\times$ $\iso{C}{13}$'' case, the production is
very strong all the way up to lead with overproduction factors much
larger than those of C11. This means that the neutron flux in this
case is very strong and the elements between iron and strontium are
depleted. We thus do not consider the two additional cases as
representative cases for the main s-process. This is why we do not
list the total uncertainties for these two cases in tables.  Rather we
use them to test the robustness of the key rates list against
variations that are larger than the variations expected to occur in
real stars.

The total uncertainties for the two additional cases are shown in
Figs. \ref{HC} and \ref{DC}. Comparing these figures to Fig.  \ref{L1}
for the standard case, we see that the same nuclides have the largest
uncertainties. We also see that uncertainties are generally small
(less than 50\%) for most nuclides in the three cases. The main
difference between the three cases is the extent of the propagation
effects. Since the flow stops around barium for the ``0.5 $\times$
$\iso{C}{13}$'' case, the propagation effect is strongest in this case for
elements around and above barium. In the ``2 $\times$ $\iso{C}{13}$'' case,
propagation effects are very small because the production easily
reaches lead.

 We list the key rates for the three cases in
Tables\, \ref{tab:key}, \ref{tab:hc} and \ref{tab:dc} in the
Appendix. Comparing Table\,\ref{tab:hc} to
Table\,\ref{tab:key}, we see that all but one key rates for the ``0.5
$\times$ $\iso{C}{13}$'' case were already key rates for the standard case. The
exception is $\iso{Bi}{209}(\mbox{n},\gamma)\iso{Bi}{210}$, which is
not important in this particular case because there is no production
beyond barium. Comparing Table\,\ref{tab:dc} to Table\,\ref{tab:key},
we see again that most key rates for the ``2 $\times$ $\iso{C}{13}$'' case were
already key rates for the standard case. The very strong flux in the
``2 $\times$ $\iso{C}{13}$'' case leads to a production, which follows a
slightly more neutron-rich path and thus to a few more key rates that
were not present in the standard case. The strong overlap in the key
reaction lists between the standard case and the other two cases
representing a very weak and very strong neutron flux shows that our
reference key reaction list is representative of the full range of
astrophysical conditions found in the $^{13}$C-pocket.

\subsection{Comparison to past sensitivity studies}
The key differences between the approach used in this study and past studies are explained in \citet{Rauscher17} and 
\citet{Rauscher16}. We summarise them here:
\begin{enumerate}
\item Instead of varying rates one-by-one, all rates involving heavy elements in the 
network are
varied simultaneously in a Monte Carlo (MC) framework.
\item Key reactions are identified by inspection of correlations in the 
simultaneous variation of all rates instead of
relying on the sensitivity of an abundance to the individual variation of a single rate.
\item Each rate is assigned an individual uncertainty which is temperature dependent and which is sampled by a different MC 
variation factor for each rate. Uncertainties do not have to be symmetric 
\item The bespoke rate uncertainties are derived and are based on both experimental data for the ground-state contributions when 
available and a theoretical uncertainty for the excited-states contributions.
\end{enumerate}

Varying rates one-by-one may result in an incorrect assessment of
total uncertainties as well as the importance of the selected
rates. This is due to the fact that the combined action of several
reactions can cover or enhance uncertainties in each single rate. We
rather define a key reaction as a reaction dominating the uncertainty
of the final abundance of a given nuclide. This means that this
abundance uncertainty will be considerably reduced when better
constraining the corresponding key reaction. Key reactions are
specific to a nuclide and it is possible that no key reaction can be
found for a given nuclide when many reactions are contributing to its
abundance.

As explained in the Introduction, \citet{Koloczek16} (Ko16) recently
reviewed the impact on the main s-process of current nuclear
uncertainties considering both the $^{13}$C-pocket and TP
conditions. Ko16 varied reaction rates one-by-one so it is interesting
to compare our results to theirs. Note that this study focused on
intermediate and heavy elements and therefore this is the atomic mass
range that we will compare. The uncertainties for rates involving
light elements is generally well established and we refer the reader
to the Ko16 and \citet{Kappeler2011} studies (and references therein)
concerning nuclear uncertainties linked to light elements (e.g.,
neutron sources and neutron poisons). Ko16 provide a list of the
strongest globally affecting reactions during both the TP (their Table
A) and the $^{13}$C-pocket (their Table B), which is very valuable
information. Since rates were varied individually, however, it is not
clear whether or not the rates in question dominate the uncertainties
for all the nuclides affected by that reaction. Re-measuring the rates
listed in Tables A \& B of Ko16 may thus not reduce the uncertainties
in predicted production of all the nuclides affected. Our definition
of key rates gives exactly this information since a rate is only key
if it dominates the uncertainty of a given nuclide. Our study shows
that in many cases, the key rates dominating the nuclear uncertainties
are the neutron captures either directly producing or destroying the
nuclide in question. Nevertheless, all the rates involving heavy
elements listed for the $^{13}$C-pocket conditions by Ko16 (Table B)
appear as key rates for at least one nuclide  in our the standard
  $^{13}$C-pocket case. There are special rates,
neutron captures on $\iso{Fe}{56}$, $\iso{Ni}{64}$, and
$\iso{Ba}{138}$, which we discuss below.
We did not find in our TP phase the same reactions as
  found by Ko16 (c.f. their Table A). The different methodologies and
  the limited cases studied are likely responsible.  
  Nevertheless, we note that most of the reactions found by Ko16 for
  the TP conditions are actually key rates for the standard case or
  the ``2 $\times$ $\iso{C}{13}$'' case (which correspond to higher
  neutron densities compared to the standard $^{13}$C-pocket case).
  The only significant rates we found in the TP condition, that are
  not either directly producing or destroying the nuclide in question
  (or very close by nuclei) are $\iso{Fe}{57}(\mbox{n},\gamma)$ and
  $\iso{Fe}{56}(\mbox{n},\gamma)$. These rates, however, only appear
  for one nuclide (148Nd) at level 2 so should be treated as any other
  level 2 key reactions (i.e. only be considered after all level 1 key
  rates have been improved). Finally, for the TP phase, we obtained
  several more $\beta$-decay reactions as key rates for the selected
  nuclei, compared to the $^{13}$C-pocket conditions. However, as
  before, most of the uncertainty in the final abundances is caused by
  the uncertainty in the neutron capture rates.

%
\subsubsection{Neutron captures on $\iso{Fe}{56}$, $\iso{Ni}{64}$, and $\iso{Ba}{138}$}
\label{sec:trio}

\begin{figure}
\includegraphics[width=\columnwidth]{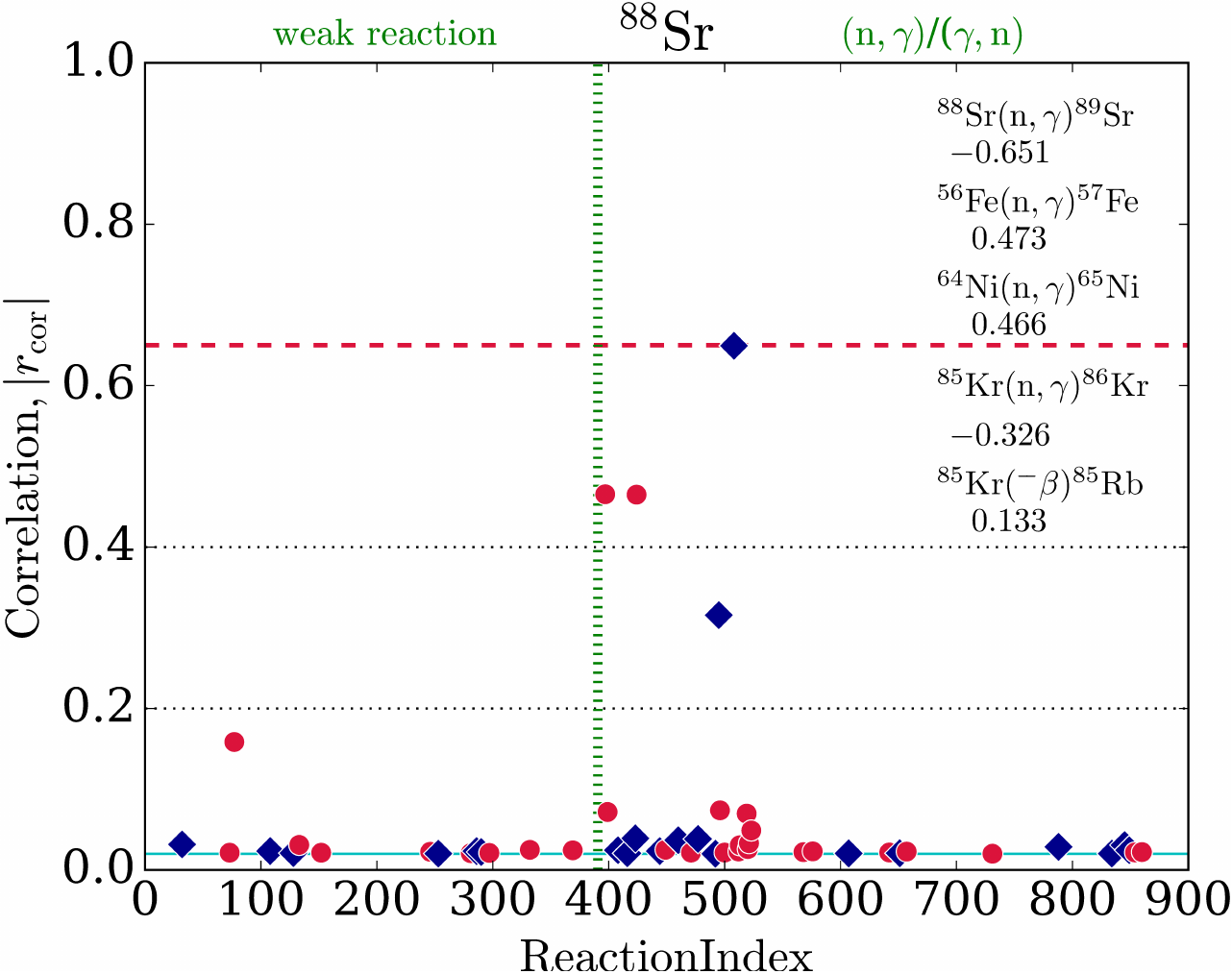}
\caption{The correlation coefficients of reactions with respect to an
  abundance change of $\iso{Sr}{88}$  during the $^{13}$C-pocket
    conditions.  The absolute values of the coefficients are plotted
  against a reaction index number. Red circles stand for positive
  correlation and blue squares for negative correlation,
  respectively. Reaction indices in the range of $1$--$390$ denote
  weak reactions and those in the range $391$--$900$ identify neutron
  captures.  The five reactions with the highest correlations are
  listed in the upper right corner.  Note that, for better
  readability, reactions with correlation factors
  $|r_{\rm cor}| < 0.02$ are omitted from this plot.}
\label{Sr_cor}
\end{figure}

\begin{figure}
\includegraphics[width=\columnwidth]{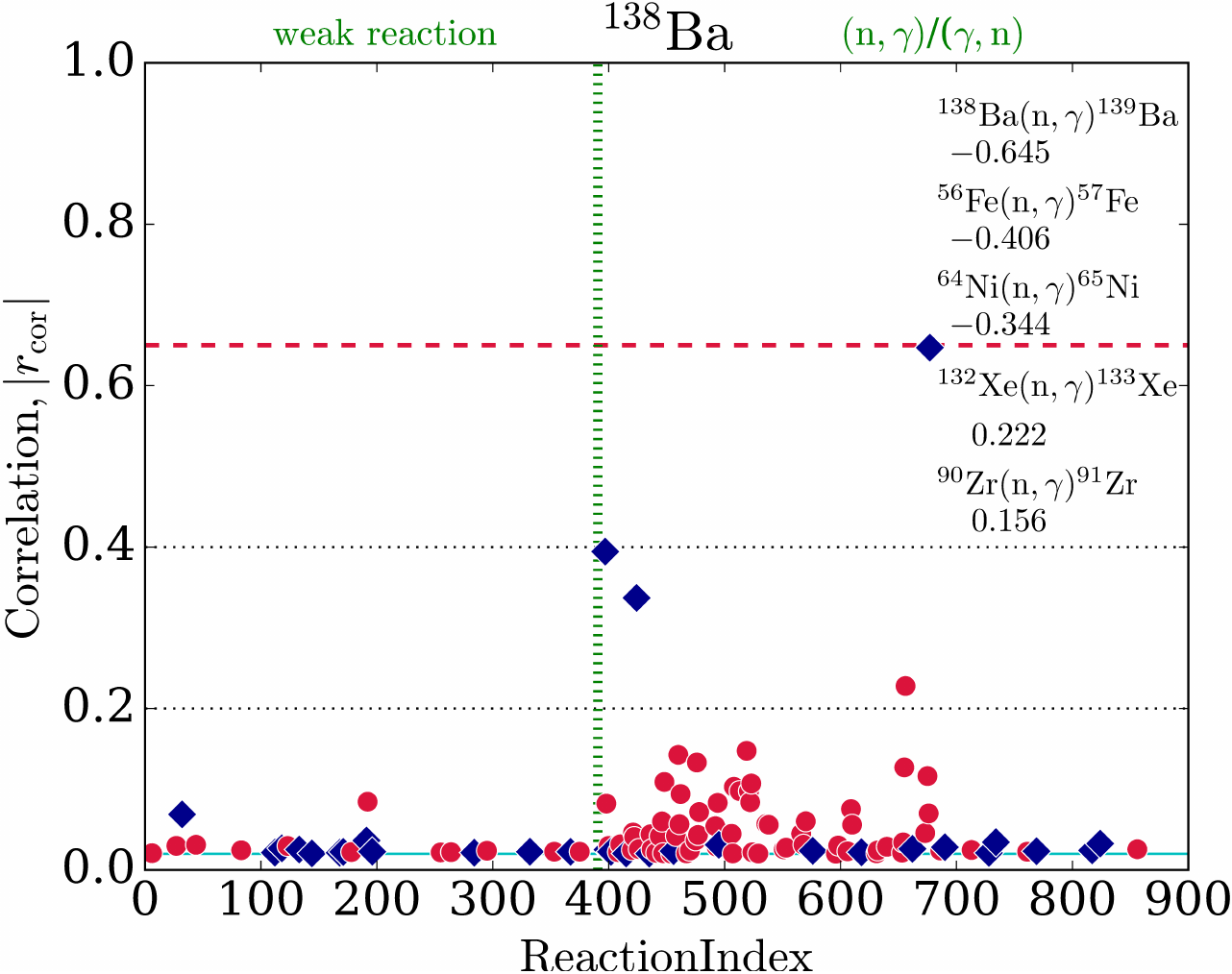}
\caption{Same as for Fig.\,\ref{Sr_cor} for $\iso{Ba}{138}$}
\label{Ba_cor}
\end{figure}

\begin{figure}
\includegraphics[width=\columnwidth]{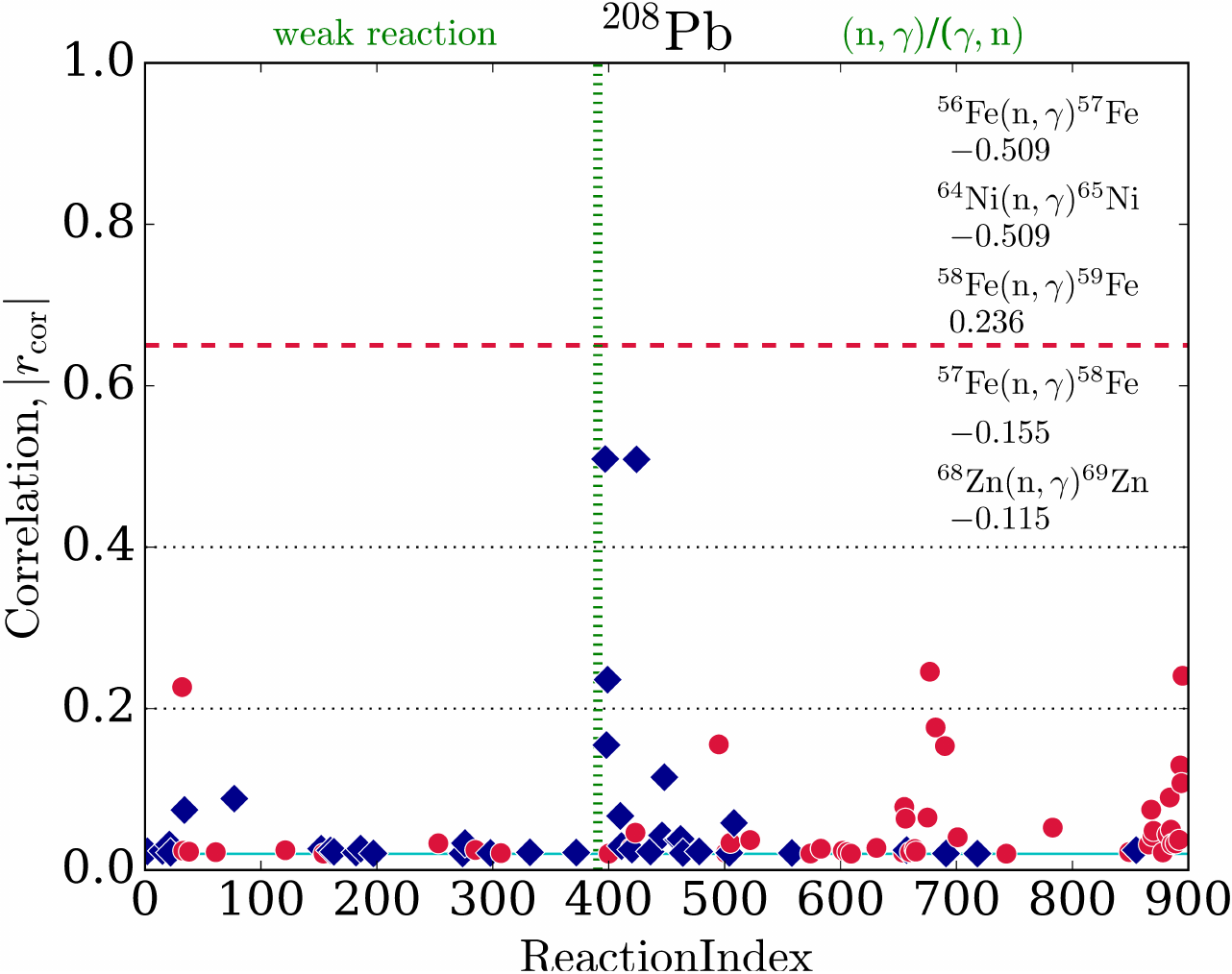}
\caption{Same as for Fig.\,\ref{Sr_cor} for $\iso{Pb}{208}$}
\label{Pb_cor}
\end{figure}

\begin{figure}
\includegraphics[width=\columnwidth]{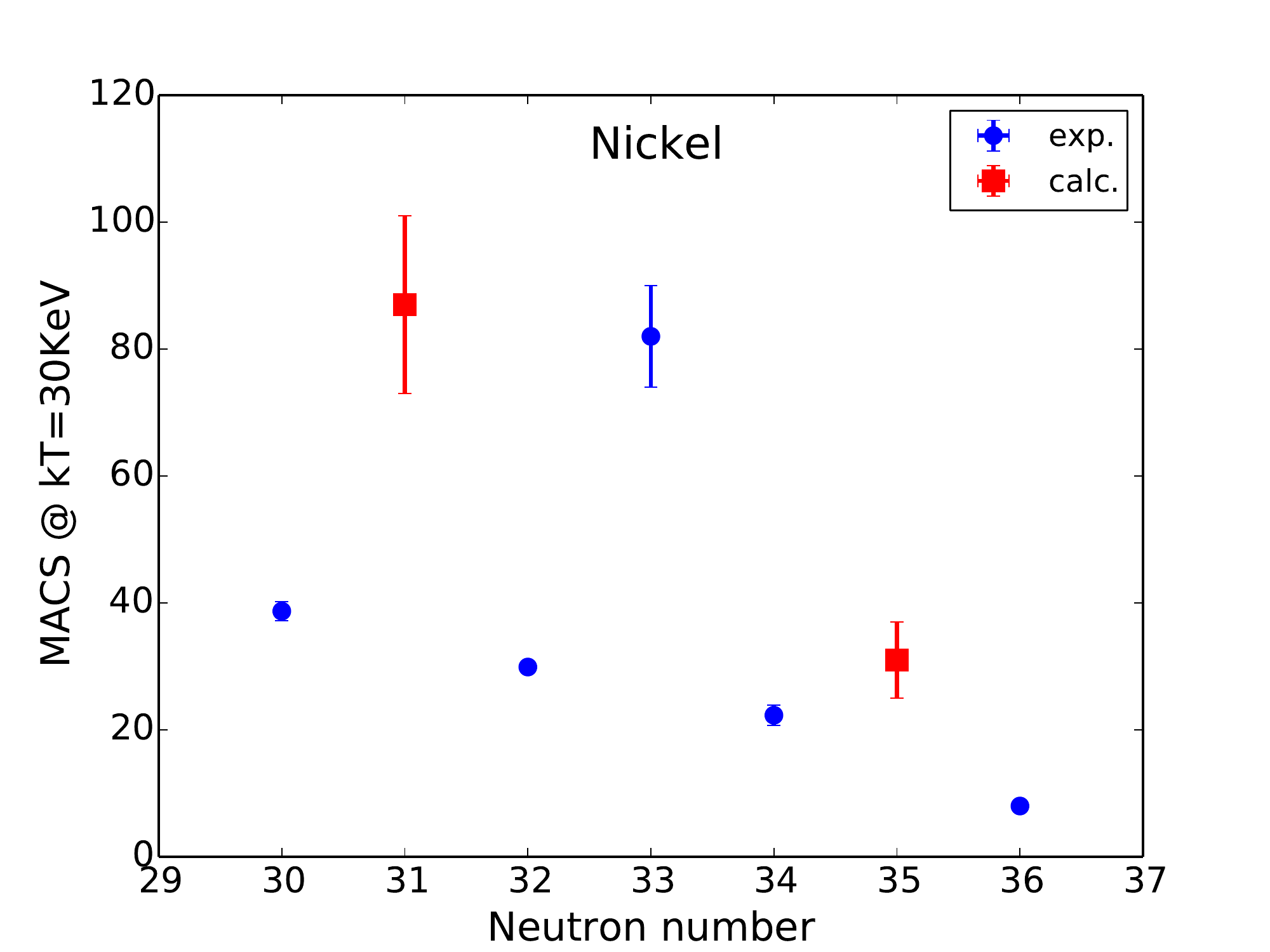}
\caption{Maxwellian averaged cross sections (MACS) at 30 keV for
  Ni isotopes (data taken by the website www.kadonis.org).}
\label{macs}
\end{figure}

As explained above, in most cases, key rates dominating the nuclear
uncertainties are the neutron captures either directly producing or
destroying the nuclide in question. There are, however, three
neutron-capture rates that play a significant role in the uncertainty
for many nuclides  during the $^{13}$C-pocket conditions. These
are the neutron capture rates on $\iso{Fe}{56}$, $\iso{Ni}{64}$, and
$\iso{Ba}{138}$.  Neutron capture rates on $\iso{Fe}{56}$,
$\iso{Ni}{64}$, and $\iso{Ba}{138}$ appear as level 2 key rates for
many nuclides in Table\,\ref{tab:key}.  This means that for many
nuclides local neutron capture rates are still the dominant source of
uncertainty but the importance of these three neutron capture rates
becomes evident by looking the correlation plots for a few key
nuclides: $\iso{Sr}{88}$ (Fig.\,\ref{Sr_cor}), $\iso{Ba}{138}$
(Fig.\,\ref{Ba_cor}), and $\iso{Pb}{208}$ (Fig.\,\ref{Pb_cor}); 
  we have selected these isotopes because they are the most abundant
  for the three main peaks of the s-process path.  In these plots, the
  correlation coefficients of the 900 reactions considered are shown,
  and the five reactions with the highest correlations are listed.
These plots explain the main reason why these two or three neutron
captures are not level 1 key rates: more than one of them contributes
to the total uncertainty. Indeed, it is very rare to have a strong
correlation with more than one rate since correlations with different
rates weaken each other.   Examination of the plots reveals that
  for all three test nuclides, the $\iso{Fe}{56}(\mbox{n},\gamma)$ and
  $\iso{Ni}{64}(\mbox{n},\gamma)$ reactions have high correlation
  factors, albeit they are below our threshold of 0.65.  Only at the
  second level do they appear as key rates, but since these three
  reactions significantly contribute to the uncertainty of so many
nuclides, it makes them priority targets for future measurements.
 For two of these neutron capture reactions, the reason for their
  importance is clear.  $\iso{Fe}{56}(\mbox{n},\gamma)$ affects the
  neutron/seed ratio, while $\iso{Ba}{138}(\mbox{n},\gamma)$ is an
  important bottleneck in the reaction chain. We elucidate the role 
  of $\iso{Ni}{64}(\mbox{n},\gamma)$ by presenting Fig. \ref{macs},
  which shows the Maxwellian averaged cross sections (MACS) at 30 keV
  for a range of Ni isotopes.  The even-neutron isotopes generally
  have smaller MACS values, reducing with increasing number of
  neutrons.  The small value for $\iso{Ni}{64}$ means that
  $\iso{Ni}{64}$ becomes an effective bottleneck in the reaction chain
  towards $\iso{Cu}{65}$ and all heavier nuclei in the s-process
  path. One further reaction that we highlight as being of possible
  interest is that of $\iso{Ce}{140}(\mbox{n},\gamma)$, which although
  identified only at level 3, is found to be a key rate at this level
  for multiple nuclei.


\subsection{Comparison to the weak s-process key rates}

In Table \ref{tab:m-w-s}, we compare the correlation coefficient for
the key reactions for the main s-process, which are also relevant for
the weak s-process \citep[see][]{Nishimura17a}. Not all of the latter
are level 1 key rates for the weak s-process but it is interesting to
know which rates uncertainties affect predictions for both the main
and weak s-process.  In particular,
$\iso{Ge}{72}(\mbox{n},\gamma)\iso{Ge}{73}$,
$\iso{Se}{78}(\mbox{n},\gamma)\iso{Se}{79}$, and
$\iso{Kr}{85}(\mbox{n},\gamma)\iso{Kr}{86}$ are key rates with very
high correlations for both the main and weak s-process. Therefore a
more precise measurement of these rates will enable more precise
nucleosynthesis predictions for both processes.

\begin{table}
\caption{Key rates dominating the production uncertainties for the
    $^{13}$C-pocket conditions and also important for the weak s-process (column 1),
nuclide for which the rate is highly correlated  during the
  $^{13}$C-pocket conditions (2), value of this correlation (3), isotopes for which the rate is correlated in the weak 
s-process production (4), and value of this correlation (5).}

\begin{tabular}{ccrcr}
\hline
Key rates & Nuclide& $r_{{\rm cor},0}$ & Nuclide&  $r_{{\rm cor},0}$ \\
             & main s- &main s-         & weak s- &    weak s-          \\
\hline
$\iso{Ge}{72}(\mbox{n},\gamma)\iso{Ge}{73}$ &$\iso{Ge}{72}$ & -0.93 & $\iso{Ge}{72}$ & -0.85 \\
$\iso{Ge}{74}(\mbox{n},\gamma)\iso{Ge}{75}$ &$\iso{Ge}{74}$ & -0.97 & $\iso{Ge}{74}$  & -0.44 \\
$\iso{As}{75}(\mbox{n},\gamma)\iso{As}{76}$ &$\iso{As}{75}$ & -0.86  & $\iso{As}{75}$ & -0.50 \\
$\iso{Se}{78}(\mbox{n},\gamma)\iso{Se}{79}$ & $\iso{Se}{78}$ & -0.96  & $\iso{Se}{78}$ & -0.71 \\
$\iso{Kr}{84}(\mbox{n},\gamma)\iso{Kr}{85}$ & $\iso{Kr}{84}$ & -0.99 &$\iso{Kr}{84}$ &  -0.49\\
$\iso{Kr}{85}(\mbox{n},\gamma)\iso{Kr}{86}$ & $\iso{Kr}{86}$ & 0.88   &$\iso{Kr}{86}$  &  0.84  \\
\hline
\end{tabular}\label{tab:m-w-s}

\end{table}

\subsection{Opportunities for improved nuclear data}
\label{opp-exp}

A significant number of key reactions have been identified, which thus
become the focus for future experimental work. Of these, the vast
majority are of (n,\,$\gamma$) type, with the remainder being
beta-decays.  By the nature of the scenario being explored, all the
reactions lie along or close to the valley of stability, and
consequently the targets required for (n,\,$\gamma$) studies are
stable or long lived such that solid or gaseous targets of
sufficiently rich isotopic content may be acquired.

Table\,\ref{tab:key} lists the key reactions obtained in the present
MC study. Before embarking on an experimental investigation of any of
the listed reactions, two issues have to be considered, which are
connected to the possible impact of a measurement. The first concerns
the fact that a straightforward measurement of a cross section in the
laboratory yields the cross section for the reaction proceeding on the
ground state of the target nuclei. Depending on the plasma temperature
$T$, however, a considerable fraction of nuclei in a star are in
excited states and reactions on those have to be predicted by
theory. The ground-state contribution to the stellar rate
\citep{xfact,sensi}
\begin{equation}
X_0 (T) = \frac{2J_0+1}{G(T)} \frac{\mathcal{R}_\mathcal{g.s.}(T)}{\mathcal{R}^*(T)}
\end{equation}
quantifies the fraction of the stellar rate which can be constrained
by such a cross section measurement. Here, $J_0$ is the spin of the
ground state and $G(T)$ is the nuclear partition function of the
target nucleus. The reaction rate obtained by energy-averaging the
ground-state cross sections is denoted by $\mathcal{R}_\mathrm{g.s.}$
and the full stellar rate, including reactions on excited states, by
$\mathcal{R}^*$. As described in \citet{Rauscher16}, the $X_0$ were
also used to construct the temperature dependence of the rate
uncertainties. An experiment will only be able to significantly reduce
uncertainties for reactions with large ground-state contributions to
the stellar rate.  Although the stellar temperatures encountered in
the s-process are comparatively low, it has been shown in
\citet{xfact,xfactsproc} that non-negligible excited-state
contributions appear for a number of nuclei also in the s-process,
especially in the rare-earth region.

For convenience, the ground-state contributions $X_0$ at two s-process
temperatures are given for each key neutron capture in Table
\ref{tab:key}.  Most of the reactions have a ground-state contribution
of unity, meaning a laboratory experiment may provide the relevant
nuclear data.

The other issue to be considered before selecting a target for a
measurement is that key rates in our definition are identified by the
strength of the correlation factor, which identifies reactions that
contribute most to the uncertainty of a particular nuclide's abundance
relative to the contributions of all other reactions. It is important
to remember that this does not indicate whether that
abundance-uncertainty itself is large or small, and hence whether it
is of acute interest for improvement -- for this, one must
cross-reference with Table\,\ref{tab-C13-unc} or Fig.\,\ref{L1} to
identify the nuclides having the largest uncertainties in their
abundance.  Doing so reveals where there is scope for updates to the
reaction rate library to be useful. In several cases, there are
already new data published or presently under analysis that are not
yet included; these are detailed below. For others, new precision data
are encouraged.

The Ge(n,\,$\gamma$) reactions have recently been subjected to 
measurements by the n$\_$TOF collaboration~\citep{Ledererntof}
and the data are presently being analysed.   
The $^{78}$Se(n\,$\gamma$) reaction is the subject of a near-future study~\citep{LedererMurphy}
that is motivated in part by previous work~\citep{Nishimura17a} from
this paper's authorship.  $^{79}$Se(n\,$\gamma$) is a well known
branching point for the main s-process and is the topic of another
near-future n$\_$TOF study~\citep{Cesarntof}.
There are also established intentions to pursue Kr(n\,$\gamma$) experiments~\citep{ReneERC}.
An n$\_$TOF study of $^{93}$Zr(n\,$\gamma$) is already published~\citep{tagliente2013}, but it is noted that the conclusions drawn were
limited by the relatively low enrichment (c. 20\%) of the target that was available. 

Our Monte Carlo process reveals a cluster of Rh and Pd nuclides with
slightly increased abundance uncertainties around mass 105; the
associated level-1 key reaction rates are also identified. New
experimental time of flight data for neutron captures on Pd isotopes,
covering the 15-100 keV region, have been provided
by~\cite{Terada}. These report uncertainties improved now to the level
of <6\%.  The data for $^{106}$Pd are interesting as they appear to
show a significant (15-22\%) reduction compared to previous data.
$^{115}$In has a raised abundance uncertainty, identified here as due
to the uncertainty in the $^{115}$In(n,\,$\gamma$) reaction rate.  New
data are available here also~\citep{Katabuchi} that show agreement
with another earlier data set but which disagree (at the level of
$\sim$17\%) with other data sets and evaluations. Further
clarification is required.

For the $^{132}$Xe(n,\,$\gamma$) reaction, the accepted rate is based on the activation study of~\citet{Beer1991}
that has an experimental uncertainty $\sim$8.5\% in the neutron capture cross section at $kT=30~keV$.
In the case of the $^{159}$Tb(n,\,$\gamma$) reaction, the reaction rate used here, and its uncertainty, are based on an average of 
the ENDFB71 and 
JENDL40 evaluated libraries, that in turn are based on several data sets that themselves show some disagreement (see e.g.~\cite{MoxonRae,Mizumoto}). 
The $^{166}$Er, $^{168}$Er and $^{169}$Tm(n,\,$\gamma$) reactions see a similar situation. 
Precision neutron capture data are needed. 

Laboratory measurements of neutron capture cross sections are
typically constrained to investigation of capture to ground
states. Consequently, despite precision measurements, the possibility
of capture on thermally excited states leads to overall greater
uncertainties. Such is the case for the $^{169}$Tm(n,\,$\gamma$),
$^{181}$Ta(n,\,$\gamma$) and $^{187}$Os(n,\,$\gamma$) reactions that
are identified as the level 1 key rates responsible for the increased
uncertainties in the $^{169}$Tm, $^{181}$Ta and $^{187}$Os
abundances. Despite relatively well measured neutron capture cross
sections, excited states at 8.4, 6.2 and 9.8~keV, respectively, lead
to the abundance variations seen in the current study that will be
hard to improve upon by experiment.

Three other nuclides are determined to have poorly constrained
abundances: $^{192,194}$Pt and $^{205}$Pb. The reaction rate library
used throughout this study provides only theoretical rates for the
associated key reactions, for which our approach has been to
consistently assign an uncertainty factor of two.  In fact, recent
experimental data now exist for the $^{192,194}$Pt(n,\,$\gamma$)
level-1 key rates~\citep{Koehler} and thus the new abundance
uncertainties for $^{192,194}$Pt are expected to best represented by
figure~\ref{L2}. This provides a useful illustration of the
improvement that new data can provide.


Several further reactions are of particular interest because of their
broader impact: $^{56}$Fe(n\,$\gamma$), $^{64}$Ni(n\,$\gamma$),
$^{138}$Ba(n\,$\gamma$) and $^{140}$Ce(n\,$\gamma$) are identified as
level 2 and 3 reactions for a large number of nuclear abundances.  For
the first of these, there are a number of published data
sets~\citep{Macklin64, Allen76, Allen82, Wang09}, resulting in an
uncertainty of around~10\%, but given the role of neutron capture on
seed $^{56}$Fe nuclei in this and other nucleosynthesis environments,
greater precision is still needed.  For the $^{64}$Ni(n\,$\gamma$)
reaction, a recent measurement of the thermal neutron capture cross
section has been made~\citep{Shivashankar} and an experiment is
approved at the n$\_$TOF facility~\citep{Heil}. In the case of
$^{138}$Ba(n\,$\gamma$),~\citet{Heil_Ba}, using the $^{18}$O(p,\,n)
reaction that produces neutrons with a 5~keV thermal energy
distribution, measured the Maxwellian-averaged neutron capture cross
section to a precision of about 4\%, in fair agreement with previous
work~\citep{Beer97}. The neutron capture on $^{140}$Ce will be the subject
of another near-future n$\_$TOF measurement \citep{Amaducci18}. Its cross
section, in fact, albeit having been precisely measured at 25 keV
\citep{Kaeppeler96}, needs more precise data at lower energies,
where the dominant resonance at 2.5 keV is poorly constrained.


Further experimental progress is anticipated thanks to new and planned
facilities. At CERN, the second experimental area at
n$\_$TOF~\citep{EAR2} has a shorter flight path to deliver higher
neutron fluxes, while the FRANZ~\citep{franz} facility at the
University of Frankfurt (Germany) and SARAF~\citep{saraf} at the Soreq
research centre (Israel), should soon deliver significantly higher
fluxes, and thus sensitivity and precision.

\section{Conclusions}

For the first time we have performed a comprehensive, large-scale MC
study for the main s$-$process in low mass stars,
varying reactions on targets from Fe to Bi. Temperature-dependent
stellar reaction rate uncertainties were individually assigned to the
reactions, allowing a quantification of the uncertainties in final abundances.

We found that $\beta$-decay rate uncertainties affect only a few
nuclei near s-process branchings, whereas most of the uncertainty in
the final abundances is caused by uncertainties in neutron capture
rates either directly producing or destroying the nuclide of interest.
Combined total nuclear uncertainties due to reactions on heavy
elements are in general small (less than 50\%). This means that
nuclear uncertainties for the main s-process will be dominated by
uncertainties in well known reactions involving light elements, such
as neutron source, e.\,g., $^{13}$C($\alpha$,n) $^{16}$O, and neutron
poisons.

We studied the dependence of the uncertainties and key rates on the
astrophysical conditions found in stars of different masses or
metallicities (neutron to seed ratio) by varying the initial abundance
of $^{13}$C. We found that the key reaction list established is
relevant for the full range of conditions studied. We compared our
results and method to past sensitivity studies focusing on the main
s-process, in particular the comprehensive study of
\citet{Koloczek16}. Our approach clearly determines the key rates that
dominate the total uncertainties in the nucleosynthesis predictions
(rather than showing that a reaction has an impact on a certain number
of nuclides). This is important to ensure that the (re-)measurement of
a key rate will significantly reduce the uncertainties in the final
abundances. While the strongest globally affecting reactions found by
Ko16 are almost all identified as key rates for a few nuclides, they
only dominate the total uncertainties for a few nuclides. The main
exceptions are three key reactions which stand out because they
significantly affect the uncertainties of a larger number of
nuclides. These are $^{56}$Fe(n,$\gamma$), $^{64}$Ni(n,$\gamma$), and
$^{138}$Ba(n,$\gamma$). Improved data for these reactions will lead to
a strong global reduction in prediction uncertainties.

We also compared our key reaction list to the one we determined for
the weak s-process \citep{Nishimura17a}. In particular,
$\iso{Ge}{72}(\mbox{n},\gamma)\iso{Ge}{73}$,
$\iso{Se}{78}(\mbox{n},\gamma)\iso{Se}{79}$, and
$\iso{Kr}{85}(\mbox{n},\gamma)\iso{Kr}{86}$ are key rates with very
high correlations for both the main and weak s-process. Therefore a
more precise measurement of these rates will enable more precise
nucleosynthesis predictions for both processes.

Finally, we discussed the prospect of reducing uncertainties in the
key reactions identified in this study with future experiments. Since
the key rates are for nuclides along the valley of stability, many
have already been measured, which explains the small total
uncertainties. Nevertheless, new improved measurements are feasible
and several are already underway.

\section*{Acknowledgements}

We thank U. Frischknecht for his initial help in the development of
the MC framework and C. Lederer-Woods for her input regards the status
of several experimental works. 
This
work was partially supported by the European Research Council (grants
GA 321263-FISH and EU-FP7-ERC-2012-St Grant 306901), the EU COST
Action CA16117 (ChETEC) and the UK Science and Technology Facilities
Council (grants ST/M000958/1, ST/M001067/1).  This work used the DIRAC
Shared Memory Processing system at the University of Cambridge,
operated by the COSMOS Project at the Department of Applied
Mathematics and Theoretical Physics on behalf of the STFC DiRAC HPC
Facility (www.dirac.ac.uk).  This equipment was funded by BIS National
E-infrastructure capital grant ST/J005673/1, STFC capital grant
ST/H008586/1, and STFC DiRAC Operations grant ST/K00333X/1. DiRAC is
part of the National E-Infrastructure. 
G.C. acknowledges financial support from the European Union Horizon
  2020 research and innovation programme under the Marie Sk\l odowska-Curie grant agreement No. 664931.
The University of Edinburgh is
a charitable body, registered in Scotland, with Registration No.\
SC005336.





\bibliographystyle{mnras}
\bibliography{nucli}



%
%


\appendix

\section{Key rates for double and half of the initial $^{13}$C abundance}
\label{sec:key:nn}

As explained in Sect.\,\ref{sec:sens}, the tables in this Appendix are
provided to assess the sensitivity of the key rate list to the
astrophysical conditions.  We list in Table \ref{tab:key}, the key
  rates for the ``standard'' $\iso{C}{13}$-pocket case and the
  corresponding total uncertainties are shown in Fig. \ref{L1}. In
  Table \ref{tab:hc}, we list the key rates for the ``0.5 $\times$
  $\iso{C}{13}$'' case and the corresponding total uncertainties are
  shown in Fig. \ref{HC}. Similarly, in Table \ref{tab:dc}, we list
  the key rates for the ``2 $\times$ $\iso{C}{13}$'' case and the
  corresponding total uncertainties are shown in Fig. \ref{DC}.  The
  reference key reaction list if that of the ``standard'' case given
  in Table\,\ref{tab:key}. The other two tables are presented for
  discussion and reference only and should not be used to extract key
  rates. Finally, in Table \ref{tab:TP}, the key rates for the TP
  phase are presented and the corresponding total uncertainties are
  shown in Fig. \ref{MC_TP}.

\begin{table*}
\caption{The key reaction rates for the standard model. Key rates in levels $1-3$ are shown, along with their correlation factors $r_{cor 0}$, $r_{cor 1}$ and $r_{cor 2}$, respectively.
Not all s-process nuclides analysed are listed but only those for which key rates were found. Also shown for each rate are
the ground state contributions $X_0$ to the stellar rate of the (n,$\gamma$) reaction and uncertainty factors of the $\beta$-decay 
rate at two plasma temperatures, respectively.}\label{tab:key}
\begin{tabular}{crrrcccccc}
\hline
Nuclide & $r_{{\rm cor},0}$ & $r_{{\rm cor},1}$ & $r_{{\rm cor},2}$&
                                                                     Key rate & Key rate & Key rate& $X_0$ & Weak rate uncertainty 
factor\\
&&&& Level 1 & Level 2 & Level 3 & $(8$, $30~{\rm keV})$ &  $(8$, $30~{\rm keV})$\\
$\iso{Ga}{69}$ & \underline{-0.77} &       &       & $\iso{Ga}{69}(\mbox{n},\gamma)\iso{Ga}{70}$& $ $& $ $&1.00, 1.00&\\
     & -0.34 & \underline{-0.67} &       & $ $& $\iso{Ni}{64}(\mbox{n},\gamma)\iso{Ni}{65}$& $ $&1.00, 1.00&\\
$\iso{Ga}{71}$ & \underline{-0.89} &       &       & $\iso{Ga}{71}(\mbox{n},\gamma)\iso{Ga}{72}$& $ $& $ $&1.00, 1.00&\\
$\iso{Ge}{70}$ & \underline{-0.87} &       &       & $\iso{Ge}{70}(\mbox{n},\gamma)\iso{Ge}{71}$& $ $& $ $&1.00, 1.00&\\
     & -0.27 & \underline{-0.66} &       & $ $& $\iso{Ni}{64}(\mbox{n},\gamma)\iso{Ni}{65}$& $ $&1.00, 1.00&\\
$\iso{Ge}{72}$ & \underline{-0.93} &       &       & $\iso{Ge}{72}(\mbox{n},\gamma)\iso{Ge}{73}$& $ $& $ $&1.00, 1.00&\\
$\iso{Ge}{74}$ & \underline{-0.97} &       &       & $\iso{Ge}{74}(\mbox{n},\gamma)\iso{Ge}{75}$& $ $& $ $&1.00, 1.00&\\
$\iso{As}{75}$ & \underline{-0.86} &       &       & $\iso{As}{75}(\mbox{n},\gamma)\iso{As}{76}$& $ $& $ $&1.00, 1.00&\\
$\iso{Se}{76}$ & \underline{-0.89} &       &       & $\iso{Se}{76}(\mbox{n},\gamma)\iso{Se}{77}$& $ $& $ $&1.00, 1.00&\\
$\iso{Se}{78}$ & \underline{-0.97} &       &       & $\iso{Se}{78}(\mbox{n},\gamma)\iso{Se}{79}$& $ $& $ $&1.00, 1.00&\\
$\iso{Se}{80}$ & \underline{-0.96} &       &       & $\iso{Se}{80}(\mbox{n},\gamma)\iso{Se}{81}$& $ $& $ $&1.00, 1.00&\\
$\iso{Br}{79}$ & \underline{-0.94} &       &       & $\iso{Se}{79}(\mbox{n},\gamma)\iso{Se}{80}$& $ $& $ $&1.00, 1.00&\\
$\iso{Br}{81}$ & \underline{-0.74} &       &       & $\iso{Br}{81}(\mbox{n},\gamma)\iso{Br}{82}$& $ $& $ $&1.00, 1.00&\\
$\iso{Kr}{80}$ & \underline{-0.90} &       &       & $\iso{Se}{79}(\mbox{n},\gamma)\iso{Se}{80}$& $ $& $ $&1.00, 1.00&\\
     &  0.24 & \underline{ 0.85} &       & $ $&
                                                $\iso{Se}{79}(\beta^-)\iso{Br}{79}$& & &1.30, 1.49\\
$\iso{Kr}{82}$ & \underline{-0.97} &       &       & $\iso{Kr}{82}(\mbox{n},\gamma)\iso{Kr}{83}$& $ $& $ $&1.00, 1.00&\\
$\iso{Kr}{84}$ & \underline{-0.98} &       &       & $\iso{Kr}{84}(\mbox{n},\gamma)\iso{Kr}{85}$& $ $& $ $&1.00, 1.00&\\
$\iso{Kr}{86}$ & \underline{ 0.88} &       &       & $\iso{Kr}{85}(\mbox{n},\gamma)\iso{Kr}{86}$& $ $& $ $&1.00, 1.00&\\
     & -0.43 & \underline{-0.95} &       & $ $&
                                                $\iso{Kr}{85}(\beta^-)\iso{Rb}{85}$&
                                                                                                   & & 1.30, 1.30\\
     & -0.12 & -0.28 & \underline{-1.00} & $ $& $ $& $\iso{Kr}{86}(\mbox{n},\gamma)\iso{Kr}{87}$&1.00, 1.00&\\
$\iso{Rb}{85}$ & \underline{-0.86} &       &       & $\iso{Rb}{85}(\mbox{n},\gamma)\iso{Rb}{86}$& $ $& $ $&1.00, 1.00&\\
$\iso{Rb}{87}$ & \underline{ 0.86} &       &       & $\iso{Kr}{85}(\mbox{n},\gamma)\iso{Kr}{86}$& $ $& $ $&1.00, 1.00&\\
     & -0.41 & \underline{-0.85} &       & $ $&
                                                $\iso{Kr}{85}(\beta^-)\iso{Rb}{85}$& & &1.30, 1.30\\
     &  0.20 &  0.39 & \underline{ 0.77} & $ $& $ $& $\iso{Kr}{86}(\mbox{n},\gamma)\iso{Kr}{87}$&1.00, 1.00&\\
$\iso{Sr}{86}$ & \underline{-0.94} &       &       & $\iso{Sr}{86}(\mbox{n},\gamma)\iso{Sr}{87}$& $ $& $ $&1.00, 1.00&\\
$\iso{Sr}{87}$ & \underline{-0.92} &       &       & $\iso{Sr}{87}(\mbox{n},\gamma)\iso{Sr}{88}$& $ $& $ $&1.00, 1.00&\\
$\iso{Sr}{88}$ & \underline{-0.65} &       &       & $\iso{Sr}{88}(\mbox{n},\gamma)\iso{Sr}{89}$& $ $& $ $&1.00, 1.00&\\
     &  0.47 & \underline{ 0.69} &       & $ $& $\iso{Fe}{56}(\mbox{n},\gamma)\iso{Fe}{57}$& $ $&1.00, 1.00&\\
     &  0.47 & \underline{ 0.68} &       & $ $& $\iso{Ni}{64}(\mbox{n},\gamma)\iso{Ni}{65}$& $ $&1.00, 1.00&\\
     &  0.06 &  0.11 & \underline{ 0.65} & $ $& $ $& $\iso{Fe}{58}(\mbox{n},\gamma)\iso{Fe}{59}$&1.00, 1.00&\\
$\iso{Y}{89}$ & \underline{-0.83} &       &       & $\iso{Y}{89}(\mbox{n},\gamma)\iso{Y}{90}$& $ $& $ $&1.00, 1.00&\\
     &  0.33 & \underline{ 0.67} &       & $ $& $\iso{Fe}{56}(\mbox{n},\gamma)\iso{Fe}{57}$& $ $&1.00, 1.00&\\
     &  0.34 & \underline{ 0.68} &       & $ $& $\iso{Ni}{64}(\mbox{n},\gamma)\iso{Ni}{65}$& $ $&1.00, 1.00&\\
     &  0.07 &  0.15 & \underline{ 0.67} & $ $& $ $& $\iso{Fe}{58}(\mbox{n},\gamma)\iso{Fe}{59}$&1.00, 1.00&\\
$\iso{Zr}{90}$ & \underline{-0.89} &       &       & $\iso{Zr}{90}(\mbox{n},\gamma)\iso{Zr}{91}$& $ $& $ $&1.00, 1.00&\\
     &  0.28 & \underline{ 0.68} &       & $ $& $\iso{Ni}{64}(\mbox{n},\gamma)\iso{Ni}{65}$& $ $&1.00, 1.00&\\
$\iso{Zr}{92}$ & \underline{-0.92} &       &       & $\iso{Zr}{92}(\mbox{n},\gamma)\iso{Zr}{93}$& $ $& $ $&1.00, 1.00&\\
     &  0.22 & \underline{ 0.67} &       & $ $& $\iso{Ni}{64}(\mbox{n},\gamma)\iso{Ni}{65}$& $ $&1.00, 1.00&\\
$\iso{Zr}{94}$ & \underline{-0.86} &       &       & $\iso{Zr}{94}(\mbox{n},\gamma)\iso{Zr}{95}$& $ $& $ $&1.00, 1.00&\\
     &  0.30 & \underline{ 0.65} &       & $ $& $\iso{Ni}{64}(\mbox{n},\gamma)\iso{Ni}{65}$& $ $&1.00, 1.00&\\
$\iso{Nb}{93}$ & \underline{-0.97} &       &       & $\iso{Zr}{93}(\mbox{n},\gamma)\iso{Zr}{94}$& $ $& $ $&1.00, 1.00&\\
     &  0.14 & \underline{ 0.67} &       & $ $& $\iso{Ni}{64}(\mbox{n},\gamma)\iso{Ni}{65}$& $ $&1.00, 1.00&\\
$\iso{Mo}{95}$ & \underline{-0.85} &       &       & $\iso{Mo}{95}(\mbox{n},\gamma)\iso{Mo}{96}$& $ $& $ $&1.00, 1.00&\\
     &  0.29 & \underline{ 0.65} &       & $ $& $\iso{Ni}{64}(\mbox{n},\gamma)\iso{Ni}{65}$& $ $&1.00, 1.00&\\
$\iso{Mo}{96}$ & \underline{-0.94} &       &       & $\iso{Mo}{96}(\mbox{n},\gamma)\iso{Mo}{97}$& $ $& $ $&1.00, 1.00&\\
$\iso{Mo}{97}$ & \underline{-0.87} &       &       & $\iso{Mo}{97}(\mbox{n},\gamma)\iso{Mo}{98}$& $ $& $ $&1.00, 1.00&\\
$\iso{Mo}{98}$ & \underline{-0.94} &       &       & $\iso{Mo}{98}(\mbox{n},\gamma)\iso{Mo}{99}$& $ $& $ $&1.00, 1.00&\\
$\iso{Ru}{99}$ & \underline{-0.91} &       &       & $\iso{Tc}{99}(\mbox{n},\gamma)\iso{Tc}{100}$& $ $& $ $&1.00, 1.00&\\
$\iso{Ru}{100}$ & \underline{-0.93} &       &       & $\iso{Ru}{100}(\mbox{n},\gamma)\iso{Ru}{101}$& $ $& $ $&1.00, 1.00&\\
$\iso{Ru}{102}$ & \underline{-0.86} &       &       & $\iso{Ru}{102}(\mbox{n},\gamma)\iso{Ru}{103}$& $ $& $ $&1.00, 1.00&\\
$\iso{Rh}{103}$ & \underline{-0.95} &       &       & $\iso{Rh}{103}(\mbox{n},\gamma)\iso{Rh}{104}$& $ $& $ $&0.95, 0.80&\\
$\iso{Pd}{104}$ & \underline{-0.97} &       &       & $\iso{Pd}{104}(\mbox{n},\gamma)\iso{Pd}{105}$& $ $& $ $&1.00, 1.00&\\
$\iso{Pd}{106}$ & \underline{-0.97} &       &       & $\iso{Pd}{106}(\mbox{n},\gamma)\iso{Pd}{107}$& $ $& $ $&1.00, 1.00&\\
$\iso{Pd}{108}$ & \underline{-0.96} &       &       & $\iso{Pd}{108}(\mbox{n},\gamma)\iso{Pd}{109}$& $ $& $ $&1.00, 1.00&\\
$\iso{Ag}{107}$ & \underline{-0.81} &       &       & $\iso{Pd}{107}(\mbox{n},\gamma)\iso{Pd}{108}$& $ $& $ $&1.00, 1.00&\\
$\iso{Ag}{109}$ & \underline{-0.80} &       &       & $\iso{Ag}{109}(\mbox{n},\gamma)\iso{Ag}{110}$& $ $& $ $&1.00, 1.00&\\
$\iso{Cd}{110}$ & -0.41 & -0.48 & \underline{-0.71} & $ $& $ $& $\iso{Cd}{110}(\mbox{n},\gamma)\iso{Cd}{111}$&1.00, 1.00&\\
$\iso{Cd}{112}$ & -0.40 & -0.45 & \underline{-0.69} & $ $& $ $& $\iso{Cd}{112}(\mbox{n},\gamma)\iso{Cd}{113}$&1.00, 1.00&\\
$\iso{Cd}{114}$ & -0.36 & -0.43 & \underline{-0.65} & $ $& $ $& $\iso{Cd}{114}(\mbox{n},\gamma)\iso{Cd}{115}$&1.00, 1.00&\\
$\iso{In}{115}$ & \underline{-0.97} &       &       & $\iso{In}{115}(\mbox{n},\gamma)\iso{In}{116}$& $ $& $ $&1.00, 1.00&\\
\hline
\end{tabular}
\end{table*}

\begin{table*}

\begin{tabular}{crrrcccccc}
\hline
Nuclide & $r_{{\rm cor},0}$ & $r_{{\rm cor},1}$ & $r_{{\rm cor},2}$&
                                                                     Key rate & Key rate & Key rate& $X_0$ & Weak rate uncertainty 
factor\\
&&&& Level 1 & Level 2 & level 3 & $(8$, $30~{\rm keV})$& $(8$, $30~{\rm keV})$\\
\hline
$\iso{Sn}{116}$ & -0.51 & -0.58 & \underline{-0.78} & $ $& $ $& $\iso{Sn}{116}(\mbox{n},\gamma)\iso{Sn}{117}$&1.00, 1.00&\\
$\iso{Sn}{118}$ & -0.59 & \underline{-0.67} &       & $ $& $\iso{Sn}{118}(\mbox{n},\gamma)\iso{Sn}{119}$& $ $&1.00, 1.00&\\
$\iso{Sn}{120}$ & -0.57 & \underline{-0.67} &       & $ $& $\iso{Sn}{120}(\mbox{n},\gamma)\iso{Sn}{121}$& $ $&1.00, 1.00&\\
$\iso{Sb}{121}$ & \underline{-0.92} &       &       & $\iso{Sb}{121}(\mbox{n},\gamma)\iso{Sb}{122}$& $ $& $ $&0.98, 0.93&\\
$\iso{Te}{124}$ & -0.53 & -0.65 & \underline{-0.76} & $ $& $ $& $\iso{Te}{124}(\mbox{n},\gamma)\iso{Te}{125}$&1.00, 1.00&\\
$\iso{Te}{126}$ & \underline{-0.69} &       &       & $\iso{Te}{126}(\mbox{n},\gamma)\iso{Te}{127}$& $ $& $ $&1.00, 1.00&\\
$\iso{I}{127}$ & \underline{-0.92} &       &       & $\iso{I}{127}(\mbox{n},\gamma)\iso{I}{128}$& $ $& $ $&1.00, 0.99&\\
$\iso{Xe}{128}$ & \underline{ 0.66} &       &       &
                                                      $\iso{I}{128}(\beta^-)\iso{Xe}{128}$&
                                                                                            $ $& & & 1.64, 5.42\\
$\iso{Xe}{130}$ & -0.57 & \underline{-0.71} &       & $ $& $\iso{Xe}{130}(\mbox{n},\gamma)\iso{Xe}{131}$& $ $&1.00, 1.00&\\
$\iso{Xe}{132}$ & \underline{-0.97} &       &       & $\iso{Xe}{132}(\mbox{n},\gamma)\iso{Xe}{133}$& $ $& $ $&1.00, 1.00&\\
$\iso{Cs}{133}$ & \underline{-0.89} &       &       & $\iso{Cs}{133}(\mbox{n},\gamma)\iso{Cs}{134}$& $ $& $ $&1.00, 1.00&\\
$\iso{Ba}{134}$ & \underline{-0.85} &       &       & $\iso{Ba}{134}(\mbox{n},\gamma)\iso{Ba}{135}$& $ $& $ $&1.00, 1.00&\\
$\iso{Ba}{136}$ & \underline{-0.88} &       &       & $\iso{Ba}{136}(\mbox{n},\gamma)\iso{Ba}{137}$& $ $& $ $&1.00, 1.00&\\
$\iso{Ba}{137}$ & \underline{-0.84} &       &       & $\iso{Ba}{137}(\mbox{n},\gamma)\iso{Ba}{138}$& $ $& $ $&1.00, 1.00&\\
$\iso{Ba}{138}$ & -0.65 & \underline{-0.73} &       & $ $& $\iso{Ba}{138}(\mbox{n},\gamma)\iso{Ba}{139}$& $ $&1.00, 1.00&\\
$\iso{La}{139}$ & \underline{-0.87} &       &       & $\iso{La}{139}(\mbox{n},\gamma)\iso{La}{140}$& $ $& $ $&1.00, 1.00&\\
     &  0.36 & \underline{ 0.83} &       & $ $& $\iso{Ba}{138}(\mbox{n},\gamma)\iso{Ba}{139}$& $ $&1.00, 1.00&\\
$\iso{Ce}{140}$ &  0.59 & \underline{ 0.65} &       & $ $& $\iso{Ba}{138}(\mbox{n},\gamma)\iso{Ba}{139}$& $ $&1.00, 1.00&\\
     & -0.39 & -0.42 & \underline{-0.90} & $ $& $ $& $\iso{Ce}{140}(\mbox{n},\gamma)\iso{Ce}{141}$&1.00, 1.00&\\
$\iso{Pr}{141}$ &  0.59 & \underline{ 0.65} &       & $ $& $\iso{Ba}{138}(\mbox{n},\gamma)\iso{Ba}{139}$& $ $&1.00, 1.00&\\
     &  0.31 &  0.33 & \underline{ 0.85} & $ $& $ $& $\iso{Ce}{140}(\mbox{n},\gamma)\iso{Ce}{141}$&1.00, 1.00&\\
$\iso{Nd}{142}$ & -0.31 & -0.34 & \underline{-0.67} & $ $& $ $& $\iso{Nd}{142}(\mbox{n},\gamma)\iso{Nd}{143}$&1.00, 1.00&\\
$\iso{Nd}{146}$ &  0.28 &  0.30 & \underline{ 0.76} & $ $& $ $& $\iso{Ce}{140}(\mbox{n},\gamma)\iso{Ce}{141}$&1.00, 1.00&\\
$\iso{Sm}{147}$ &  0.28 &  0.30 & \underline{ 0.74} & $ $& $ $& $\iso{Ce}{140}(\mbox{n},\gamma)\iso{Ce}{141}$&1.00, 1.00&\\
$\iso{Sm}{148}$ &  0.28 &  0.29 & \underline{ 0.74} & $ $& $ $& $\iso{Ce}{140}(\mbox{n},\gamma)\iso{Ce}{141}$&1.00, 1.00&\\
$\iso{Sm}{150}$ &  0.28 &  0.29 & \underline{ 0.76} & $ $& $ $& $\iso{Ce}{140}(\mbox{n},\gamma)\iso{Ce}{141}$&1.00, 1.00&\\
$\iso{Eu}{151}$ & \underline{-0.70} &       &       & $\iso{Eu}{151}(\mbox{n},\gamma)\iso{Eu}{152}$& $ $& $ $&0.89, 0.79&\\
     &  0.19 &  0.29 & \underline{ 0.68} & $ $& $ $& $\iso{Ce}{140}(\mbox{n},\gamma)\iso{Ce}{141}$&1.00, 1.00&\\
$\iso{Gd}{152}$ &  0.59 &  0.61 & \underline{ 0.79} & $ $& $ $&
                                                                $\iso{Sm}{151}(\beta^-)\iso{Eu}{151}$&
                                                                                                                                 & 3.60, 5.42\\
$\iso{Gd}{154}$ &  0.27 &  0.29 & \underline{ 0.71} & $ $& $ $& $\iso{Ce}{140}(\mbox{n},\gamma)\iso{Ce}{141}$&1.00, 1.00&\\
$\iso{Gd}{156}$ &  0.27 &  0.28 & \underline{ 0.67} & $ $& $ $& $\iso{Ce}{140}(\mbox{n},\gamma)\iso{Ce}{141}$&1.00, 1.00&\\
$\iso{Tb}{159}$ & \underline{-0.79} &       &       & $\iso{Tb}{159}(\mbox{n},\gamma)\iso{Tb}{160}$& $ $& $ $&1.00, 0.98&\\
     &  0.16 &  0.29 & \underline{ 0.74} & $ $& $ $& $\iso{Ce}{140}(\mbox{n},\gamma)\iso{Ce}{141}$&1.00, 1.00&\\
$\iso{Dy}{164}$ & -0.35 & -0.38 & \underline{-0.71} & $ $& $ $& $\iso{Dy}{164}(\mbox{n},\gamma)\iso{Dy}{165}$&1.00, 0.97&\\
$\iso{Ho}{165}$ & \underline{-0.68} &       &       & $\iso{Ho}{165}(\mbox{n},\gamma)\iso{Ho}{166}$& $ $& $ $&1.00, 1.00&\\
     &  0.19 &  0.28 & \underline{ 0.72} & $ $& $ $& $\iso{Ce}{140}(\mbox{n},\gamma)\iso{Ce}{141}$&1.00, 1.00&\\
$\iso{Er}{166}$ & \underline{-0.81} &       &       & $\iso{Er}{166}(\mbox{n},\gamma)\iso{Er}{167}$& $ $& $ $&1.00, 0.98&\\
     &  0.15 &  0.28 & \underline{ 0.72} & $ $& $ $& $\iso{Ce}{140}(\mbox{n},\gamma)\iso{Ce}{141}$&1.00, 1.00&\\
$\iso{Er}{167}$ & \underline{-0.78} &       &       & $\iso{Er}{167}(\mbox{n},\gamma)\iso{Er}{168}$& $ $& $ $&1.00, 1.00&\\
     &  0.17 &  0.28 & \underline{ 0.72} & $ $& $ $& $\iso{Ce}{140}(\mbox{n},\gamma)\iso{Ce}{141}$&1.00, 1.00&\\
$\iso{Er}{168}$ & \underline{-0.86} &       &       & $\iso{Er}{168}(\mbox{n},\gamma)\iso{Er}{169}$& $ $& $ $&1.00, 0.98&\\
     &  0.11 &  0.28 & \underline{ 0.72} & $ $& $ $& $\iso{Ce}{140}(\mbox{n},\gamma)\iso{Ce}{141}$&1.00, 1.00&\\
$\iso{Tm}{169}$ & \underline{-0.90} &       &       & $\iso{Tm}{169}(\mbox{n},\gamma)\iso{Tm}{170}$& $ $& $ $&0.51, 0.42&\\
     &  0.10 &  0.28 & \underline{ 0.71} & $ $& $ $& $\iso{Ce}{140}(\mbox{n},\gamma)\iso{Ce}{141}$&1.00, 1.00&\\
$\iso{Lu}{175}$ &  0.26 &  0.27 & \underline{ 0.69} & $ $& $ $& $\iso{Ce}{140}(\mbox{n},\gamma)\iso{Ce}{141}$&1.00, 1.00&\\
$\iso{Lu}{176}$ &  0.25 &  0.26 & \underline{ 0.65} & $ $& $ $& $\iso{Ce}{140}(\mbox{n},\gamma)\iso{Ce}{141}$&1.00, 1.00&\\
$\iso{Hf}{176}$ &  0.61 &  0.63 & \underline{ 0.89} & $ $& $ $& $\iso{Lu}{176}(\beta^-)\iso{Hf}{176}$&&1.30, 1.33\\
$\iso{Ta}{181}$ & \underline{-0.84} &       &       & $\iso{Ta}{181}(\mbox{n},\gamma)\iso{Ta}{182}$& $ $& $ $&0.61, 0.55&\\
     &  0.12 &  0.26 & \underline{ 0.67} & $ $& $ $& $\iso{Ce}{140}(\mbox{n},\gamma)\iso{Ce}{141}$&1.00, 1.00&\\
$\iso{W}{183}$ & -0.49 & -0.51 & \underline{-0.82} & $ $& $ $& $\iso{W}{183}(\mbox{n},\gamma)\iso{W}{184}$&0.99, 0.93&\\
$\iso{Os}{187}$ & \underline{-0.86} &       &       & $\iso{Os}{187}(\mbox{n},\gamma)\iso{Os}{188}$& $ $& $ $&0.57, 0.46&\\
$\iso{Os}{188}$ & -0.34 & -0.35 & \underline{-0.67} & $ $& $ $& $\iso{Os}{188}(\mbox{n},\gamma)\iso{Os}{189}$&1.00, 1.00&\\
$\iso{Ir}{193}$ & -0.58 & -0.59 & \underline{-0.84} & $ $& $ $& $\iso{Ir}{193}(\mbox{n},\gamma)\iso{Ir}{194}$&1.00, 0.99&\\
$\iso{Pt}{192}$ & \underline{-0.89} &       &       & $\iso{Pt}{192}(\mbox{n},\gamma)\iso{Pt}{193}$& $ $& $ $&1.00, 1.00&\\
$\iso{Pt}{194}$ & \underline{-0.90} &       &       & $\iso{Pt}{194}(\mbox{n},\gamma)\iso{Pt}{195}$& $ $& $ $&1.00, 1.00&\\
$\iso{Pt}{196}$ & -0.63 & -0.65 & \underline{-0.89} & $ $& $ $& $\iso{Pt}{196}(\mbox{n},\gamma)\iso{Pt}{197}$&1.00, 1.00&\\
$\iso{Hg}{198}$ & -0.63 & \underline{-0.65} &       & $ $& $\iso{Hg}{198}(\mbox{n},\gamma)\iso{Hg}{199}$& $ $&1.00, 1.00&\\
$\iso{Hg}{200}$ & \underline{-0.67} &       &       & $\iso{Hg}{200}(\mbox{n},\gamma)\iso{Hg}{201}$& $ $& $ $&1.00, 1.00&\\
$\iso{Tl}{203}$ & -0.48 & -0.49 & \underline{-0.78} & $ $& $ $& $\iso{Tl}{203}(\mbox{n},\gamma)\iso{Tl}{204}$&1.00, 1.00&\\
$\iso{Tl}{205}$ & \underline{-0.87} &       &       & $\iso{Pb}{205}(\mbox{n},\gamma)\iso{Pb}{206}$& $ $& $ $&0.83, 0.82&\\
$\iso{Bi}{209}$ &  0.53 &  0.56 & \underline{ 0.71} & $ $& $ $& $\iso{Pb}{208}(\mbox{n},\gamma)\iso{Pb}{209}$&1.00, 1.00&\\

\hline
\end{tabular}
\end{table*}

\begin{table*}
\caption{Key reaction rates for the model with half the initial
  $^{13}$C abundance  (``0.5 $\times$ $\iso{C}{13}$'' case).
 Key rates in levels $1-3$ are shown, along with their correlation factors $r_{cor 0}$, $r_{cor 1}$ and $r_{cor 2}$, respectively.
Not all s-process nuclides analysed are listed but only those for which key rates were found. Also shown for each rate are
the ground-state contributions of the (n,$\gamma$) reaction to the stellar
rate and uncertainty factors of the $\beta$-decay rate at two plasma
temperatures, respectively. 
} \label{tab:hc}
\begin{tabular}{crrrcccccc}
\hline
Nuclide & $r_{{\rm cor},0}$ & $r_{{\rm cor},1}$ & $r_{{\rm cor},2}$& Key rate & Key rate & Key rate& $X_0$ $(8$, $30~{\rm keV})$ & $\beta$-decay \\
&&&& Level 1 & Level 2 & Level 3 & & \\
\hline
$\iso{Ga}{69}$ & \underline{-0.88} &       &       & $\iso{Ga}{69}(\mbox{n},\gamma)\iso{Ga}{70}$& $ $& $ $&1.00, 1.00&\\
     &  0.33 & \underline{ 0.73} &       & $ $& $\iso{Fe}{56}(\mbox{n},\gamma)\iso{Fe}{57}$& $ $&1.00, 1.00&\\
     &  0.29 & \underline{ 0.65} &       & $ $& $\iso{Ni}{64}(\mbox{n},\gamma)\iso{Ni}{65}$& $ $&1.00, 1.00&\\
$\iso{Ga}{71}$ & \underline{-0.93} &       &       & $\iso{Ga}{71}(\mbox{n},\gamma)\iso{Ga}{72}$& $ $& $ $&1.00, 1.00&\\
     &  0.26 & \underline{ 0.72} &       & $ $& $\iso{Fe}{56}(\mbox{n},\gamma)\iso{Fe}{57}$& $ $&1.00, 1.00&\\
     &  0.25 & \underline{ 0.67} &       & $ $& $\iso{Ni}{64}(\mbox{n},\gamma)\iso{Ni}{65}$& $ $&1.00, 1.00&\\
     &  0.04 &  0.12 & \underline{ 0.72} & $ $& $ $& $\iso{Fe}{58}(\mbox{n},\gamma)\iso{Fe}{59}$&1.00, 1.00&\\
$\iso{Ge}{70}$ & \underline{-0.92} &       &       & $\iso{Ge}{70}(\mbox{n},\gamma)\iso{Ge}{71}$& $ $& $ $&1.00, 1.00&\\
     &  0.25 & \underline{ 0.73} &       & $ $& $\iso{Fe}{56}(\mbox{n},\gamma)\iso{Fe}{57}$& $ $&1.00, 1.00&\\
     &  0.25 & \underline{ 0.66} &       & $ $& $\iso{Ni}{64}(\mbox{n},\gamma)\iso{Ni}{65}$& $ $&1.00, 1.00&\\
     &  0.05 &  0.11 & \underline{ 0.67} & $ $& $ $& $\iso{Fe}{58}(\mbox{n},\gamma)\iso{Fe}{59}$&1.00, 1.00&\\
$\iso{Ge}{72}$ & \underline{-0.93} &       &       & $\iso{Ge}{72}(\mbox{n},\gamma)\iso{Ge}{73}$& $ $& $ $&1.00, 1.00&\\
     &  0.06 & \underline{ 0.71} &       & $ $& $\iso{Fe}{56}(\mbox{n},\gamma)\iso{Fe}{57}$& $ $&1.00, 1.00&\\
     &  0.06 & \underline{ 0.67} &       & $ $& $\iso{Ni}{64}(\mbox{n},\gamma)\iso{Ni}{65}$& $ $&1.00, 1.00&\\
     &  0.02 &  0.14 & \underline{ 0.76} & $ $& $ $& $\iso{Fe}{58}(\mbox{n},\gamma)\iso{Fe}{59}$&1.00, 1.00&\\
$\iso{Ge}{74}$ & -0.96 & \underline{-0.96} &       & $ $& $\iso{Ge}{74}(\mbox{n},\gamma)\iso{Ge}{75}$& $ $&1.00, 1.00&\\
     &  0.05 &  0.06 & \underline{ 0.73} & $ $& $ $& $\iso{Fe}{58}(\mbox{n},\gamma)\iso{Fe}{59}$&1.00, 1.00&\\
$\iso{As}{75}$ & \underline{-0.83} &       &       & $\iso{As}{75}(\mbox{n},\gamma)\iso{As}{76}$& $ $& $ $&1.00, 1.00&\\
     &  0.37 & \underline{ 0.69} &       & $ $& $\iso{Fe}{56}(\mbox{n},\gamma)\iso{Fe}{57}$& $ $&1.00, 1.00&\\
     &  0.36 & \underline{ 0.68} &       & $ $& $\iso{Ni}{64}(\mbox{n},\gamma)\iso{Ni}{65}$& $ $&1.00, 1.00&\\
     &  0.10 &  0.18 & \underline{ 0.73} & $ $& $ $& $\iso{Fe}{58}(\mbox{n},\gamma)\iso{Fe}{59}$&1.00, 1.00&\\
$\iso{Se}{76}$ & \underline{-0.86} &       &       & $\iso{Se}{76}(\mbox{n},\gamma)\iso{Se}{77}$& $ $& $ $&1.00, 1.00&\\
     &  0.35 & \underline{ 0.69} &       & $ $& $\iso{Fe}{56}(\mbox{n},\gamma)\iso{Fe}{57}$& $ $&1.00, 1.00&\\
     &  0.34 & \underline{ 0.68} &       & $ $& $\iso{Ni}{64}(\mbox{n},\gamma)\iso{Ni}{65}$& $ $&1.00, 1.00&\\
     &  0.11 &  0.19 & \underline{ 0.71} & $ $& $ $& $\iso{Fe}{58}(\mbox{n},\gamma)\iso{Fe}{59}$&1.00, 1.00&\\
$\iso{Se}{78}$ & \underline{-0.97} &       &       & $\iso{Se}{78}(\mbox{n},\gamma)\iso{Se}{79}$& $ $& $ $&1.00, 1.00&\\
     &  0.12 & \underline{ 0.67} &       & $ $& $\iso{Fe}{56}(\mbox{n},\gamma)\iso{Fe}{57}$& $ $&1.00, 1.00&\\
     &  0.12 & \underline{ 0.68} &       & $ $& $\iso{Ni}{64}(\mbox{n},\gamma)\iso{Ni}{65}$& $ $&1.00, 1.00&\\
     &  0.05 &  0.21 & \underline{ 0.67} & $ $& $ $& $\iso{Fe}{58}(\mbox{n},\gamma)\iso{Fe}{59}$&1.00, 1.00&\\
$\iso{Se}{80}$ & \underline{-0.91} &       &       & $\iso{Se}{80}(\mbox{n},\gamma)\iso{Se}{81}$& $ $& $ $&1.00, 1.00&\\
     &  0.25 & \underline{ 0.66} &       & $ $& $\iso{Ni}{64}(\mbox{n},\gamma)\iso{Ni}{65}$& $ $&1.00, 1.00&\\
$\iso{Br}{79}$ & \underline{-0.94} &       &       & $\iso{Se}{79}(\mbox{n},\gamma)\iso{Se}{80}$& $ $& $ $&1.00, 1.00&\\
     &  0.05 & \underline{ 0.65} &       & $ $& $\iso{Ni}{64}(\mbox{n},\gamma)\iso{Ni}{65}$& $ $&1.00, 1.00&\\
$\iso{Br}{81}$ & -0.58 & -0.59 & \underline{-0.83} & $ $& $ $& $\iso{Br}{81}(\mbox{n},\gamma)\iso{Br}{82}$&1.00, 1.00&\\
$\iso{Kr}{80}$ & \underline{-0.90} &       &       & $\iso{Se}{79}(\mbox{n},\gamma)\iso{Se}{80}$& $ $& $ $&1.00, 1.00&\\
     &  0.25 & \underline{ 0.80} &       & $ $& $\iso{Se}{79}(\beta^-)\iso{Br}{79}$& $ $& & 1.30, 1.49\\
$\iso{Kr}{82}$ & \underline{-0.91} &       &       & $\iso{Kr}{82}(\mbox{n},\gamma)\iso{Kr}{83}$& $ $& $ $&1.00, 1.00&\\
     &  0.25 & \underline{ 0.66} &       & $ $& $\iso{Ni}{64}(\mbox{n},\gamma)\iso{Ni}{65}$& $ $&1.00, 1.00&\\
$\iso{Kr}{84}$ & \underline{-0.96} &       &       & $\iso{Kr}{84}(\mbox{n},\gamma)\iso{Kr}{85}$& $ $& $ $&1.00, 1.00&\\
     &  0.17 & \underline{ 0.65} &       & $ $& $\iso{Ni}{64}(\mbox{n},\gamma)\iso{Ni}{65}$& $ $&1.00, 1.00&\\
$\iso{Kr}{86}$ & \underline{ 0.88} &       &       & $\iso{Kr}{85}(\mbox{n},\gamma)\iso{Kr}{86}$& $ $& $ $&1.00, 1.00&\\
     & -0.43 & \underline{-0.97} &       & $ $& $\iso{Kr}{85}(\beta^-)\iso{Rb}{85}$& $ $& & 1.30, 1.30\\
     & -0.07 & -0.12 & \underline{-0.93} & $ $& $ $& $\iso{Kr}{86}(\mbox{n},\gamma)\iso{Kr}{87}$&1.00, 1.00&\\
$\iso{Rb}{85}$ & \underline{-0.66} &       &       & $\iso{Rb}{85}(\mbox{n},\gamma)\iso{Rb}{86}$& $ $& $ $&1.00, 1.00&\\
     &  0.47 & \underline{ 0.65} &       & $ $& $\iso{Ni}{64}(\mbox{n},\gamma)\iso{Ni}{65}$& $ $&1.00, 1.00&\\
$\iso{Rb}{87}$ & \underline{ 0.84} &       &       & $\iso{Kr}{85}(\mbox{n},\gamma)\iso{Kr}{86}$& $ $& $ $&1.00, 1.00&\\
     & -0.42 & \underline{-0.82} &       & $ $& $\iso{Kr}{85}(\beta^-)\iso{Rb}{85}$& $ $& & 1.30, 1.30\\
     &  0.24 &  0.48 & \underline{ 0.88} & $ $& $ $& $\iso{Kr}{86}(\mbox{n},\gamma)\iso{Kr}{87}$&1.00, 1.00&\\
$\iso{Sr}{86}$ & \underline{-0.85} &       &       & $\iso{Sr}{86}(\mbox{n},\gamma)\iso{Sr}{87}$& $ $& $ $&1.00, 1.00&\\
$\iso{Sr}{87}$ & \underline{-0.81} &       &       & $\iso{Sr}{87}(\mbox{n},\gamma)\iso{Sr}{88}$& $ $& $ $&1.00, 1.00&\\
$\iso{Y}{89}$ & \underline{-0.69} &       &       & $\iso{Y}{89}(\mbox{n},\gamma)\iso{Y}{90}$& $ $& $ $&1.00, 1.00&\\
$\iso{Zr}{90}$ & \underline{-0.73} &       &       & $\iso{Zr}{90}(\mbox{n},\gamma)\iso{Zr}{91}$& $ $& $ $&1.00, 1.00&\\
$\iso{Zr}{92}$ & \underline{-0.79} &       &       & $\iso{Zr}{92}(\mbox{n},\gamma)\iso{Zr}{93}$& $ $& $ $&1.00, 1.00&\\
$\iso{Zr}{94}$ & -0.57 & \underline{-0.71} &       & $ $& $\iso{Zr}{94}(\mbox{n},\gamma)\iso{Zr}{95}$& $ $&1.00, 1.00&\\
$\iso{Nb}{93}$ & \underline{-0.92} &       &       & $\iso{Zr}{93}(\mbox{n},\gamma)\iso{Zr}{94}$& $ $& $ $&1.00, 1.00&\\
$\iso{Mo}{95}$ & -0.64 & \underline{-0.76} &       & $ $& $\iso{Mo}{95}(\mbox{n},\gamma)\iso{Mo}{96}$& $ $&1.00, 1.00&\\
$\iso{Mo}{96}$ & \underline{-0.81} &       &       & $\iso{Mo}{96}(\mbox{n},\gamma)\iso{Mo}{97}$& $ $& $ $&1.00, 1.00&\\
     & -0.25 & \underline{-0.65} &       & $ $& $\iso{Fe}{56}(\mbox{n},\gamma)\iso{Fe}{57}$& $ $&1.00, 1.00&\\

\hline

\end{tabular}
\end{table*}

\begin{table*}
\begin{tabular}{crrrcccccc}
\hline
Nuclide & $r_{{\rm cor},0}$ & $r_{{\rm cor},1}$ & $r_{{\rm cor},2}$& Key rate & Key rate & Key rate& $X_0$ $(8$, $30~{\rm keV})$ & $\beta$-decay \\
&&&& Level 1 & Level 2 & Level 3 & & \\
\hline
$\iso{Mo}{97}$ & \underline{-0.65} &       &       & $\iso{Mo}{97}(\mbox{n},\gamma)\iso{Mo}{98}$& $ $& $ $&1.00, 1.00&\\
     & -0.34 & \underline{-0.65} &       & $ $& $\iso{Fe}{56}(\mbox{n},\gamma)\iso{Fe}{57}$& $ $&1.00, 1.00&\\
$\iso{Mo}{98}$ & \underline{-0.79} &       &       & $\iso{Mo}{98}(\mbox{n},\gamma)\iso{Mo}{99}$& $ $& $ $&1.00, 1.00&\\
     & -0.29 & \underline{-0.67} &       & $ $& $\iso{Fe}{56}(\mbox{n},\gamma)\iso{Fe}{57}$& $ $&1.00, 1.00&\\
$\iso{Ru}{99}$ & \underline{-0.71} &       &       & $\iso{Tc}{99}(\mbox{n},\gamma)\iso{Tc}{100}$& $ $& $ $&1.00, 1.00&\\
     & -0.33 & \underline{-0.67} &       & $ $& $\iso{Fe}{56}(\mbox{n},\gamma)\iso{Fe}{57}$& $ $&1.00, 1.00&\\
$\iso{Ru}{100}$ & \underline{-0.73} &       &       & $\iso{Ru}{100}(\mbox{n},\gamma)\iso{Ru}{101}$& $ $& $ $&1.00, 1.00&\\
     & -0.32 & \underline{-0.68} &       & $ $& $\iso{Fe}{56}(\mbox{n},\gamma)\iso{Fe}{57}$& $ $&1.00, 1.00&\\
$\iso{Ru}{102}$ & -0.57 & \underline{-0.68} &       & $ $& $\iso{Ru}{102}(\mbox{n},\gamma)\iso{Ru}{103}$& $ $&1.00, 1.00&\\
$\iso{Rh}{103}$ & \underline{-0.79} &       &       & $\iso{Rh}{103}(\mbox{n},\gamma)\iso{Rh}{104}$& $ $& $ $&0.95, 0.80&\\
     & -0.29 & \underline{-0.69} &       & $ $& $\iso{Fe}{56}(\mbox{n},\gamma)\iso{Fe}{57}$& $ $&1.00, 1.00&\\
$\iso{Pd}{104}$ & \underline{-0.86} &       &       & $\iso{Pd}{104}(\mbox{n},\gamma)\iso{Pd}{105}$& $ $& $ $&1.00, 1.00&\\
     & -0.26 & \underline{-0.69} &       & $ $& $\iso{Fe}{56}(\mbox{n},\gamma)\iso{Fe}{57}$& $ $&1.00, 1.00&\\
$\iso{Pd}{106}$ & \underline{-0.85} &       &       & $\iso{Pd}{106}(\mbox{n},\gamma)\iso{Pd}{107}$& $ $& $ $&1.00, 1.00&\\
     & -0.26 & \underline{-0.70} &       & $ $& $\iso{Fe}{56}(\mbox{n},\gamma)\iso{Fe}{57}$& $ $&1.00, 1.00&\\
$\iso{Pd}{108}$ & \underline{-0.83} &       &       & $\iso{Pd}{108}(\mbox{n},\gamma)\iso{Pd}{109}$& $ $& $ $&1.00, 1.00&\\
     & -0.30 & \underline{-0.70} &       & $ $& $\iso{Fe}{56}(\mbox{n},\gamma)\iso{Fe}{57}$& $ $&1.00, 1.00&\\
$\iso{Ag}{107}$ & -0.48 & -0.58 & \underline{-0.84} & $ $& $ $& $\iso{Pd}{107}(\mbox{n},\gamma)\iso{Pd}{108}$&1.00, 1.00&\\
$\iso{Ag}{109}$ & -0.46 & -0.56 & \underline{-0.82} & $ $& $ $& $\iso{Ag}{109}(\mbox{n},\gamma)\iso{Ag}{110}$&1.00, 1.00&\\
$\iso{Cd}{110}$ & -0.52 & \underline{-0.69} &       & $ $& $\iso{Fe}{56}(\mbox{n},\gamma)\iso{Fe}{57}$& $ $&1.00, 1.00&\\
$\iso{Cd}{112}$ & -0.53 & \underline{-0.70} &       & $ $& $\iso{Fe}{56}(\mbox{n},\gamma)\iso{Fe}{57}$& $ $&1.00, 1.00&\\
$\iso{Cd}{114}$ & -0.54 & \underline{-0.71} &       & $ $& $\iso{Fe}{56}(\mbox{n},\gamma)\iso{Fe}{57}$& $ $&1.00, 1.00&\\
$\iso{In}{115}$ & \underline{-0.83} &       &       & $\iso{In}{115}(\mbox{n},\gamma)\iso{In}{116}$& $ $& $ $&1.00, 1.00&\\
     & -0.30 & \underline{-0.71} &       & $ $& $\iso{Fe}{56}(\mbox{n},\gamma)\iso{Fe}{57}$& $ $&1.00, 1.00&\\
$\iso{Sn}{116}$ & -0.55 & \underline{-0.70} &       & $ $& $\iso{Fe}{56}(\mbox{n},\gamma)\iso{Fe}{57}$& $ $&1.00, 1.00&\\
$\iso{Sn}{118}$ & -0.57 & \underline{-0.70} &       & $ $& $\iso{Fe}{56}(\mbox{n},\gamma)\iso{Fe}{57}$& $ $&1.00, 1.00&\\
$\iso{Sn}{120}$ & -0.59 & \underline{-0.71} &       & $ $& $\iso{Fe}{56}(\mbox{n},\gamma)\iso{Fe}{57}$& $ $&1.00, 1.00&\\
$\iso{Sb}{121}$ & -0.50 & -0.55 & \underline{-0.83} & $ $& $ $& $\iso{Sb}{121}(\mbox{n},\gamma)\iso{Sb}{122}$&0.98, 0.93&\\
$\iso{Te}{122}$ & -0.60 & \underline{-0.71} &       & $ $& $\iso{Fe}{56}(\mbox{n},\gamma)\iso{Fe}{57}$& $ $&1.00, 1.00&\\
$\iso{Te}{123}$ & -0.59 & \underline{-0.71} &       & $ $& $\iso{Fe}{56}(\mbox{n},\gamma)\iso{Fe}{57}$& $ $&1.00, 1.00&\\
$\iso{Te}{124}$ & -0.60 & \underline{-0.71} &       & $ $& $\iso{Fe}{56}(\mbox{n},\gamma)\iso{Fe}{57}$& $ $&1.00, 1.00&\\
$\iso{Te}{126}$ & -0.60 & \underline{-0.70} &       & $ $& $\iso{Fe}{56}(\mbox{n},\gamma)\iso{Fe}{57}$& $ $&1.00, 1.00&\\
$\iso{I}{127}$ & -0.44 & -0.47 & \underline{-0.77} & $ $& $ $& $\iso{I}{127}(\mbox{n},\gamma)\iso{I}{128}$&1.00, 0.99&\\
$\iso{Xe}{128}$ & -0.59 & \underline{-0.68} &       & $ $& $\iso{Fe}{56}(\mbox{n},\gamma)\iso{Fe}{57}$& $ $&1.00, 1.00&\\
$\iso{Xe}{130}$ & -0.61 & \underline{-0.70} &       & $ $& $\iso{Fe}{56}(\mbox{n},\gamma)\iso{Fe}{57}$& $ $&1.00, 1.00&\\
$\iso{Xe}{132}$ & -0.53 & -0.55 & \underline{-0.81} & $ $& $ $& $\iso{Xe}{132}(\mbox{n},\gamma)\iso{Xe}{133}$&1.00, 1.00&\\
$\iso{Cs}{133}$ & -0.58 & \underline{-0.65} &       & $ $& $\iso{Fe}{56}(\mbox{n},\gamma)\iso{Fe}{57}$& $ $&1.00, 1.00&\\
$\iso{Ba}{134}$ & -0.60 & \underline{-0.66} &       & $ $& $\iso{Fe}{56}(\mbox{n},\gamma)\iso{Fe}{57}$& $ $&1.00, 1.00&\\
$\iso{Ba}{136}$ & -0.60 & \underline{-0.66} &       & $ $& $\iso{Fe}{56}(\mbox{n},\gamma)\iso{Fe}{57}$& $ $&1.00, 1.00&\\
$\iso{Ba}{137}$ & -0.60 & \underline{-0.66} &       & $ $& $\iso{Fe}{56}(\mbox{n},\gamma)\iso{Fe}{57}$& $ $&1.00, 1.00&\\
$\iso{Ba}{138}$ & -0.62 & \underline{-0.66} &       & $ $& $\iso{Fe}{56}(\mbox{n},\gamma)\iso{Fe}{57}$& $ $&1.00, 1.00&\\
$\iso{Tb}{159}$ & -0.54 & -0.58 & \underline{-0.72} & $ $& $ $& $\iso{Tb}{159}(\mbox{n},\gamma)\iso{Tb}{160}$&1.00, 0.98&\\
$\iso{Er}{166}$ & -0.59 & -0.63 & \underline{-0.76} & $ $& $ $& $\iso{Er}{166}(\mbox{n},\gamma)\iso{Er}{167}$&1.00, 0.98&\\
$\iso{Er}{167}$ & -0.55 & -0.59 & \underline{-0.73} & $ $& $ $& $\iso{Er}{167}(\mbox{n},\gamma)\iso{Er}{168}$&1.00, 1.00&\\
$\iso{Er}{168}$ & \underline{-0.68} &       &       & $\iso{Er}{168}(\mbox{n},\gamma)\iso{Er}{169}$& $ $& $ $&1.00, 0.98&\\
$\iso{Tm}{169}$ & \underline{-0.77} &       &       & $\iso{Tm}{169}(\mbox{n},\gamma)\iso{Tm}{170}$& $ $& $ $&0.51, 0.42&\\
$\iso{Ta}{181}$ & \underline{-0.74} &       &       & $\iso{Ta}{181}(\mbox{n},\gamma)\iso{Ta}{182}$& $ $& $ $&0.61, 0.55&\\
$\iso{Os}{187}$ & \underline{-0.81} &       &       & $\iso{Os}{187}(\mbox{n},\gamma)\iso{Os}{188}$& $ $& $ $&0.57, 0.46&\\
$\iso{Ir}{193}$ & -0.55 & -0.59 & \underline{-0.70} & $ $& $ $& $\iso{Ir}{193}(\mbox{n},\gamma)\iso{Ir}{194}$&1.00, 0.99&\\
$\iso{Pt}{192}$ & \underline{-0.88} &       &       & $\iso{Pt}{192}(\mbox{n},\gamma)\iso{Pt}{193}$& $ $& $ $&1.00, 1.00&\\
$\iso{Pt}{194}$ & \underline{-0.89} &       &       & $\iso{Pt}{194}(\mbox{n},\gamma)\iso{Pt}{195}$& $ $& $ $&1.00, 1.00&\\
$\iso{Pt}{196}$ & \underline{-0.65} &       &       & $\iso{Pt}{196}(\mbox{n},\gamma)\iso{Pt}{197}$& $ $& $ $&1.00, 1.00&\\
$\iso{Hg}{198}$ & \underline{-0.66} &       &       & $\iso{Hg}{198}(\mbox{n},\gamma)\iso{Hg}{199}$& $ $& $ $&1.00, 1.00&\\
$\iso{Hg}{200}$ & \underline{-0.76} &       &       & $\iso{Hg}{200}(\mbox{n},\gamma)\iso{Hg}{201}$& $ $& $ $&1.00, 1.00&\\
$\iso{Tl}{203}$ & \underline{-0.66} &       &       & $\iso{Tl}{203}(\mbox{n},\gamma)\iso{Tl}{204}$& $ $& $ $&1.00, 1.00&\\
$\iso{Tl}{205}$ & \underline{-0.92} &       &       & $\iso{Pb}{205}(\mbox{n},\gamma)\iso{Pb}{206}$& $ $& $ $&0.83, 0.82&\\
$\iso{Pb}{207}$ & -0.61 & -0.62 & \underline{-0.67} & $ $& $ $& $\iso{Pb}{207}(\mbox{n},\gamma)\iso{Pb}{208}$&1.00, 1.00&\\
$\iso{Bi}{209}$ & \underline{-0.77} &       &       & $\iso{Bi}{209}(\mbox{n},\gamma)\iso{Bi}{210}$& $ $& $ $&1.00, 1.00&\\
     &  0.58 & \underline{ 0.92} &       & $ $& $\iso{Pb}{208}(\mbox{n},\gamma)\iso{Pb}{209}$& $ $&1.00, 1.00&\\
\hline

\end{tabular}
\end{table*}

\begin{table*}
  \caption{Key reaction rates for the model with double the initial
    $^{13}$C abundance  (``2 $\times$ $\iso{C}{13}$'' case). 
 Key rates in levels $1-3$ are shown, along with their correlation factors $r_{cor 0}$, $r_{cor 1}$ and $r_{cor 2}$, respectively.
    Not all s-process nuclides analysed are listed but only those for which key rates were found. Also shown for each rate are
    the ground-state contributions of the (n,$\gamma$) reaction to the
    stellar rate and uncertainty factors of the $\beta$-decay rate
 at two plasma temperatures, respectively. 
}\label{tab:dc}

\begin{tabular}{crrrcccccc}
\hline
Nuclide & $r_{{\rm cor},0}$ & $r_{{\rm cor},1}$ & $r_{{\rm cor},2}$& Key rate & Key rate & Key rate& $X_0$ $(8$, $30~{\rm keV})$ & $\beta$-decay \\
&&&& Level 1 & Level 2 & Level 3 & & \\
\hline
$\iso{Ga}{69}$ & -0.42 & -0.43 & \underline{-0.92} & $ $& $ $& $\iso{Ga}{69}(\mbox{n},\gamma)\iso{Ga}{70}$&1.00, 1.00&\\
$\iso{Ga}{71}$ & -0.57 & -0.58 & \underline{-0.96} & $ $& $ $& $\iso{Ga}{71}(\mbox{n},\gamma)\iso{Ga}{72}$&1.00, 1.00&\\
$\iso{Ge}{70}$ & -0.53 & -0.55 & \underline{-0.95} & $ $& $ $& $\iso{Ge}{70}(\mbox{n},\gamma)\iso{Ge}{71}$&1.00, 1.00&\\
$\iso{Ge}{72}$ & \underline{-0.89} &       &       & $\iso{Ge}{72}(\mbox{n},\gamma)\iso{Ge}{73}$& $ $& $ $&1.00, 1.00&\\
     & -0.19 & \underline{-0.69} &       & $ $& $\iso{Fe}{56}(\mbox{n},\gamma)\iso{Fe}{57}$& $ $&1.00, 1.00&\\
     & -0.17 & \underline{-0.68} &       & $ $& $\iso{Ni}{64}(\mbox{n},\gamma)\iso{Ni}{65}$& $ $&1.00, 1.00&\\
$\iso{Ge}{74}$ & \underline{-0.76} &       &       & $\iso{Ge}{74}(\mbox{n},\gamma)\iso{Ge}{75}$& $ $& $ $&1.00, 1.00&\\
     & -0.40 & \underline{-0.68} &       & $ $& $\iso{Fe}{56}(\mbox{n},\gamma)\iso{Fe}{57}$& $ $&1.00, 1.00&\\
     & -0.41 & \underline{-0.68} &       & $ $& $\iso{Ni}{64}(\mbox{n},\gamma)\iso{Ni}{65}$& $ $&1.00, 1.00&\\
$\iso{As}{75}$ & -0.41 & -0.46 & \underline{-0.91} & $ $& $ $& $\iso{As}{75}(\mbox{n},\gamma)\iso{As}{76}$&1.00, 1.00&\\
$\iso{Se}{76}$ & -0.46 & -0.49 & \underline{-0.93} & $ $& $ $& $\iso{Se}{76}(\mbox{n},\gamma)\iso{Se}{77}$&1.00, 1.00&\\
$\iso{Se}{78}$ & \underline{-0.87} &       &       & $\iso{Se}{78}(\mbox{n},\gamma)\iso{Se}{79}$& $ $& $ $&1.00, 1.00&\\
     & -0.27 & \underline{-0.67} &       & $ $& $\iso{Fe}{56}(\mbox{n},\gamma)\iso{Fe}{57}$& $ $&1.00, 1.00&\\
     & -0.29 & \underline{-0.69} &       & $ $& $\iso{Ni}{64}(\mbox{n},\gamma)\iso{Ni}{65}$& $ $&1.00, 1.00&\\
$\iso{Se}{80}$ & \underline{-0.67} &       &       & $\iso{Se}{80}(\mbox{n},\gamma)\iso{Se}{81}$& $ $& $ $&1.00, 1.00&\\
     & -0.46 & \underline{-0.67} &       & $ $& $\iso{Fe}{56}(\mbox{n},\gamma)\iso{Fe}{57}$& $ $&1.00, 1.00&\\
     & -0.47 & \underline{-0.69} &       & $ $& $\iso{Ni}{64}(\mbox{n},\gamma)\iso{Ni}{65}$& $ $&1.00, 1.00&\\
$\iso{Br}{79}$ & \underline{-0.90} &       &       & $\iso{Se}{79}(\mbox{n},\gamma)\iso{Se}{80}$& $ $& $ $&1.00, 1.00&\\
     & -0.19 & \underline{-0.67} &       & $ $& $\iso{Fe}{56}(\mbox{n},\gamma)\iso{Fe}{57}$& $ $&1.00, 1.00&\\
     & -0.19 & \underline{-0.68} &       & $ $& $\iso{Ni}{64}(\mbox{n},\gamma)\iso{Ni}{65}$& $ $&1.00, 1.00&\\
$\iso{Br}{81}$ & -0.61 & \underline{-0.66} &       & $ $& $\iso{Ni}{64}(\mbox{n},\gamma)\iso{Ni}{65}$& $ $&1.00, 1.00&\\
     & -0.26 & -0.28 & \underline{-0.75} & $ $& $ $& $\iso{Br}{81}(\mbox{n},\gamma)\iso{Br}{82}$&1.00, 1.00&\\
$\iso{Kr}{80}$ & \underline{-0.86} &       &       & $\iso{Se}{79}(\mbox{n},\gamma)\iso{Se}{80}$& $ $& $ $&1.00, 1.00&\\
     &  0.24 &  0.64 & \underline{ 0.84} & $ $& $ $& $\iso{Se}{79}(\beta^-)\iso{Br}{79}$& & 1.30, 1.49\\
$\iso{Kr}{82}$ & -0.64 & \underline{-0.68} &       & $ $& $\iso{Kr}{82}(\mbox{n},\gamma)\iso{Kr}{83}$& $ $&1.00, 1.00&\\
$\iso{Kr}{84}$ & \underline{-0.80} &       &       & $\iso{Kr}{84}(\mbox{n},\gamma)\iso{Kr}{85}$& $ $& $ $&1.00, 1.00&\\
     & -0.36 & \underline{-0.66} &       & $ $& $\iso{Fe}{56}(\mbox{n},\gamma)\iso{Fe}{57}$& $ $&1.00, 1.00&\\
     & -0.37 & \underline{-0.69} &       & $ $& $\iso{Ni}{64}(\mbox{n},\gamma)\iso{Ni}{65}$& $ $&1.00, 1.00&\\
$\iso{Kr}{86}$ & \underline{ 0.85} &       &       & $\iso{Kr}{85}(\mbox{n},\gamma)\iso{Kr}{86}$& $ $& $ $&1.00, 1.00&\\
     & -0.41 & \underline{-0.84} &       & $ $& $\iso{Kr}{85}(\beta^-)\iso{Rb}{85}$& $ $& & 1.30, 1.30\\
     & -0.27 & -0.52 & \underline{-1.00} & $ $& $ $& $\iso{Kr}{86}(\mbox{n},\gamma)\iso{Kr}{87}$&1.00, 1.00&\\
$\iso{Rb}{85}$ & -0.60 & \underline{-0.65} &       & $ $& $\iso{Ni}{64}(\mbox{n},\gamma)\iso{Ni}{65}$& $ $&1.00, 1.00&\\
     & -0.32 & -0.36 & \underline{-0.82} & $ $& $ $& $\iso{Rb}{85}(\mbox{n},\gamma)\iso{Rb}{86}$&1.00, 1.00&\\
$\iso{Rb}{87}$ & \underline{ 0.87} &       &       & $\iso{Kr}{85}(\mbox{n},\gamma)\iso{Kr}{86}$& $ $& $ $&1.00, 1.00&\\
     & -0.41 & \underline{-0.89} &       & $ $& $\iso{Kr}{85}(\beta^-)\iso{Rb}{85}$& $ $& & 1.30, 1.30\\
     & -0.18 & -0.39 & \underline{-0.92} & $ $& $ $& $\iso{Rb}{87}(\mbox{n},\gamma)\iso{Rb}{88}$&1.00, 1.00&\\
$\iso{Sr}{86}$ & -0.54 & -0.58 & \underline{-0.94} & $ $& $ $& $\iso{Sr}{86}(\mbox{n},\gamma)\iso{Sr}{87}$&1.00, 1.00&\\
$\iso{Sr}{87}$ & -0.46 & -0.53 & \underline{-0.91} & $ $& $ $& $\iso{Sr}{87}(\mbox{n},\gamma)\iso{Sr}{88}$&1.00, 1.00&\\
$\iso{Sr}{88}$ & \underline{-0.66} &       &       & $\iso{Sr}{88}(\mbox{n},\gamma)\iso{Sr}{89}$& $ $& $ $&1.00, 1.00&\\
     & -0.41 & \underline{-0.66} &       & $ $& $\iso{Ni}{64}(\mbox{n},\gamma)\iso{Ni}{65}$& $ $&1.00, 1.00&\\
$\iso{Y}{89}$ & \underline{-0.81} &       &       & $\iso{Y}{89}(\mbox{n},\gamma)\iso{Y}{90}$& $ $& $ $&1.00, 1.00&\\
$\iso{Zr}{90}$ & \underline{-0.92} &       &       & $\iso{Zr}{90}(\mbox{n},\gamma)\iso{Zr}{91}$& $ $& $ $&1.00, 1.00&\\
$\iso{Zr}{92}$ & \underline{-0.94} &       &       & $\iso{Zr}{92}(\mbox{n},\gamma)\iso{Zr}{93}$& $ $& $ $&1.00, 1.00&\\
$\iso{Zr}{94}$ & \underline{-0.89} &       &       & $\iso{Zr}{94}(\mbox{n},\gamma)\iso{Zr}{95}$& $ $& $ $&1.00, 1.00&\\
$\iso{Nb}{93}$ & \underline{-0.97} &       &       & $\iso{Zr}{93}(\mbox{n},\gamma)\iso{Zr}{94}$& $ $& $ $&1.00, 1.00&\\
$\iso{Mo}{95}$ & \underline{-0.88} &       &       & $\iso{Mo}{95}(\mbox{n},\gamma)\iso{Mo}{96}$& $ $& $ $&1.00, 1.00&\\
$\iso{Mo}{96}$ & \underline{-0.96} &       &       & $\iso{Mo}{96}(\mbox{n},\gamma)\iso{Mo}{97}$& $ $& $ $&1.00, 1.00&\\
$\iso{Mo}{97}$ & \underline{-0.89} &       &       & $\iso{Mo}{97}(\mbox{n},\gamma)\iso{Mo}{98}$& $ $& $ $&1.00, 1.00&\\
$\iso{Mo}{98}$ & \underline{-0.95} &       &       & $\iso{Mo}{98}(\mbox{n},\gamma)\iso{Mo}{99}$& $ $& $ $&1.00, 1.00&\\
$\iso{Ru}{99}$ & \underline{-0.92} &       &       & $\iso{Tc}{99}(\mbox{n},\gamma)\iso{Tc}{100}$& $ $& $ $&1.00, 1.00&\\
$\iso{Ru}{100}$ & \underline{-0.94} &       &       & $\iso{Ru}{100}(\mbox{n},\gamma)\iso{Ru}{101}$& $ $& $ $&1.00, 1.00&\\
     &  0.10 &  0.48 & \underline{ 0.65} & $ $& $ $& $\iso{Kr}{86}(\mbox{n},\gamma)\iso{Kr}{87}$&1.00, 1.00&\\
$\iso{Ru}{102}$ & \underline{-0.88} &       &       & $\iso{Ru}{102}(\mbox{n},\gamma)\iso{Ru}{103}$& $ $& $ $&1.00, 1.00&\\
     &  0.15 &  0.47 & \underline{ 0.65} & $ $& $ $& $\iso{Kr}{86}(\mbox{n},\gamma)\iso{Kr}{87}$&1.00, 1.00&\\
$\iso{Rh}{103}$ & \underline{-0.95} &       &       & $\iso{Rh}{103}(\mbox{n},\gamma)\iso{Rh}{104}$& $ $& $ $&0.95, 0.80&\\
     &  0.09 &  0.47 & \underline{ 0.66} & $ $& $ $& $\iso{Kr}{86}(\mbox{n},\gamma)\iso{Kr}{87}$&1.00, 1.00&\\
$\iso{Pd}{104}$ & \underline{-0.97} &       &       & $\iso{Pd}{104}(\mbox{n},\gamma)\iso{Pd}{105}$& $ $& $ $&1.00, 1.00&\\
     &  0.07 &  0.47 & \underline{ 0.66} & $ $& $ $& $\iso{Kr}{86}(\mbox{n},\gamma)\iso{Kr}{87}$&1.00, 1.00&\\
\hline

\end{tabular}
\end{table*}

\begin{table*}
\begin{tabular}{crrrcccccc}
\hline
Nuclide & $r_{{\rm cor},0}$ & $r_{{\rm cor},1}$ & $r_{{\rm cor},2}$& Key rate & Key rate & Key rate& $X_0$ $(8$, $30~{\rm keV})$ & $\beta$-decay \\
&&&& Level 1 & Level 2 & Level 3 & & \\
\hline
$\iso{Pd}{106}$ & \underline{-0.97} &       &       & $\iso{Pd}{106}(\mbox{n},\gamma)\iso{Pd}{107}$& $ $& $ $&1.00, 1.00&\\
     &  0.07 & \underline{ 0.65} &       & $ $& $\iso{Kr}{85}(\beta^-)\iso{Rb}{85}$& $ $& & 1.30, 1.30\\
     &  0.05 &  0.46 & \underline{ 0.67} & $ $& $ $& $\iso{Kr}{86}(\mbox{n},\gamma)\iso{Kr}{87}$&1.00, 1.00&\\
$\iso{Pd}{108}$ & \underline{-0.97} &       &       & $\iso{Pd}{108}(\mbox{n},\gamma)\iso{Pd}{109}$& $ $& $ $&1.00, 1.00&\\
     &  0.10 & \underline{ 0.65} &       & $ $& $\iso{Kr}{85}(\beta^-)\iso{Rb}{85}$& $ $& & 1.30, 1.30\\
     &  0.05 &  0.45 & \underline{ 0.67} & $ $& $ $& $\iso{Kr}{86}(\mbox{n},\gamma)\iso{Kr}{87}$&1.00, 1.00&\\
$\iso{Ag}{107}$ & \underline{-0.82} &       &       & $\iso{Pd}{107}(\mbox{n},\gamma)\iso{Pd}{108}$& $ $& $ $&1.00, 1.00&\\
     &  0.19 & \underline{ 0.65} &       & $ $& $\iso{Kr}{85}(\beta^-)\iso{Rb}{85}$& $ $& & 1.30, 1.30\\
     &  0.14 &  0.46 & \underline{ 0.67} & $ $& $ $& $\iso{Kr}{86}(\mbox{n},\gamma)\iso{Kr}{87}$&1.00, 1.00&\\
$\iso{Ag}{109}$ & \underline{-0.81} &       &       & $\iso{Ag}{109}(\mbox{n},\gamma)\iso{Ag}{110}$& $ $& $ $&1.00, 1.00&\\
     &  0.19 & \underline{ 0.65} &       & $ $& $\iso{Kr}{85}(\beta^-)\iso{Rb}{85}$& $ $& & 1.30, 1.30\\
     &  0.17 &  0.45 & \underline{ 0.67} & $ $& $ $& $\iso{Kr}{86}(\mbox{n},\gamma)\iso{Kr}{87}$&1.00, 1.00&\\
$\iso{Cd}{110}$ & \underline{-0.67} &       &       & $\iso{Kr}{85}(\mbox{n},\gamma)\iso{Kr}{86}$& $ $& $ $&1.00, 1.00&\\
     & -0.42 & \underline{-0.71} &       & $ $& $\iso{Cd}{110}(\mbox{n},\gamma)\iso{Cd}{111}$& $ $&1.00, 1.00&\\
     &  0.23 &  0.31 & \underline{ 0.67} & $ $& $ $& $\iso{Kr}{86}(\mbox{n},\gamma)\iso{Kr}{87}$&1.00, 1.00&\\
$\iso{Cd}{112}$ & \underline{-0.69} &       &       & $\iso{Kr}{85}(\mbox{n},\gamma)\iso{Kr}{86}$& $ $& $ $&1.00, 1.00&\\
     & -0.41 & \underline{-0.68} &       & $ $& $\iso{Cd}{112}(\mbox{n},\gamma)\iso{Cd}{113}$& $ $&1.00, 1.00&\\
     &  0.23 &  0.33 & \underline{ 0.69} & $ $& $ $& $\iso{Kr}{86}(\mbox{n},\gamma)\iso{Kr}{87}$&1.00, 1.00&\\
$\iso{Cd}{114}$ & \underline{-0.70} &       &       & $\iso{Kr}{85}(\mbox{n},\gamma)\iso{Kr}{86}$& $ $& $ $&1.00, 1.00&\\
     & -0.36 & -0.63 & \underline{-0.81} & $ $& $ $& $\iso{Cd}{114}(\mbox{n},\gamma)\iso{Cd}{115}$&1.00, 1.00&\\
$\iso{In}{115}$ & \underline{-0.97} &       &       & $\iso{In}{115}(\mbox{n},\gamma)\iso{In}{116}$& $ $& $ $&1.00, 1.00&\\
     &  0.07 & \underline{ 0.65} &       & $ $& $\iso{Kr}{85}(\beta^-)\iso{Rb}{85}$& $ $& & 1.30, 1.30\\
     &  0.06 &  0.42 & \underline{ 0.70} & $ $& $ $& $\iso{Kr}{86}(\mbox{n},\gamma)\iso{Kr}{87}$&1.00, 1.00&\\
$\iso{Sn}{116}$ & \underline{-0.66} &       &       & $\iso{Kr}{85}(\mbox{n},\gamma)\iso{Kr}{86}$& $ $& $ $&1.00, 1.00&\\
     & -0.50 & \underline{-0.77} &       & $ $& $\iso{Sn}{116}(\mbox{n},\gamma)\iso{Sn}{117}$& $ $&1.00, 1.00&\\
     &  0.19 &  0.25 & \underline{ 0.71} & $ $& $ $& $\iso{Kr}{86}(\mbox{n},\gamma)\iso{Kr}{87}$&1.00, 1.00&\\
$\iso{Sn}{118}$ & -0.55 & \underline{-0.81} &       & $ $& $\iso{Sn}{118}(\mbox{n},\gamma)\iso{Sn}{119}$& $ $&1.00, 1.00&\\
     &  0.18 &  0.19 & \underline{ 0.74} & $ $& $ $& $\iso{Kr}{86}(\mbox{n},\gamma)\iso{Kr}{87}$&1.00, 1.00&\\
$\iso{Sn}{120}$ & \underline{-0.65} &       &       & $\iso{Kr}{85}(\mbox{n},\gamma)\iso{Kr}{86}$& $ $& $ $&1.00, 1.00&\\
     & -0.52 & \underline{-0.76} &       & $ $& $\iso{Sn}{120}(\mbox{n},\gamma)\iso{Sn}{121}$& $ $&1.00, 1.00&\\
     &  0.17 &  0.22 & \underline{ 0.79} & $ $& $ $& $\iso{Kr}{86}(\mbox{n},\gamma)\iso{Kr}{87}$&1.00, 1.00&\\
$\iso{Sb}{121}$ & \underline{-0.89} &       &       & $\iso{Sb}{121}(\mbox{n},\gamma)\iso{Sb}{122}$& $ $& $ $&0.98, 0.93&\\
     &  0.09 &  0.32 & \underline{ 0.79} & $ $& $ $& $\iso{Kr}{86}(\mbox{n},\gamma)\iso{Kr}{87}$&1.00, 1.00&\\
$\iso{Te}{122}$ & \underline{-0.71} &       &       & $\iso{Kr}{85}(\mbox{n},\gamma)\iso{Kr}{86}$& $ $& $ $&1.00, 1.00&\\
     & -0.34 & -0.54 & \underline{-0.80} & $ $& $ $& $\iso{Te}{122}(\mbox{n},\gamma)\iso{Te}{123}$&1.00, 1.00&\\
$\iso{Te}{123}$ & \underline{-0.71} &       &       & $\iso{Kr}{85}(\mbox{n},\gamma)\iso{Kr}{86}$& $ $& $ $&1.00, 1.00&\\
     &  0.35 &  0.54 & \underline{ 0.80} & $ $& $ $& $\iso{Te}{123}(\mbox{n},\gamma)\iso{Te}{124}$&1.00, 1.00&\\
$\iso{Te}{124}$ & \underline{-0.68} &       &       & $\iso{Kr}{85}(\mbox{n},\gamma)\iso{Kr}{86}$& $ $& $ $&1.00, 1.00&\\
     & -0.45 & \underline{-0.69} &       & $ $& $\iso{Te}{124}(\mbox{n},\gamma)\iso{Te}{125}$& $ $&1.00, 1.00&\\
     &  0.17 &  0.22 & \underline{ 0.81} & $ $& $ $& $\iso{Kr}{86}(\mbox{n},\gamma)\iso{Kr}{87}$&1.00, 1.00&\\
$\iso{Te}{126}$ & -0.60 & \underline{-0.80} &       & $ $& $\iso{Te}{126}(\mbox{n},\gamma)\iso{Te}{127}$& $ $&1.00, 1.00&\\
     &  0.14 &  0.17 & \underline{ 0.83} & $ $& $ $& $\iso{Kr}{86}(\mbox{n},\gamma)\iso{Kr}{87}$&1.00, 1.00&\\
$\iso{I}{127}$ & \underline{-0.87} &       &       & $\iso{I}{127}(\mbox{n},\gamma)\iso{I}{128}$& $ $& $ $&1.00, 0.99&\\
     &  0.09 &  0.28 & \underline{ 0.83} & $ $& $ $& $\iso{Kr}{86}(\mbox{n},\gamma)\iso{Kr}{87}$&1.00, 1.00&\\
$\iso{Xe}{128}$ &  0.57 & \underline{ 0.72} &       & $ $& $\iso{I}{128}(\beta^-)\iso{Xe}{128}$& $ $& & 1.64, 5.42\\
     & -0.22 & -0.27 & \underline{-0.69} & $ $& $ $& $\iso{I}{128}(\beta^+)\iso{Te}{128}$\\
$\iso{Xe}{130}$ & \underline{-0.67} &       &       & $\iso{Kr}{85}(\mbox{n},\gamma)\iso{Kr}{86}$& $ $& $ $&1.00, 1.00&\\
     & -0.44 & \underline{-0.67} &       & $ $& $\iso{Xe}{130}(\mbox{n},\gamma)\iso{Xe}{131}$& $ $&1.00, 1.00&\\
     &  0.16 &  0.19 & \underline{ 0.86} & $ $& $ $& $\iso{Kr}{86}(\mbox{n},\gamma)\iso{Kr}{87}$&1.00, 1.00&\\
$\iso{Xe}{132}$ & \underline{-0.94} &       &       & $\iso{Xe}{132}(\mbox{n},\gamma)\iso{Xe}{133}$& $ $& $ $&1.00, 1.00&\\
     &  0.05 &  0.23 & \underline{ 0.87} & $ $& $ $& $\iso{Kr}{86}(\mbox{n},\gamma)\iso{Kr}{87}$&1.00, 1.00&\\
$\iso{Cs}{133}$ & \underline{-0.80} &       &       & $\iso{Cs}{133}(\mbox{n},\gamma)\iso{Cs}{134}$& $ $& $ $&1.00, 1.00&\\
     &  0.12 &  0.23 & \underline{ 0.87} & $ $& $ $& $\iso{Kr}{86}(\mbox{n},\gamma)\iso{Kr}{87}$&1.00, 1.00&\\
$\iso{Ba}{134}$ & \underline{-0.75} &       &       & $\iso{Ba}{134}(\mbox{n},\gamma)\iso{Ba}{135}$& $ $& $ $&1.00, 1.00&\\
     &  0.10 &  0.22 & \underline{ 0.73} & $ $& $ $& $\iso{Kr}{86}(\mbox{n},\gamma)\iso{Kr}{87}$&1.00, 1.00&\\
$\iso{Ba}{136}$ & \underline{-0.78} &       &       & $\iso{Ba}{136}(\mbox{n},\gamma)\iso{Ba}{137}$& $ $& $ $&1.00, 1.00&\\
     &  0.10 &  0.20 & \underline{ 0.87} & $ $& $ $& $\iso{Kr}{86}(\mbox{n},\gamma)\iso{Kr}{87}$&1.00, 1.00&\\
$\iso{Ba}{137}$ & \underline{-0.72} &       &       & $\iso{Ba}{137}(\mbox{n},\gamma)\iso{Ba}{138}$& $ $& $ $&1.00, 1.00&\\
     &  0.12 &  0.19 & \underline{ 0.87} & $ $& $ $& $\iso{Kr}{86}(\mbox{n},\gamma)\iso{Kr}{87}$&1.00, 1.00&\\
$\iso{Ba}{138}$ & \underline{-0.87} &       &       & $\iso{Ba}{138}(\mbox{n},\gamma)\iso{Ba}{139}$& $ $& $ $&1.00, 1.00&\\
     &  0.23 & \underline{ 0.65} &       & $ $& $\iso{Ni}{64}(\mbox{n},\gamma)\iso{Ni}{65}$& $ $&1.00, 1.00&\\
$\iso{La}{139}$ & \underline{-0.96} &       &       & $\iso{La}{139}(\mbox{n},\gamma)\iso{La}{140}$& $ $& $ $&1.00, 1.00&\\
     &  0.11 & \underline{ 0.65} &       & $ $& $\iso{Ni}{64}(\mbox{n},\gamma)\iso{Ni}{65}$& $ $&1.00, 1.00&\\
\hline

\end{tabular}
\end{table*}
\begin{table*}
\begin{tabular}{crrrcccccc}
\hline
Nuclide & $r_{{\rm cor},0}$ & $r_{{\rm cor},1}$ & $r_{{\rm cor},2}$& Key rate & Key rate & Key rate& $X_0$ $(8$, $30~{\rm keV})$ & $\beta$-decay \\
&&&& Level 1 & Level 2 & Level 3 & & \\
\hline

$\iso{Ce}{140}$ & \underline{-0.84} &       &       & $\iso{Ce}{140}(\mbox{n},\gamma)\iso{Ce}{141}$& $ $& $ $&1.00, 1.00&\\
$\iso{Nd}{142}$ & \underline{-0.73} &       &       & $\iso{Nd}{142}(\mbox{n},\gamma)\iso{Nd}{143}$& $ $& $ $&1.00, 1.00&\\
$\iso{Nd}{144}$ & \underline{-0.66} &       &       & $\iso{Nd}{144}(\mbox{n},\gamma)\iso{Nd}{145}$& $ $& $ $&1.00, 1.00&\\
$\iso{Nd}{146}$ & -0.26 & -0.42 & \underline{-0.69} & $ $& $ $& $\iso{Nd}{146}(\mbox{n},\gamma)\iso{Nd}{147}$&1.00, 1.00&\\
$\iso{Sm}{147}$ & -0.31 & -0.50 & \underline{-0.76} & $ $& $ $& $\iso{Sm}{147}(\mbox{n},\gamma)\iso{Sm}{148}$&1.00, 1.00&\\
$\iso{Sm}{148}$ & -0.29 & -0.48 & \underline{-0.73} & $ $& $ $& $\iso{Sm}{148}(\mbox{n},\gamma)\iso{Sm}{149}$&1.00, 1.00&\\
$\iso{Sm}{150}$ & \underline{ 0.66} &       &       & $\iso{Ba}{138}(\mbox{n},\gamma)\iso{Ba}{139}$& $ $& $ $&1.00, 1.00&\\
$\iso{Eu}{151}$ & \underline{-0.93} &       &       & $\iso{Eu}{151}(\mbox{n},\gamma)\iso{Eu}{152}$& $ $& $ $&0.89, 0.79&\\
$\iso{Eu}{153}$ & -0.32 & -0.57 & \underline{-0.80} & $ $& $ $& $\iso{Eu}{153}(\mbox{n},\gamma)\iso{Eu}{154}$&1.00, 1.00&\\
$\iso{Gd}{152}$ & \underline{ 0.80} &       &       & $\iso{Sm}{151}(\beta^-)\iso{Eu}{151}$& $ $& $ $& & 3.60, 5.42\\
     & -0.16 & \underline{-0.74} &       & $ $& $\iso{Sm}{151}(\mbox{n},\gamma)\iso{Sm}{152}$& $ $&0.80, 0.76&\\
$\iso{Gd}{156}$ & -0.51 & \underline{-0.73} &       & $ $& $\iso{Gd}{156}(\mbox{n},\gamma)\iso{Gd}{157}$& $ $&1.00, 0.99&\\
$\iso{Gd}{158}$ & \underline{-0.65} &       &       & $\iso{Gd}{158}(\mbox{n},\gamma)\iso{Gd}{159}$& $ $& $ $&1.00, 0.98&\\
$\iso{Tb}{159}$ & \underline{-0.97} &       &       & $\iso{Tb}{159}(\mbox{n},\gamma)\iso{Tb}{160}$& $ $& $ $&1.00, 0.98&\\
$\iso{Dy}{160}$ & \underline{-0.65} &       &       & $\iso{Dy}{160}(\mbox{n},\gamma)\iso{Dy}{161}$& $ $& $ $&1.00, 0.99&\\
$\iso{Dy}{162}$ & \underline{-0.69} &       &       & $\iso{Dy}{162}(\mbox{n},\gamma)\iso{Dy}{163}$& $ $& $ $&1.00, 0.98&\\
$\iso{Dy}{164}$ & \underline{-0.80} &       &       & $\iso{Dy}{164}(\mbox{n},\gamma)\iso{Dy}{165}$& $ $& $ $&1.00, 0.97&\\
$\iso{Ho}{165}$ & \underline{-0.95} &       &       & $\iso{Ho}{165}(\mbox{n},\gamma)\iso{Ho}{166}$& $ $& $ $&1.00, 1.00&\\
$\iso{Er}{166}$ & \underline{-0.97} &       &       & $\iso{Er}{166}(\mbox{n},\gamma)\iso{Er}{167}$& $ $& $ $&1.00, 0.98&\\
$\iso{Er}{167}$ & \underline{-0.97} &       &       & $\iso{Er}{167}(\mbox{n},\gamma)\iso{Er}{168}$& $ $& $ $&1.00, 1.00&\\
$\iso{Er}{168}$ & \underline{-0.98} &       &       & $\iso{Er}{168}(\mbox{n},\gamma)\iso{Er}{169}$& $ $& $ $&1.00, 0.98&\\
$\iso{Tm}{169}$ & \underline{-0.97} &       &       & $\iso{Tm}{169}(\mbox{n},\gamma)\iso{Tm}{170}$& $ $& $ $&0.51, 0.42&\\
$\iso{Yb}{170}$ & -0.64 & \underline{-0.84} &       & $ $& $\iso{Yb}{170}(\mbox{n},\gamma)\iso{Yb}{171}$& $ $&1.00, 0.99&\\
$\iso{Yb}{172}$ & \underline{-0.71} &       &       & $\iso{Yb}{172}(\mbox{n},\gamma)\iso{Yb}{173}$& $ $& $ $&1.00, 0.98&\\
$\iso{Yb}{174}$ & \underline{-0.72} &       &       & $\iso{Yb}{174}(\mbox{n},\gamma)\iso{Yb}{175}$& $ $& $ $&1.00, 0.98&\\
$\iso{Lu}{175}$ & \underline{ 0.70} &       &       & $\iso{Ba}{138}(\mbox{n},\gamma)\iso{Ba}{139}$& $ $& $ $&1.00, 1.00&\\
$\iso{Lu}{176}$ & \underline{ 0.66} &       &       & $\iso{Ba}{138}(\mbox{n},\gamma)\iso{Ba}{139}$& $ $& $ $&1.00, 1.00&\\
$\iso{Hf}{176}$ & \underline{ 0.90} &       &       & $\iso{Lu}{176}(\beta^-)\iso{Hf}{176}$& $ $& $ $& & 1.30, 1.33\\
     & -0.11 & -0.48 & \underline{-0.69} & $ $& $ $& $\iso{Hf}{176}(\mbox{n},\gamma)\iso{Hf}{177}$&1.00, 0.99&\\
$\iso{Hf}{178}$ & -0.48 & \underline{-0.72} &       & $ $& $\iso{Hf}{178}(\mbox{n},\gamma)\iso{Hf}{179}$& $ $&1.00, 0.99&\\
$\iso{Hf}{180}$ & -0.53 & \underline{-0.78} &       & $ $& $\iso{Hf}{180}(\mbox{n},\gamma)\iso{Hf}{181}$& $ $&1.00, 0.99&\\
$\iso{Ta}{181}$ & \underline{-0.98} &       &       & $\iso{Ta}{181}(\mbox{n},\gamma)\iso{Ta}{182}$& $ $& $ $&0.61, 0.55&\\
$\iso{W}{182}$ & \underline{-0.77} &       &       & $\iso{W}{182}(\mbox{n},\gamma)\iso{W}{183}$& $ $& $ $&1.00, 1.00&\\
$\iso{W}{183}$ & \underline{-0.91} &       &       & $\iso{W}{183}(\mbox{n},\gamma)\iso{W}{184}$& $ $& $ $&0.99, 0.93&\\
$\iso{W}{184}$ & -0.58 & \underline{-0.83} &       & $ $& $\iso{W}{184}(\mbox{n},\gamma)\iso{W}{185}$& $ $&1.00, 1.00&\\
$\iso{Re}{185}$ & \underline{-0.80} &       &       & $\iso{Re}{185}(\mbox{n},\gamma)\iso{Re}{186}$& $ $& $ $&1.00, 1.00&\\
$\iso{Os}{186}$ & \underline{-0.74} &       &       & $\iso{Os}{186}(\mbox{n},\gamma)\iso{Os}{187}$& $ $& $ $&1.00, 1.00&\\
     &  0.34 & \underline{ 0.68} &       & $ $& $\iso{Re}{186}(\beta^-)\iso{Os}{186}$& $ $& & 1.30, 3.59\\
     & -0.21 & -0.43 & \underline{-0.77} & $ $& $ $& $\iso{Re}{186}(\beta^+)\iso{W}{186}$\\
$\iso{Os}{187}$ & \underline{-0.97} &       &       & $\iso{Os}{187}(\mbox{n},\gamma)\iso{Os}{188}$& $ $& $ $&0.57, 0.46&\\
     &  0.07 & \underline{ 0.67} &       & $ $& $\iso{Re}{186}(\beta^-)\iso{Os}{186}$& $ $& & 1.30, 3.59\\
     & -0.05 & -0.43 & \underline{-0.76} & $ $& $ $& $\iso{Re}{186}(\beta^+)\iso{W}{186}$\\
$\iso{Os}{188}$ & \underline{-0.84} &       &       & $\iso{Os}{188}(\mbox{n},\gamma)\iso{Os}{189}$& $ $& $ $&1.00, 1.00&\\
$\iso{Os}{190}$ & \underline{-0.78} &       &       & $\iso{Os}{190}(\mbox{n},\gamma)\iso{Os}{191}$& $ $& $ $&1.00, 1.00&\\
$\iso{Ir}{191}$ & -0.62 & \underline{-0.86} &       & $ $& $\iso{Ir}{191}(\mbox{n},\gamma)\iso{Ir}{192}$& $ $&1.00, 1.00&\\
$\iso{Ir}{193}$ & \underline{-0.90} &       &       & $\iso{Ir}{193}(\mbox{n},\gamma)\iso{Ir}{194}$& $ $& $ $&1.00, 0.99&\\
     & -0.31 & \underline{-0.91} &       & $ $& $\iso{Pt}{193}(\mbox{n},\gamma)\iso{Pt}{194}$& $ $&0.25, 0.21&\\
$\iso{Pt}{192}$ & \underline{-0.96} &       &       & $\iso{Pt}{192}(\mbox{n},\gamma)\iso{Pt}{193}$& $ $& $ $&1.00, 1.00&\\
     &  0.04 & \underline{ 0.74} &       & $ $& $\iso{Ir}{192}(\beta^-)\iso{Pt}{192}$& $ $& & 1.31, 6.36\\
$\iso{Pt}{194}$ & \underline{-0.94} &       &       & $\iso{Pt}{194}(\mbox{n},\gamma)\iso{Pt}{195}$& $ $& $ $&1.00, 1.00&\\
     & -0.04 & \underline{-0.73} &       & $ $& $\iso{Pt}{193}(\mbox{n},\gamma)\iso{Pt}{194}$& $ $&0.25, 0.21&\\
$\iso{Pt}{196}$ & \underline{-0.96} &       &       & $\iso{Pt}{196}(\mbox{n},\gamma)\iso{Pt}{197}$& $ $& $ $&1.00, 1.00&\\
$\iso{Au}{197}$ & \underline{ 0.73} &       &       & $\iso{Ba}{138}(\mbox{n},\gamma)\iso{Ba}{139}$& $ $& $ $&1.00, 1.00&\\
$\iso{Hg}{198}$ & \underline{-0.96} &       &       & $\iso{Hg}{198}(\mbox{n},\gamma)\iso{Hg}{199}$& $ $& $ $&1.00, 1.00&\\
$\iso{Hg}{200}$ & \underline{-0.97} &       &       & $\iso{Hg}{200}(\mbox{n},\gamma)\iso{Hg}{201}$& $ $& $ $&1.00, 1.00&\\
$\iso{Hg}{202}$ & \underline{-0.77} &       &       & $\iso{Hg}{202}(\mbox{n},\gamma)\iso{Hg}{203}$& $ $& $ $&1.00, 1.00&\\
$\iso{Tl}{203}$ & \underline{-0.94} &       &       & $\iso{Tl}{203}(\mbox{n},\gamma)\iso{Tl}{204}$& $ $& $ $&1.00, 1.00&\\
$\iso{Tl}{205}$ & \underline{-0.94} &       &       & $\iso{Pb}{205}(\mbox{n},\gamma)\iso{Pb}{206}$& $ $& $ $&0.83, 0.82&\\
$\iso{Pb}{204}$ & \underline{-0.84} &       &       & $\iso{Pb}{204}(\mbox{n},\gamma)\iso{Pb}{205}$& $ $& $ $&1.00, 1.00&\\
$\iso{Pb}{206}$ & -0.59 & \underline{-0.88} &       & $ $& $\iso{Pb}{206}(\mbox{n},\gamma)\iso{Pb}{207}$& $ $&1.00, 1.00&\\
$\iso{Pb}{207}$ & -0.64 & \underline{-0.81} &       & $ $& $\iso{Pb}{207}(\mbox{n},\gamma)\iso{Pb}{208}$& $ $&1.00, 1.00&\\
$\iso{Bi}{209}$ & \underline{ 0.68} &       &       & $\iso{Pb}{208}(\mbox{n},\gamma)\iso{Pb}{209}$& $ $& $ $&1.00, 1.00&\\
     & -0.36 & -0.57 & \underline{-0.95} & $ $& $ $& $\iso{Bi}{209}(\mbox{n},\gamma)\iso{Bi}{210}$&1.00, 1.00&\\
\hline

\end{tabular}
\end{table*}

\begin{table*}
\caption{The key reaction rates for the TP model. Key rates in levels $1-3$ are shown, along with their correlation factors $r_{cor 0}$, $r_{cor 1}$ and $r_{cor 2}$, respectively.
Not all s-process nuclides analysed are listed but only those for which key rates were found. Also shown for each rate are
the ground state contributions $X_0$ to the stellar rate of the (n,$\gamma$) reaction and uncertainty factors of the $\beta$-decay 
rate at two plasma temperatures, respectively.}\label{tab:TP}
\begin{tabular}{crrrcccccc}
\hline
Nuclide & $r_{{\rm cor},0}$ & $r_{{\rm cor},1}$ & $r_{{\rm cor},2}$& Key rate & Key rate & Key rate& $X_0$ $(8$, $30~{\rm keV})$ & $\beta$-decay \\
&&&& Level 1 & Level 2 & Level 3 & & \\
\hline
$\iso{Ge}{70}$ & \underline{-0.83} &       &       & $\iso{Ge}{70}(\mbox{n},\gamma)\iso{Ge}{71}$& $ $& $ $&1.00, 1.00&\\
     &  0.41 & \underline{ 0.73} &       & $ $& $\iso{Zn}{68}(\mbox{n},\gamma)\iso{Zn}{69}$& $ $&1.00, 1.00&\\
     &  0.36 & \underline{ 0.67} &       & $ $& $\iso{Ga}{69}(\mbox{n},\gamma)\iso{Ga}{70}$& $ $&1.00, 1.00&\\
     & -0.04 & -0.11 & \underline{-0.94} & $ $& $ $& $\iso{Ga}{70}(\mbox{n},\gamma)\iso{Ga}{71}$&1.00, 1.00&\\
$\iso{Se}{76}$ & \underline{ 0.76} &       &       & $\iso{Ge}{74}(\mbox{n},\gamma)\iso{Ge}{75}$& $ $& $ $&1.00, 1.00&\\
     & -0.59 & \underline{-0.90} &       & $ $& $\iso{Se}{76}(\mbox{n},\gamma)\iso{Se}{77}$& $ $&1.00, 1.00&\\
     &  0.27 &  0.37 & \underline{ 0.84} & $ $& $ $& $\iso{As}{75}(\mbox{n},\gamma)\iso{As}{76}$&1.00, 1.00&\\
$\iso{Se}{82}$ & \underline{ 0.88} &       &       & $\iso{Se}{81}(\mbox{n},\gamma)\iso{Se}{82}$& $ $& $ $&1.00, 1.00&\\
     & -0.30 & \underline{-0.83} &       & $ $& $\iso{Se}{82}(\mbox{n},\gamma)\iso{Se}{83}$& $ $&1.00, 1.00&\\
     & -0.27 & -0.46 & \underline{-0.85} & $ $& $ $& $\iso{Se}{81}(\beta^-)\iso{Br}{81}$& & 1.30, 2.17\\
$\iso{Kr}{80}$ & \underline{ 0.66} &       &       & $\iso{Br}{80}(\beta^-)\iso{Kr}{80}$& $ $& $ $& & 1.31, 4.70\\
     & -0.47 & \underline{-0.66} &       & $ $& $\iso{Kr}{80}(\mbox{n},\gamma)\iso{Kr}{81}$& $ $&1.00, 1.00&\\
$\iso{Sr}{86}$ & \underline{-0.92} &       &       & $\iso{Sr}{86}(\mbox{n},\gamma)\iso{Sr}{87}$& $ $& $ $&1.00, 1.00&\\
     & -0.27 & \underline{-0.71} &       & $ $& $\iso{Rb}{86}(\mbox{n},\gamma)\iso{Rb}{87}$& $ $&1.00, 1.00&\\
     &  0.15 &  0.39 & \underline{ 0.89} & $ $& $ $& $\iso{Rb}{85}(\mbox{n},\gamma)\iso{Rb}{86}$&1.00, 1.00&\\
$\iso{Sr}{87}$ & \underline{ 0.65} &       &       & $\iso{Sr}{86}(\mbox{n},\gamma)\iso{Sr}{87}$& $ $& $ $&1.00, 1.00&\\
     & \underline{-0.75} &       &       & $\iso{Sr}{87}(\mbox{n},\gamma)\iso{Sr}{88}$& $ $& $ $&1.00, 1.00&\\
     &  0.04 &  0.44 & \underline{ 0.76} & $ $& $ $& $\iso{Rb}{85}(\mbox{n},\gamma)\iso{Rb}{86}$&1.00, 1.00&\\
$\iso{Zr}{96}$ & \underline{-0.74} &       &       & $\iso{Zr}{95}(\beta^-)\iso{Nb}{95}$& $ $& $ $\\
     &  0.46 & \underline{ 0.98} &       & $ $& $\iso{Zr}{95}(\mbox{n},\gamma)\iso{Zr}{96}$& $ $&1.00, 1.00&\\
     &  0.06 &  0.13 & \underline{ 0.95} & $ $& $ $& $\iso{Zr}{94}(\mbox{n},\gamma)\iso{Zr}{95}$&1.00, 1.00&\\
$\iso{Mo}{94}$ & \underline{-0.80} &       &       & $\iso{Mo}{94}(\mbox{n},\gamma)\iso{Mo}{95}$& $ $& $ $&1.00, 1.00&\\
     & -0.43 & \underline{-0.71} &       & $ $& $\iso{Nb}{94}(\mbox{n},\gamma)\iso{Nb}{95}$& $ $&0.99, 0.97&\\
     &  0.33 &  0.56 & \underline{ 0.81} & $ $& $ $& $\iso{Nb}{94}(\beta^-)\iso{Mo}{94}$& & 1.35, 3.22\\
$\iso{Mo}{96}$ & \underline{-0.66} &       &       & $\iso{Mo}{96}(\mbox{n},\gamma)\iso{Mo}{97}$& $ $& $ $&1.00, 1.00&\\
$\iso{Mo}{100}$ & \underline{ 0.85} &       &       & $\iso{Mo}{99}(\mbox{n},\gamma)\iso{Mo}{100}$& $ $& $ $&1.00, 1.00&\\
     & -0.42 & \underline{-0.82} &       & $ $& $\iso{Mo}{99}(\beta^-)\iso{Tc}{99}$& $ $& & 1.30, 2.13\\
     & -0.14 & -0.32 & \underline{-1.00} & $ $& $ $& $\iso{Mo}{100}(\mbox{n},\gamma)\iso{Mo}{101}$&1.00, 1.00&\\
$\iso{Ru}{100}$ & \underline{-0.79} &       &       & $\iso{Ru}{100}(\mbox{n},\gamma)\iso{Ru}{101}$& $ $& $ $&1.00, 1.00&\\
     &  0.57 & \underline{ 0.95} &       & $ $& $\iso{Mo}{98}(\mbox{n},\gamma)\iso{Mo}{99}$& $ $&1.00, 1.00&\\
     &  0.18 &  0.29 & \underline{ 0.91} & $ $& $ $& $\iso{Tc}{99}(\mbox{n},\gamma)\iso{Tc}{100}$&1.00, 1.00&\\
$\iso{Ru}{104}$ & \underline{ 0.65} &       &       & $\iso{Ru}{103}(\mbox{n},\gamma)\iso{Ru}{104}$& $ $& $ $&0.45, 0.41&\\
     & -0.49 & \underline{-0.67} &       & $ $& $\iso{Ru}{103}(\beta^-)\iso{Rh}{103}$& $ $& & 5.76, 6.34\\
     & -0.06 & -0.06 & \underline{-0.98} & $ $& $ $& $\iso{Ru}{104}(\mbox{n},\gamma)\iso{Ru}{105}$&1.00, 1.00&\\
$\iso{Pd}{104}$ & \underline{-0.95} &       &       & $\iso{Pd}{104}(\mbox{n},\gamma)\iso{Pd}{105}$& $ $& $ $&1.00, 1.00&\\
     &  0.20 & \underline{ 0.69} &       & $ $& $\iso{Ru}{102}(\mbox{n},\gamma)\iso{Ru}{103}$& $ $&1.00, 1.00&\\
     &  0.20 & \underline{ 0.66} &       & $ $& $\iso{Rh}{103}(\mbox{n},\gamma)\iso{Rh}{104}$& $ $&0.95, 0.80&\\
$\iso{Pd}{110}$ & \underline{ 0.86} &       &       & $\iso{Pd}{109}(\mbox{n},\gamma)\iso{Pd}{110}$& $ $& $ $&1.00, 1.00&\\
$\iso{Cd}{110}$ & \underline{ 0.87} &       &       & $\iso{Pd}{108}(\mbox{n},\gamma)\iso{Pd}{109}$& $ $& $ $&1.00, 1.00&\\
     &  0.31 & \underline{ 0.75} &       & $ $& $\iso{Pd}{106}(\mbox{n},\gamma)\iso{Pd}{107}$& $ $&1.00, 1.00&\\
     &  0.20 &  0.47 & \underline{ 0.72} & $ $& $ $& $\iso{Ag}{109}(\mbox{n},\gamma)\iso{Ag}{110}$&1.00, 1.00&\\
$\iso{Cd}{116}$ & \underline{ 0.96} &       &       & $\iso{Cd}{115}(\mbox{n},\gamma)\iso{Cd}{116}$& $ $& $ $&1.00, 1.00&\\
     & -0.26 & \underline{-0.96} &       & $ $& $\iso{Cd}{115}(\beta^-)\iso{In}{115}$& $ $& & 1.30, 1.44\\
$\iso{Sn}{114}$ & \underline{ 0.66} &       &       & $\iso{Sn}{113}(\mbox{n},\gamma)\iso{Sn}{114}$& $ $& $ $&1.00, 0.99&\\
     &  0.55 & \underline{ 0.77} &       & $ $& $\iso{Sn}{112}(\mbox{n},\gamma)\iso{Sn}{113}$& $ $&1.00, 1.00&\\
$\iso{Sn}{115}$ & \underline{ 0.67} &       &       & $\iso{Sn}{113}(\mbox{n},\gamma)\iso{Sn}{114}$& $ $& $ $&1.00, 0.99&\\
     & -0.45 & -0.59 & \underline{-0.66} & $ $& $ $& $\iso{Sn}{115}(\mbox{n},\gamma)\iso{Sn}{116}$&1.00, 1.00&\\
$\iso{Sn}{116}$ & \underline{ 0.81} &       &       & $\iso{In}{115}(\mbox{n},\gamma)\iso{In}{116}$& $ $& $ $&1.00, 1.00&\\
     & -0.27 & \underline{-0.82} &       & $ $& $\iso{Sn}{116}(\mbox{n},\gamma)\iso{Sn}{117}$& $ $&1.00, 1.00&\\
     & -0.15 & -0.42 & \underline{-0.89} & $ $& $ $& $\iso{In}{116}(\mbox{n},\gamma)\iso{In}{117}$&1.00, 1.00&\\
$\iso{Sn}{124}$ & \underline{ 0.96} &       &       & $\iso{Sn}{123}(\mbox{n},\gamma)\iso{Sn}{124}$& $ $& $ $&0.98, 0.96&\\
$\iso{Te}{122}$ & \underline{ 0.74} &       &       & $\iso{Sb}{121}(\mbox{n},\gamma)\iso{Sb}{122}$& $ $& $ $&0.98, 0.93&\\
$\iso{Te}{123}$ & -0.60 & \underline{-0.78} &       & $ $& $\iso{Te}{123}(\mbox{n},\gamma)\iso{Te}{124}$& $ $&1.00, 1.00&\\
$\iso{Te}{124}$ & \underline{ 0.87} &       &       & $\iso{Sb}{121}(\mbox{n},\gamma)\iso{Sb}{122}$& $ $& $ $&0.98, 0.93&\\
     & -0.37 & \underline{-0.71} &       & $ $& $\iso{Te}{124}(\mbox{n},\gamma)\iso{Te}{125}$& $ $&1.00, 1.00&\\
$\iso{Te}{130}$ & \underline{-0.78} &       &       & $\iso{I}{130}(\beta^-)\iso{Xe}{130}$& $ $& $ $& & 1.31, 4.97\\
$\iso{Xe}{128}$ & \underline{ 0.75} &       &       & $\iso{I}{128}(\beta^-)\iso{Xe}{128}$& $ $& $ $& & 1.64, 5.42\\
     &  0.31 & \underline{ 0.66} &       & $ $& $\iso{I}{127}(\mbox{n},\gamma)\iso{I}{128}$& $ $&1.00, 0.99&\\
$\iso{Xe}{130}$ & \underline{ 0.83} &       &       & $\iso{Xe}{129}(\mbox{n},\gamma)\iso{Xe}{130}$& $ $& $ $&0.98, 0.90&\\
     &  0.32 & \underline{ 0.67} &       & $ $& $\iso{I}{127}(\mbox{n},\gamma)\iso{I}{128}$& $ $&1.00, 0.99&\\
     & -0.25 & -0.58 & \underline{-0.79} & $ $& $ $& $\iso{Xe}{130}(\mbox{n},\gamma)\iso{Xe}{131}$&1.00, 1.00&\\

\hline
\end{tabular}
\end{table*}

\begin{table*}
\begin{tabular}{crrrcccccc}
\hline
Nuclide & $r_{{\rm cor},0}$ & $r_{{\rm cor},1}$ & $r_{{\rm cor},2}$& Key rate & Key rate & Key rate& $X_0$ $(8$, $30~{\rm keV})$ & $\beta$-decay \\
&&&& Level 1 & Level 2 & Level 3 & & \\
\hline
$\iso{Xe}{134}$ & \underline{-0.68} &       &       & $\iso{Cs}{134}(\beta^-)\iso{Ba}{134}$& $ $& $ $& & 3.24, 5.52\\
     &  0.39 & \underline{ 0.83} &       & $ $& $\iso{Xe}{133}(\mbox{n},\gamma)\iso{Xe}{134}$& $ $&1.00, 1.00&\\
     & -0.10 & -0.20 & \underline{-0.70} & $ $& $ $& $\iso{Xe}{133}(\beta^-)\iso{Cs}{133}$& & 1.30, 1.30\\
     &  0.08 &  0.15 & \underline{ 0.66} & $ $& $ $& $\iso{Xe}{131}(\mbox{n},\gamma)\iso{Xe}{132}$&1.00, 1.00&\\
$\iso{Xe}{136}$ & \underline{-0.66} &       &       & $\iso{Cs}{134}(\beta^-)\iso{Ba}{134}$& $ $& $ $& & 3.24, 5.52\\
     &  0.39 & \underline{ 0.86} &       & $ $& $\iso{Cs}{134}(\mbox{n},\gamma)\iso{Cs}{135}$& $ $&0.78, 0.68&\\
     & -0.19 & -0.40 & \underline{-0.90} & $ $& $ $& $\iso{Cs}{136}(\beta^-)\iso{Ba}{136}$\\
$\iso{Ba}{134}$ &  0.64 & \underline{ 0.72} &       & $ $& $\iso{Xe}{132}(\mbox{n},\gamma)\iso{Xe}{133}$& $ $&1.00, 1.00&\\
     & -0.38 & -0.40 & \underline{-0.77} & $ $& $ $& $\iso{Ba}{134}(\mbox{n},\gamma)\iso{Ba}{135}$&1.00, 1.00&\\
$\iso{Ba}{136}$ & \underline{-0.82} &       &       & $\iso{Ba}{136}(\mbox{n},\gamma)\iso{Ba}{137}$& $ $& $ $&1.00, 1.00&\\
     &  0.37 &  0.64 & \underline{ 0.71} & $ $& $ $& $\iso{Ba}{134}(\mbox{n},\gamma)\iso{Ba}{135}$&1.00, 1.00&\\
$\iso{La}{138}$ & \underline{-0.94} &       &       & $\iso{La}{138}(\mbox{n},\gamma)\iso{La}{139}$& $ $& $ $&1.00, 1.00&\\
     &  0.14 & \underline{ 0.66} &       & $ $& $\iso{La}{137}(\mbox{n},\gamma)\iso{La}{138}$& $ $&0.87, 0.80&\\
     &  0.07 &  0.32 & \underline{ 0.82} & $ $& $ $& $\iso{Ce}{136}(\mbox{n},\gamma)\iso{Ce}{137}$&1.00, 1.00&\\
$\iso{Ce}{142}$ & \underline{-0.69} &       &       & $\iso{Ce}{141}(\beta^-)\iso{Pr}{141}$& $ $& $ $\\
     &  0.40 & \underline{ 0.93} &       & $ $& $\iso{Ce}{141}(\mbox{n},\gamma)\iso{Ce}{142}$& $ $&1.00, 1.00&\\
$\iso{Nd}{148}$ & \underline{-0.68} &       &       & $\iso{Nd}{147}(\beta^-)\iso{Pm}{147}$& $ $& $ $& & 1.30, 3.03\\
     & \underline{ 0.65} &       &       & $\iso{Nd}{147}(\mbox{n},\gamma)\iso{Nd}{148}$& $ $& $ $&1.00, 0.98&\\
     & -0.05 & \underline{-0.66} &       & $ $& $\iso{Fe}{56}(\mbox{n},\gamma)\iso{Fe}{57}$& $ $&1.00, 1.00&\\
     & -0.02 & \underline{-0.67} &       & $ $& $\iso{Fe}{57}(\mbox{n},\gamma)\iso{Fe}{58}$& $ $&0.73, 0.59&\\
$\iso{Sm}{148}$ &  0.48 & \underline{ 0.65} &       & $ $& $\iso{Pm}{148}(\beta^-)\iso{Sm}{148}$& $ $& & 1.30, 2.77\\
     &  0.35 & \underline{ 0.70} &       & $ $& $\iso{Pm}{147}(\mbox{n},\gamma)\iso{Pm}{148}$& $ $&1.00, 1.00&\\
$\iso{Sm}{150}$ & \underline{ 0.66} &       &       & $\iso{Pm}{148}(\mbox{n},\gamma)\iso{Pm}{149}$& $ $& $ $&1.00, 1.00&\\
     &  0.47 & \underline{ 0.73} &       & $ $& $\iso{Pm}{147}(\mbox{n},\gamma)\iso{Pm}{148}$& $ $&1.00, 1.00&\\
     &  0.07 &  0.13 & \underline{ 0.91} & $ $& $ $& $\iso{Sm}{149}(\mbox{n},\gamma)\iso{Sm}{150}$&0.97, 0.93&\\
$\iso{Sm}{152}$ & \underline{-0.76} &       &       & $\iso{Sm}{152}(\mbox{n},\gamma)\iso{Sm}{153}$& $ $& $ $&1.00, 1.00&\\
     &  0.13 &  0.31 & \underline{ 0.95} & $ $& $ $& $\iso{Sm}{149}(\mbox{n},\gamma)\iso{Sm}{150}$&0.97, 0.93&\\
$\iso{Gd}{152}$ & \underline{ 0.93} &       &       & $\iso{Sm}{151}(\beta^-)\iso{Eu}{151}$& $ $& $ $& & 3.60, 5.42\\
     & -0.31 & \underline{-0.89} &       & $ $& $\iso{Sm}{151}(\mbox{n},\gamma)\iso{Sm}{152}$& $ $&0.80, 0.76&\\
$\iso{Gd}{154}$ & \underline{-0.75} &       &       & $\iso{Gd}{154}(\mbox{n},\gamma)\iso{Gd}{155}$& $ $& $ $&1.00, 1.00&\\
$\iso{Gd}{160}$ & \underline{ 0.81} &       &       & $\iso{Gd}{159}(\mbox{n},\gamma)\iso{Gd}{160}$& $ $& $ $&1.00, 0.97&\\
$\iso{Dy}{160}$ & \underline{-0.85} &       &       & $\iso{Dy}{160}(\mbox{n},\gamma)\iso{Dy}{161}$& $ $& $ $&1.00, 0.99&\\
$\iso{Er}{170}$ & \underline{-0.66} &       &       & $\iso{Er}{169}(\beta^-)\iso{Tm}{169}$& $ $& $ $& & 1.30, 4.46\\
     &  0.55 & \underline{ 0.90} &       & $ $& $\iso{Er}{169}(\mbox{n},\gamma)\iso{Er}{170}$& $ $&1.00, 0.98&\\
     &  0.21 &  0.34 & \underline{ 0.80} & $ $& $ $& $\iso{Er}{168}(\mbox{n},\gamma)\iso{Er}{169}$&1.00, 0.98&\\
$\iso{Yb}{170}$ & \underline{-0.85} &       &       & $\iso{Tm}{170}(\mbox{n},\gamma)\iso{Tm}{171}$& $ $& $ $&0.98, 0.91&\\
$\iso{Yb}{176}$ & \underline{ 0.90} &       &       & $\iso{Yb}{175}(\mbox{n},\gamma)\iso{Yb}{176}$& $ $& $ $&1.00, 1.00&\\
     & -0.26 & \underline{-0.84} &       & $ $& $\iso{Yb}{175}(\beta^-)\iso{Lu}{175}$& $ $& & 1.30, 1.58\\
     & -0.15 & -0.48 & \underline{-0.98} & $ $& $ $& $\iso{Yb}{176}(\mbox{n},\gamma)\iso{Yb}{177}$&1.00, 0.98&\\
$\iso{Lu}{176}$ & \underline{ 0.85} &       &       & $\iso{Yb}{174}(\mbox{n},\gamma)\iso{Yb}{175}$& $ $& $ $&1.00, 0.98&\\
     & -0.37 & \underline{-0.72} &       & $ $& $\iso{Lu}{176}(\beta^-)\iso{Hf}{176}$& $ $& & 1.30, 1.33\\
     &  0.30 &  0.55 & \underline{ 0.83} & $ $& $ $& $\iso{Yb}{172}(\mbox{n},\gamma)\iso{Yb}{173}$&1.00, 0.98&\\
$\iso{W}{186}$ & \underline{-0.83} &       &       & $\iso{W}{185}(\beta^-)\iso{Re}{185}$& $ $& $ $& & 1.44, 3.87\\
     &  0.31 & \underline{ 0.71} &       & $ $& $\iso{W}{185}(\mbox{n},\gamma)\iso{W}{186}$& $ $&0.98, 0.95&\\
$\iso{Re}{187}$ & -0.54 & -0.62 & \underline{-0.68} & $ $& $ $& $\iso{Re}{186}(\beta^-)\iso{Os}{186}$& & 1.30, 3.59\\
$\iso{Os}{186}$ & \underline{ 0.72} &       &       & $\iso{W}{185}(\beta^-)\iso{Re}{185}$& $ $& $ $& & 1.44, 3.87\\
     & -0.43 & \underline{-0.67} &       & $ $& $\iso{Os}{186}(\mbox{n},\gamma)\iso{Os}{187}$& $ $&1.00, 1.00&\\
$\iso{Os}{187}$ & \underline{-0.88} &       &       & $\iso{Os}{187}(\mbox{n},\gamma)\iso{Os}{188}$& $ $& $ $&0.57, 0.46&\\
     &  0.14 &  0.58 & \underline{ 0.67} & $ $& $ $& $\iso{Re}{186}(\beta^-)\iso{Os}{186}$& & 1.30, 3.59\\
$\iso{Os}{192}$ & \underline{ 0.85} &       &       & $\iso{Os}{191}(\mbox{n},\gamma)\iso{Os}{192}$& $ $& $ $&1.00, 1.00&\\
     & -0.44 & \underline{-0.82} &       & $ $& $\iso{Os}{191}(\beta^-)\iso{Ir}{191}$& $ $& & 1.30, 1.76\\
     & -0.17 & -0.35 & \underline{-0.71} & $ $& $ $& $\iso{Os}{192}(\mbox{n},\gamma)\iso{Os}{193}$&1.00, 1.00&\\
$\iso{Pt}{192}$ & \underline{-0.69} &       &       & $\iso{Pt}{192}(\mbox{n},\gamma)\iso{Pt}{193}$& $ $& $ $&1.00, 1.00&\\
     & -0.56 & \underline{-0.81} &       & $ $& $\iso{Ir}{192}(\mbox{n},\gamma)\iso{Ir}{193}$& $ $&0.64, 0.51&\\
     &  0.33 &  0.50 & \underline{ 0.90} & $ $& $ $& $\iso{Ir}{192}(\beta^-)\iso{Pt}{192}$& & 1.31, 6.36\\
$\iso{Pt}{195}$ & \underline{-0.91} &       &       & $\iso{Pt}{195}(\mbox{n},\gamma)\iso{Pt}{196}$& $ $& $ $&1.00, 1.00&\\
     &  0.22 & \underline{ 0.88} &       & $ $& $\iso{Pt}{194}(\mbox{n},\gamma)\iso{Pt}{195}$& $ $&1.00, 1.00&\\
$\iso{Pt}{198}$ & \underline{ 0.91} &       &       & $\iso{Pt}{197}(\mbox{n},\gamma)\iso{Pt}{198}$& $ $& $ $&0.99, 0.94&\\
     & -0.27 & \underline{-0.71} &       & $ $& $\iso{Pt}{197}(\beta^-)\iso{Au}{197}$& $ $& & 1.31, 4.90\\
     & -0.06 & -0.20 & \underline{-1.00} & $ $& $ $& $\iso{Pt}{198}(\mbox{n},\gamma)\iso{Pt}{199}$&1.00, 1.00&\\
$\iso{Hg}{198}$ & \underline{-0.70} &       &       & $\iso{Hg}{198}(\mbox{n},\gamma)\iso{Hg}{199}$& $ $& $ $&1.00, 1.00&\\
     &  0.49 & \underline{ 0.78} &       & $ $& $\iso{Pt}{196}(\mbox{n},\gamma)\iso{Pt}{197}$& $ $&1.00, 1.00&\\
     & -0.11 & -0.17 & \underline{-0.76} & $ $& $ $& $\iso{Au}{198}(\mbox{n},\gamma)\iso{Au}{199}$&1.00, 1.00&\\
$\iso{Pb}{204}$ & \underline{ 0.76} &       &       & $\iso{Tl}{203}(\mbox{n},\gamma)\iso{Tl}{204}$& $ $& $ $&1.00, 1.00&\\
     & -0.47 & \underline{-0.74} &       & $ $& $\iso{Tl}{204}(\mbox{n},\gamma)\iso{Tl}{205}$& $ $&1.00, 1.00&\\
     & -0.38 & -0.59 & \underline{-0.89} & $ $& $ $& $\iso{Pb}{204}(\mbox{n},\gamma)\iso{Pb}{205}$&1.00, 1.00&\\
$\iso{Bi}{209}$ & \underline{ 0.94} &       &       & $\iso{Pb}{208}(\mbox{n},\gamma)\iso{Pb}{209}$& $ $& $ $&1.00, 1.00&\\
     & -0.32 & \underline{-0.91} &       & $ $& $\iso{Bi}{209}(\mbox{n},\gamma)\iso{Bi}{210}$& $ $&1.00, 1.00&\\
\hline
\end{tabular}
\end{table*}

%

\bsp	
\label{lastpage}
\end{document}